\documentstyle[12pt,supertab,rotating]{article}
\input{epsf}
\newlength{\dinwidth}
\newlength{\dinmargin}
\newlength{\savelen}
\begin{document}
\pagestyle{empty}
\thispagestyle{empty}
\vspace{1 cm}
\newcommand{\perc}      {~\mbox{\hspace*{-0.3em}\%}}
\newcommand{\Gev}       {\mbox{${\rm GeV}$}}
\newcommand{\Gevsq}     {\mbox{${\rm GeV}^2$}}
\newcommand{\Fem}       {\mbox{$F_2^{em}$}}
\newcommand{\Fint}      {\mbox{$F_2^{int}$}}
\newcommand{\Fwk}       {\mbox{$F_2^{wk}$}}
\newcommand{\Ft}        {\mbox{$F_2$}}
\newcommand{\Fd}        {\mbox{$F_3$}}
\newcommand{\FL}        {\mbox{$F_{_{L}}$}}
\newcommand{\minpk}     {\mbox{\rule[0.4ex]{1ex}{0.2pt}}}
\newcommand{\syst}[2]   {\mbox{\raisebox{-0.6ex}[0.9em][0.2em]
{$\stackrel{\scriptstyle #1}{\scriptstyle #2}$}}}
\newcommand{\mva}[6]    {{$\!\! {#1}({#2}-{#3})10^{\minpk {#4}}$}&{$\!\! {#5}\!\!$}&{$\!\! {#6}\!\!$}&}

\newcommand{\mvd}[5]    {{$\!\! #1 \pm#2\syst{+#3}{-#4}$}&{$\!\! {#5}\!\!$} \\}
\newcommand{\mvdl}[5]   {{$\!\! #1 \pm#2\syst{+#3}{-#4}$}&{$\!\! {#5}\!\!$} }

\newcommand{\pkd}[6]    {{$\!\! #1 \!\!$}&{$\!\! #2 \pm#3\syst{+#4}{-#5}$}&{$\!\! {#6}\!\!$} \\}

\newcommand{\mvqbin}[3] {\hline \hline \multicolumn{5}{||c||}{$Q^2={#1} \; ({#2} -{ #3}) \;\Gevsq $} \\ \hline } 
\newcommand{\mvqbinhq}[3] {\hline \hline \multicolumn{6}{||c||}{$Q^2={#1} \; ({#2} -{ #3}) \;\Gevsq $} \\ \hline } 
\newcommand{\mvqbinr}[3] {\hline \hline \multicolumn{5}{|c||}{$Q^2={#1} \; ({#2} -{ #3}) \;\Gevsq $} \\ \hline }

\newcommand{\cpt} {\multicolumn{5}{l}{Table \thetable\ (continued): {\it The measured $F_2$ values. }} \\  }
\newcommand{\head}      {\hline {$x(range)$}&{$N_{ev}$}&{$N_{bg}$}&{$F_2\pm stat^{+sys}_{-sys}$}&{$\delta_L\perc$} \\
\hline}

\newcommand{\headhq} {\hline {$x(range)$}&{$N_{ev}$}&{$N_{bg}$}&{\Ft}&{$F_2^{em}\pm stat^{+sys}_{-sys}$}&{$\delta_L\perc$} \\}

\newcommand{\qsd}       {\mbox{${Q^2}$}}
\newcommand{\x}         {\mbox{${\it x}$}}
\newcommand{\y}         {\mbox{${\it y}$}}
\newcommand{\ye}        {\mbox{${y_{e}}$}}
\newcommand{\smallqsd}  {\mbox{${q^2}$}}
\newcommand{\ra}        {\mbox{$ \rightarrow $}}
\newcommand{\ygen}      {\mbox{${y_{gen}}$}}
\newcommand{\yjb}       {\mbox{${y_{_{JB}}}$}}
\newcommand{\yda}       {\mbox{${y_{_{DA}}}$}}
\newcommand{\qda}       {\mbox{${Q^2_{_{DA}}}$}}
\newcommand{\qjb}       {\mbox{${Q^2_{_{JB}}}$}}
\newcommand{\ypt}       {\mbox{${y_{_{PT}}}$}}
\renewcommand{\xpt}     {\mbox{${x_{_{PT}}}$}}
\newcommand{\qpt}       {\mbox{${Q^2_{_{PT}}}$}}
\newcommand{\ypr}       {\mbox{${y_{(1)}}$}}
\newcommand{\yprpr}     {\mbox{${y_{(2)}}$}}
\newcommand{\ypps}      {\mbox{${y_{(2)}^2}$}}
\newcommand{\gammah}    {\mbox{$\gamma_{_{H}}$}}
\newcommand{\gammahc}   {\mbox{$\gamma_{_{PT}}$}}
\newcommand{\gap}       {\hspace{0.5cm}}
\newcommand{\gsim}      {\mbox{\raisebox{-0.4ex}{$\;\stackrel{>}{\scriptstyle \sim}\;$}}}
\newcommand{\lsim}      {\mbox{\raisebox{-0.4ex}{$\;\stackrel{<}{\scriptstyle \sim}\;$}}}
\newcommand{\ptrat}     {\mbox{$\frac{p_{Th}}{p_{Te}}$}}
\newcommand{\yjbrat}    {\mbox{$\frac{y_{_{JB}}}{y_{gen}}$}}
\newcommand{\ydarat}    {\mbox{$\frac{y_{_{DA}}}{y_{gen}}$}}
\newcommand{\yptrat}    {\mbox{$\frac{y_{_{PT}}}{y_{gen}}$}}
\newcommand{\yprrat}    {\mbox{$\frac{y_{(1)}}{y_{gen}}$}}
\newcommand{\yprprrat}  {\mbox{$\frac{y_{(2)}}{y_{gen}}$}}
\newcommand{\yerat}     {\mbox{$\frac{y_{e}}{y_{gen}}$}}
\newcommand{\yptye}      {\mbox{$\frac{y_{_{PT}}}{y_{e}}$}}
\newcommand{\qptrat}    {\mbox{$\frac{Q^2_{PT}}{Q^2_{gen}}$}}
\newcommand{\qdarat}    {\mbox{$\frac{Q^2_{DA}}{Q^2_{gen}}$}}
\newcommand{\qerat}     {\mbox{$\frac{Q^2_{e}}{Q^2_{gen}}$}}
\newcommand{\qprprrat}  {\mbox{$\frac{Q^2_{(2)}}{Q^2_{gen}}$}}
\def\3{\ss}                                                                                        

\def\ctr#1{{\it #1}\\\vspace{10pt}}

\renewcommand{\thefootnote}{\arabic{footnote}}
\date{}
\title {
\bf  Measurement of the \Ft\ structure function in deep inelastic
$e^+p$ scattering using 1994 data from the ZEUS detector at HERA} 
\author{ZEUS Collaboration\\ }
\maketitle
\vspace{5 cm}
\begin{abstract}
We present measurements of the structure function \Ft\  in $e^+p$ scattering
at HERA in the range 
$3.5\;\Gevsq < \qsd < 5000\;\Gevsq$. A new reconstruction method
has allowed a significant improvement in the resolution of the 
kinematic variables and an extension of the kinematic region covered
by the experiment. At $ \qsd < 35 \;\Gevsq$ the range in $x$ now spans
$6.3\cdot 10^{-5} < x < 0.08$ providing overlap  
with measurements
from fixed target experiments. At values of $Q^2$ above 1000 GeV$^2$
the $x$ range extends to 0.5. Systematic errors 
below 5\perc\ have been achieved 
for most of the kinematic region.
The structure function rises as \x\ decreases;
the rise becomes more pronounced as \qsd\ increases. The behaviour of 
the structure function
data is  well 
described by next-to-leading order perturbative QCD as implemented in
the DGLAP evolution equations.
 
\end{abstract}
\thispagestyle{empty}
\newpage
\pagestyle{plain}
\setcounter{page}{1}
\pagenumbering{Roman}                                                                              
\begin{center}
{                      \large  The ZEUS Collaboration              }                               
\end{center}
  M.~Derrick,                                                                                      
  D.~Krakauer,                                                                                     
  S.~Magill,                                                                                       
  D.~Mikunas,                                                                                      
  B.~Musgrave,                                                                                     
  J.R.~Okrasinski,                                                                                 
  J.~Repond,                                                                                       
  R.~Stanek,                                                                                       
  R.L.~Talaga,                                                                                     
  H.~Zhang  \\                                                                                     
 {\it Argonne National Laboratory, Argonne, IL, USA}~$^{p}$
% ...
\newpage                                                                                           
% ...
$^{\    2}$ now at Univ. of Salerno and INFN Napoli, Italy \\                                      
% ...
$^{  20}$ now at ATLAS Collaboration, Univ. of Munich \\                                           
% ...
                                                           %
                                                           %
\newpage   
                                                           %
                                                           %
\begin{tabular}[h]{rp{14cm}}                                                                       
$^{a}$ &  supported by the Natural Sciences and Engineering Research                               
          Council of Canada (NSERC)  \\                                                            
$^{b}$ &  supported by the FCAR of Qu\'ebec, Canada  \\                                            
$^{c}$ &  supported by the German Federal Ministry for Education and                               
          Science, Research and Technology (BMBF), under contract                                  
          numbers 057BN19P, 057FR19P, 057HH19P, 057HH29P, 057SI75I \\                              
$^{d}$ &  supported by the MINERVA Gesellschaft f\"ur Forschung GmbH,                              
          the Israel Academy of Science and the U.S.-Israel Binational                             
          Science Foundation \\                                                                    
$^{e}$ &  supported by the German Israeli Foundation, and                                          
          by the Israel Academy of Science  \\                                                     
$^{f}$ &  supported by the Italian National Institute for Nuclear Physics                          
          (INFN) \\                                                                                
$^{g}$ &  supported by the Japanese Ministry of Education, Science and                             
          Culture (the Monbusho) and its grants for Scientific Research \\                         
$^{h}$ &  supported by the Korean Ministry of Education and Korea Science                          
          and Engineering Foundation  \\                                                           
$^{i}$ &  supported by the Netherlands Foundation for Research on                                  
          Matter (FOM) \\                                                                          
$^{j}$ &  supported by the Polish State Committee for Scientific                                   
          Research, grants No.~115/E-343/SPUB/P03/109/95, 2P03B 244                                
          08p02, p03, p04 and p05, and the Foundation for Polish-German                            
          Collaboration (proj. No. 506/92) \\                                                      
$^{k}$ &  supported by the Polish State Committee for Scientific                                   
          Research (grant No. 2 P03B 083 08) and Foundation for                                    
          Polish-German Collaboration  \\                                                          
$^{l}$ &  partially supported by the German Federal Ministry for                                   
          Education and Science, Research and Technology (BMBF)  \\                                
$^{m}$ &  supported by the German Federal Ministry for Education and                               
          Science, Research and Technology (BMBF), and the Fund of                                 
          Fundamental Research of Russian Ministry of Science and                                  
          Education and by INTAS-Grant No. 93-63 \\                                                
$^{n}$ &  supported by the Spanish Ministry of Education                                           
          and Science through funds provided by CICYT \\                                           
$^{o}$ &  supported by the Particle Physics and                                                    
          Astronomy Research Council \\                                                            
$^{p}$ &  supported by the US Department of Energy \\                                              
$^{q}$ &  supported by the US National Science Foundation \\                                       
\end{tabular}                                                                                      
                                                           %
                                                           %


\pagestyle{plain}

\newpage
\pagenumbering{arabic}                   
\setcounter{page}{1}

\section{\bf Introduction}
\label{s:Intro}

Measurements of neutral current (NC) deep inelastic scattering (DIS)
at HERA~\cite{b:ZEUS_F2,b:H1_F2} have revealed a rapid rise of the 
proton structure function \Ft\ as Bjorken-\x\ decreases below $10^{-2}$.
Extensions of these measurements to low \qsd\ have shown that this 
rise persists down to \qsd\ values 
as low as 1.5~\Gevsq ~\cite{b:LOWQ,b:H11994}.

In this report we present a new measurement of \Ft\ from a
DIS event sample eight times larger than that used in our previous 
analysis~\cite{b:ZEUS_F2}.
 The increase in statistics combined with a new method
that provides a substantially more precise reconstruction of the event 
kinematics has allowed us to extend the range of 
the measurements in \x\ by an order of 
magnitude and
to decrease the systematic uncertainties of \Ft\ by roughly a factor of three. 
 
New data on \Ft\ with similar statistics in a similar kinematic range
have recently  been presented by H1~\cite{b:H11994}.  

\section{\bf Experimental conditions}
\label{s:detector}
The data presented here were taken with the ZEUS detector at HERA in 1994.
HERA operated with 153 colliding bunches of 820~GeV protons and 27.5 GeV 
positrons, with a time between bunches of 96~ns. 
Additional unpaired positron (15) and proton (17) bunches 
circulated,
which are used to determine beam related backgrounds. 
The proton bunch length was approximately 20 cm (r.m.s.) while 
the positron bunch length was negligible in comparison which, together
with run-to-run variations of the mean interaction position, leads to
a length of the interaction region 
of 12 cm (r.m.s.) centered around
$Z$=+6 cm.~\footnote{The ZEUS coordinate system
is defined as right handed with the $Z$ axis pointing in the proton beam
direction, and the $X$ axis horizontal, pointing towards the centre of
HERA.}  
The data of this analysis correspond to a 
luminosity of 2.50$\pm$0.04~~${\rm pb^{-1}}$. 
Approximately 5\% of the proton current was contained 
in satellite bunches, which were shifted by 4.8~ns with respect 
to the primary bunch crossing time, resulting in a fraction of 
the $ep$ interactions occurring at $\langle Z \rangle =+78~{\rm cm}$. 

A description of the ZEUS detector can be 
found in~\cite{b:sigtot_photoprod,b:Detector}.
The components used in this analysis are briefly discussed. 
The uranium-scintillator calorimeter (CAL)~\cite{b:CAL} covers
99.7\% of the total solid angle. It consists of the barrel calorimeter (BCAL)
covering the range $36.7^\circ~<~\theta~<~129.1^\circ$ in polar angle, the forward calorimeter 
(FCAL) covering $2.6^\circ<\theta<36.7^\circ$ and the rear calorimeter (RCAL) covering
 $129.1^\circ<\theta<176.2^\circ$. The FCAL and RCAL are divided in two halves to 
allow retraction during beam injection. 
Each calorimeter part is segmented 
in electromagnetic (EMC) and hadronic (HAC) sections.
Each section is further subdivided into cells of typically 
$5 \times 20$ cm$^2$ ($10 \times 20$~cm$^2$ 
in the RCAL) for the EMC  
and $20 \times 20$~cm$^2$ for the HAC sections.
Each cell is viewed by two photomultipliers (PMTs).
Under test beam conditions the calorimeter has an energy resolution of 
$\sigma/E$~=~18\%/$\sqrt{E ({\rm GeV})}$ 
for electrons and 
$\sigma/E$~=~35\%/$\sqrt {E({\rm GeV})}$ for hadrons.
The timing resolution of a calorimeter cell is less  than 1 ns for energy
deposits greater than 4.5~\Gev .  
In order to minimise the effects of noise due to the uranium radioactivity
 on the measurements
all EMC(HAC) cells with an energy deposit of less than 60(110)~MeV 
are discarded from the
analysis. For cells with isolated energy deposits
 this cut was increased to 100(150)~MeV. 

The tracking system consists of a vertex detector (VXD)~\cite{b:VXD}, 
a central tracking chamber (CTD)~\cite{b:CTD}, and a rear tracking 
detector (RTD)~\cite{b:Detector}  
enclosed in a 1.43 T solenoidal magnetic field. The interaction vertex
is measured with a typical resolution along (transverse to) the beam direction 
of 0.4~(0.1)~cm. For high momentum  tracks ($p>5\;\Gev$)
the extrapolated position on the inner face of the 
calorimeters is known  with a typical resolution of 0.3~cm. 

The position of positrons scattered at small angles 
to the positron beam direction
is measured using the small angle rear tracking detector (SRTD)
 which is attached to the front face of the RCAL.
The SRTD consists of two planes of scintillator strips, 1 cm wide and 
0.5 cm thick, arranged in orthogonal directions and read out 
via optical fibers and photo-multiplier tubes. 
It covers the region of 68~$\times$~68~${\rm cm^2}$ in $X$  
and $Y$ and is positioned at $Z = -148$~cm.
A hole of 20~$\times$~20~${\rm cm^2}$ at the center of the RCAL and SRTD
accommodates the beampipe. 
The SRTD signals resolve single minimum ionizing particles 
and provide a position resolution of 0.3~cm.
The time resolution is less than 2 ns for a minimum ionizing particle. 

The luminosity is measured via the positron-proton 
brems\-strahlung process, $ep \rightarrow e \gamma p$,
using a lead-scintillator calorimeter (LUMI)~\cite{b:LUMI}
which accepts photons at angles~$\le$~0.5~mrad with respect to the 
positron beam direction. 
 The LUMI photon calorimeter is also used to tag 
photons from initial state radiation in DIS events.
It is positioned at $Z = -107$~m and 
has an energy resolution of $\sigma/E$ = 18\%/$\sqrt {E ({\rm GeV})}$
 under test beam conditions.
In its operating position, however, it is shielded from synchrotron radiation 
by a carbon-lead filter and has an energy resolution of 
$\sigma/E$ = 26.5\%/$\sqrt {E ({\rm GeV})}$, as determined
from bremsstrahlung data. The position resolution is 0.2~cm in $X$ and $Y$. 
In addition, an electromagnetic calorimeter positioned at
$Z = -35$~m is used for tagging positrons scattered at small angles.

\subsection{Triggering}

Events were filtered online by a three level trigger system~\cite{b:Detector}.
At the first level, the events are selected by the logical OR of the
following conditions (further details can be found in~\cite{b:CALFLT}):

\begin{itemize}
 \item {Total EMC energy deposit in the BCAL is greater than 4.8 GeV.}
 \item {Total EMC energy deposit in the RCAL, excluding
        the region closest to the rear beampipe, is greater than 3.4 GeV.}
 \item {In the RCAL, the isolated positron condition (ISO-e) is fulfilled.
        The ISO-e condition requires that the isolated EMC energy deposit
        be greater than 2.5 GeV and that the corresponding HAC energy
        be less than 0.95 GeV or no more than a third of the EMC energy.
        The above condition is ANDed with the requirement that the total
        energy deposit in  RCAL EMC is greater than 3.8 GeV.}
\end{itemize}
Events where the positron is scattered in the direction of 
the FCAL are triggered efficiently, by the hadronic final state,
with the above requirements.

Backgrounds from protons interacting outside the detector were rejected
using the time measurement of the energy deposits in upstream veto counters
and the SRTD.  The average trigger efficiency for events that pass
the off-line selection cuts (see Sect.~\ref{ANA}) 
is above 99\%, as determined from
independent triggers and MC simulation. 

In the second level trigger (SLT), background was further reduced 
using the measured times of energy deposits and the summed energies 
from the calorimeter. The events were accepted if:
\begin{equation}
  \delta_{SLT} \equiv \sum_i E_i(1-\cos\theta_i) > (24 - 2E_{\gamma})\gap 
(\Gev )
\end{equation}
where $E_i$ and $\theta_i$ are the energies and polar angles (with respect
to $X=Y=Z=0$ cm) of calorimeter cells, and $E_{\gamma}$
is the energy deposit measured in the LUMI photon calorimeter.
For perfect detector resolution, 
$\delta_{SLT}$ is twice the positron beam energy (55~\Gev ) 
for fully contained DIS events while for photoproduction events,
where the scattered positron escapes down the beampipe,
$\delta_{SLT}$ peaks at much lower values. Proton beam-gas events 
which originate from inside the detector have
energy flows which are concentrated
in the forward direction and so also have small values of $\delta_{SLT}$.

The full event information was available at the third level trigger (TLT).
Tighter timing cuts as well as algorithms to remove beam-halo muons 
and cosmic muons were applied.
The quantity $\delta_{TLT}$ was determined in the same way as 
$\delta_{SLT}$. The events were required to have
\mbox{$\delta_{TLT}>(25 - 2E_{\gamma})\;(\Gev )$}.
Finally, events were accepted if
a scattered positron candidate of energy greater than 4~GeV was found.
In total 900853 NC DIS candidates satisfied the above trigger conditions.

\section{Monte Carlo simulation}
\label{section:MC}

Monte Carlo (MC) event simulation is used to correct for detector
acceptance and smearing effects. The detector simulation is based on 
the GEANT program~\cite{b:GEANT} and incorporates our understanding
of the detector, the trigger and test beam results.
Neutral current DIS events are simulated  using
the HERACLES program~\cite{b:HCL} which includes photon and $Z^0$ exchanges
and first order 
electroweak radiative corrections.
The hadronic final state is simulated using the
colour-dipole model CDMBGF~\cite{b:CDM} including all leading order QCD 
diagrams as 
implemented in ARIADNE~\cite{b:ARIADNE} for the QCD cascade and 
JETSET~\cite{b:JETSET} for the hadronisation.
The ARIADNE model provides the best description of the observed 
DIS non-diffractive hadronic final state~\cite{b:HAD_PAPERS}. 
Diffractive events which have been observed in the 
data~\cite{b:LRG,b:H1LRG} by the occurrence of a large 
rapidity gap in the detector
are simulated within ARIADNE by assuming that the 
struck quark belongs to a colourless state having only a small 
fraction of the proton's momentum. The parameters of the model are 
adjusted to be consistent with recent ZEUS measurements~\cite{b:F2D}.
The MRSA~\cite{b:MRSA} parton density parameterisations, modified at 
low $Q^2$ as described in~\cite{b:modifiedMRSA}, are used. For the final
acceptance corrections, the events are reweighted with the help of
a NLO QCD fit 
to the data in an iterative procedure as 
described in Sect.~\ref{section:rew}. 
For systematic checks, a sample of events was also generated using
the LEPTO Matrix Element and Parton Shower (MEPS)~\cite{b:LEPTO} final state
simulation.

The shape of the vertex distribution used in the simulation
is taken from non-diffractive photoproduction events, since for these 
events the vertex reconstruction efficiency is found to be 
both high~($\sim 90$\perc ) and  
independent of the $Z$ position of the interaction.

The effects of the uranium radioactivity were simulated 
using distributions obtained from  randomly triggered events. 

A MC event sample corresponding to an integrated luminosity of 2.5$~{\rm pb^{-1}}$
for $\qsd > 1.8~\Gevsq$ was generated. This sample was supplemented by an
additional sample equivalent to 2.5$~{\rm pb^{-1}}$ with  $\qsd > 20~\Gevsq$.

The main source of background in the data 
comes from the few photoproduction interactions which lead
 to the detection of a
fake scattered positron. `Minimum bias' photoproduction
events were simulated using PYTHIA~\cite{b:PYTHIA} with cross
sections according to the ALLM parameterisation~\cite{b:ALLM}. 
Photoproduction events corresponding to an integrated luminosity of 
250 ${\rm nb^{-1}}$ were generated with a photon-proton center-of-mass energy
$W\,\lower 2pt \hbox{$\stackrel{>}{\scriptstyle\sim}$ }\!190\,{\rm GeV}$.
Events with smaller $W$ values do not contribute to the photoproduction
background due to the requirement on $\delta$ (see Sect.~6).

\section{Energy and angle measurements}
\label{section:CALIB}

In the determination of the DIS kinematics, the CAL energy deposits
are separated into those associated with  
the identified scattered positron, and
all other energy deposits.
The latter is defined as energy from the
hadronic system, or hadronic energy.
The kinematics of the event is then determined from: 
\begin{itemize}
\item The energy ($E_e^{\prime}$) and polar angle ($\theta_e$) of the 
scattered positron.
\item The hadronic energy in terms of 
\begin{equation} 
\delta_h \equiv \sum_h (E_h - p_{Zh})  \label{DH}
\end{equation} 
and
\begin{equation} 
p_{Th} \equiv \sqrt{(\sum_h p_{Xh})^2 +(\sum_h p_{Yh})^2}
\end{equation} 
where the sums run over all CAL energy deposits not associated with the 
scattered positron.

\end{itemize}
The hadronic energy flow is characterised
by the angle \gammah\  defined by:
\begin{eqnarray}
\cos \gammah &=& \frac{p_{Th}^2 - \delta_h^2}{p_{Th}^2 + \delta_h^2} .
  \label{COSG}
\end{eqnarray}
In the naive quark parton model picture of DIS
the angle \gammah\ corresponds to the polar angle of the struck quark.

\subsection{Positron identification and efficiency}\label{s:PIE}
The positron identification algorithm (SINISTRA94) is based on a neural network
using information from the CAL~\cite{b:NN}.
The network separates deposits in the calorimeter which are due 
to electromagnetically showering particles from those which are of hadronic
origin, by assigning a `positron-probability'. A cut on this probability
allows a clean identification of the scattered positron.\footnote{ Contrary 
to the previous analysis~\cite{b:LOWQ}, in this 
analysis we allow the cut on the probability to vary with the calorimeter
energy associated with the positron.}
 The efficiency for finding the scattered positron
was determined from MC simulations to increase from 80\% at 10~\Gev\ to
greater than 99\perc\ for energies above 15~\Gev . 
The efficiency was checked by using QED-Compton events and by 
comparing results from this algorithm 
with other positron finding algorithms~\cite{b:ZEUS_F2} and found to be
consistent with expectations.

\subsection{Positron angle measurement}

The scattering angle of the positron is determined either by measurement
of its impact position on the CAL inner face together with
the event vertex, or, 
for $\theta_e<135^\circ$, from the 
parameters of a reconstructed track matched with 
the positron impact position.

\begin{figure}[p]
\begin{center}
\setlength{\savelen}{1.0\unitlength}
\setlength{\unitlength}{1cm}
\begin{picture}(16,16)
\put(4,4){e)}
\put(12,4){f)}
\put(6,14){a)}
\put(6,11){b)}
\put(14,14){c)}
\put(14,11){d)}
\put(6.5,16){\large {\bf ZEUS 1994}}
\put(0,0){\mbox{\epsfxsize=7.5cm\epsffile{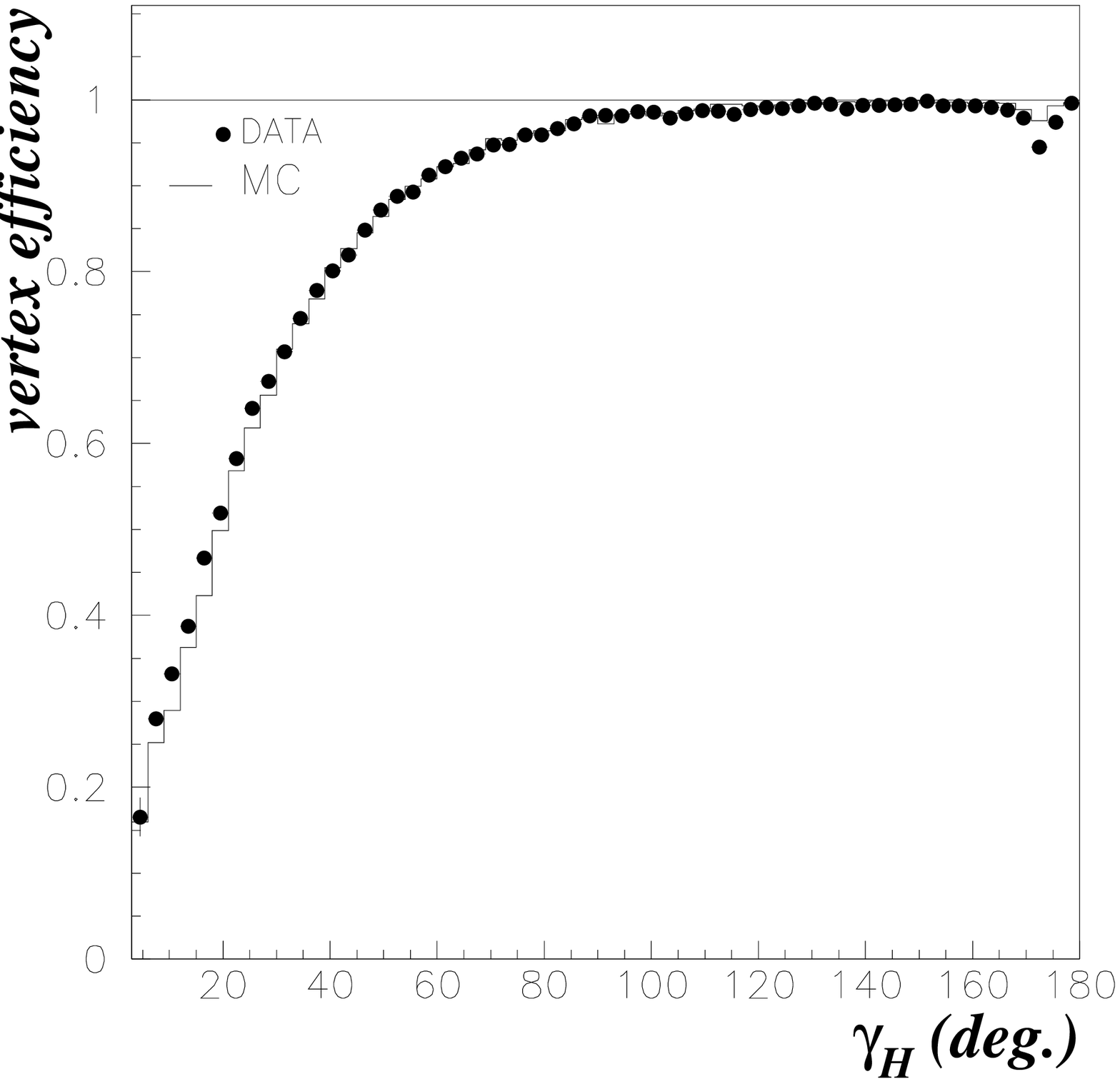}}}
\put(8,0){\mbox{\epsfxsize=7.5cm\epsffile{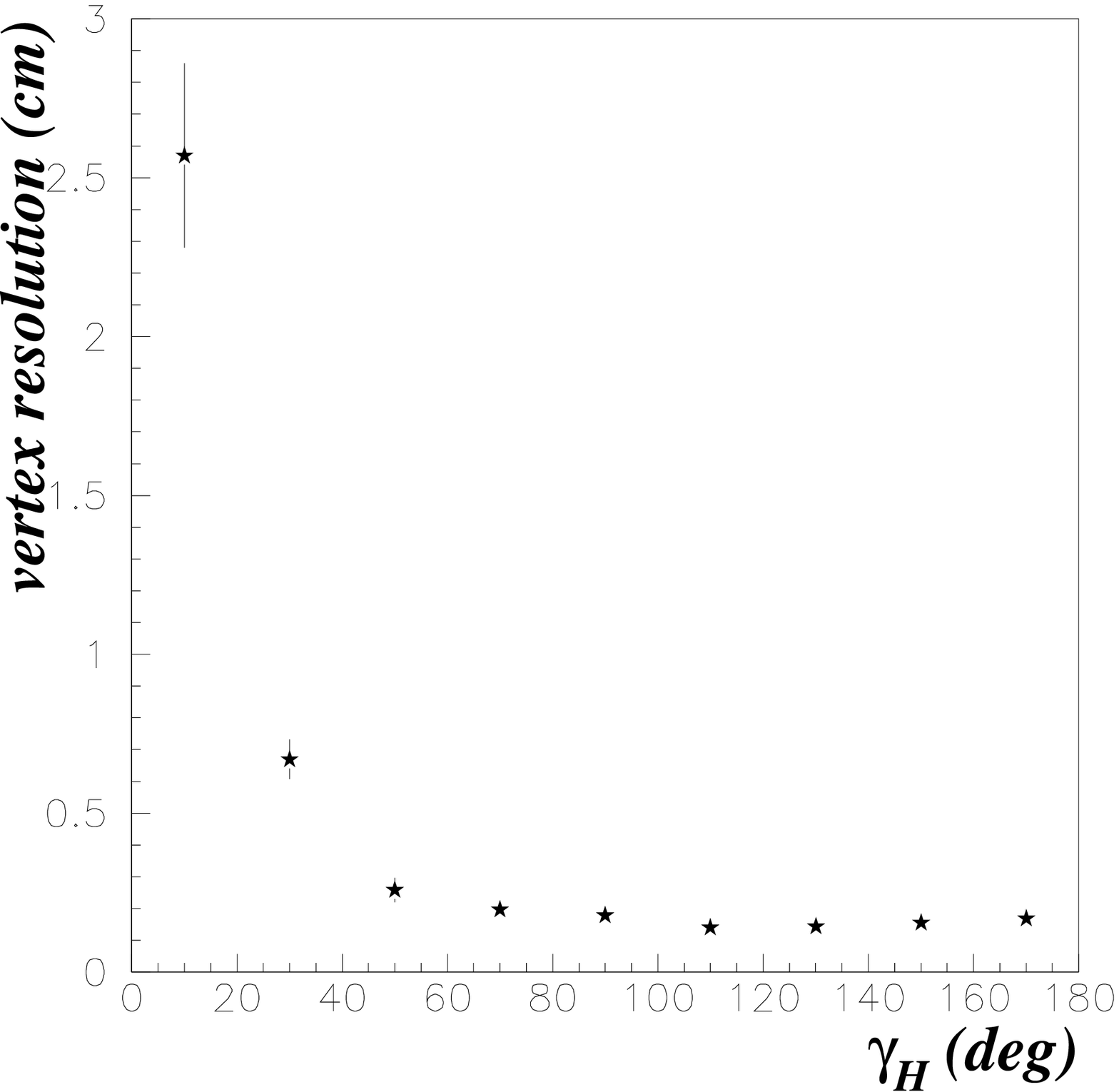}}}
\put(8,8){\mbox{\epsfxsize=7.5cm\epsffile{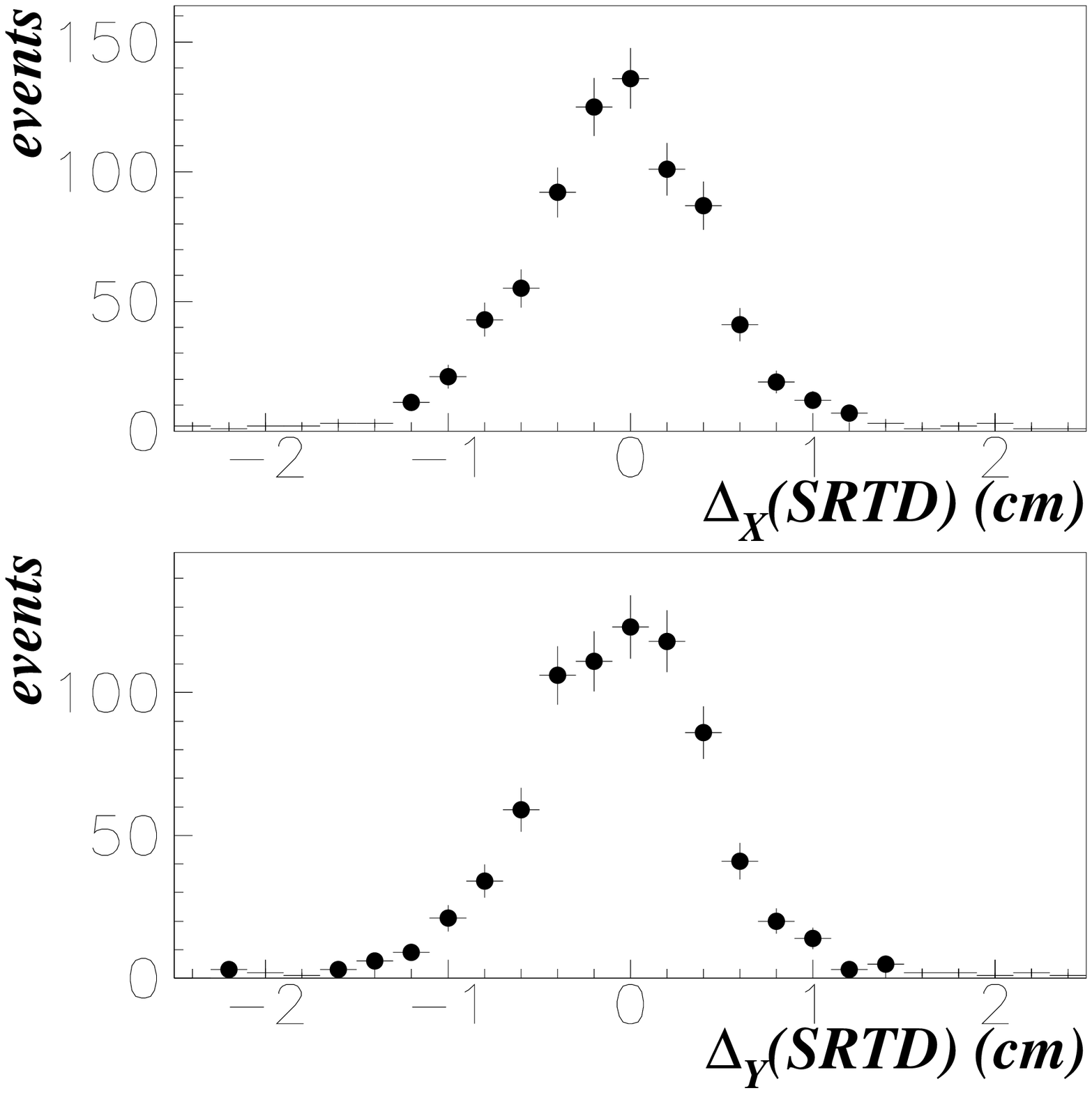}}}
\put(0,8){\mbox{\epsfxsize=7.5cm\epsffile{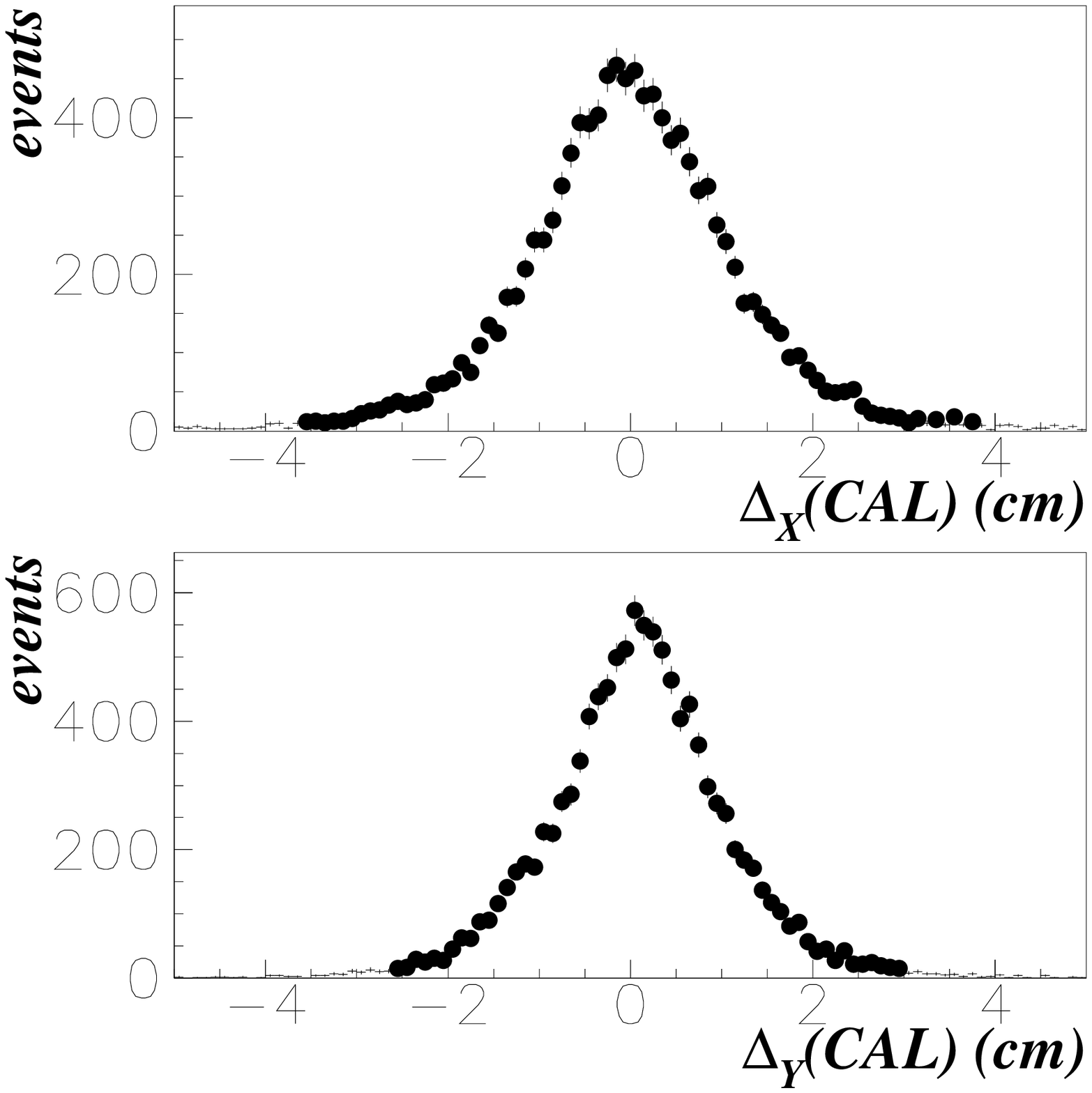}}}
\end{picture}
\setlength{\unitlength}{1\savelen}

\end{center}
\caption[]
{\it 
Distance (in cm) between track and RCAL position a) in X, b) and Y.
Distance (in cm) between track and SRTD position c) in X, d) and Y.
e) Vertex reconstruction 
efficiency as a function of $\gammah$ (points: data; histogram:
MC). f) Vertex 
resolution along the beam (in cm) as a function of $\gammah$.

}
\label{f:verteff}
\end{figure}

The impact position of the scattered positron at the calorimeter is  
reconstructed using  the CAL or the SRTD. When
both CAL and SRTD information are available, the SRTD reconstruction
is used.
\begin{itemize}
\item CAL -- The position of a shower in the CAL can be determined using
the pulse-height sharing between left and right PMTs and between neighbouring
cells. The position resolution on the face of the CAL is 
about 1~cm.  
The position measurement was checked 
by matching tracks and positron candidates.
The resolution of the track-CAL matching is found to be 
 1.1~cm and systematic biases 
are less than 2~mm (see Fig.~\ref{f:verteff}a and b).

\item SRTD -- The position in the SRTD is determined from the centres of
gravity of the pulse-heights of the strips in the $X$ and $Y$ planes of the SRTD. 
The measured SRTD position resolution is 0.3~cm.  
For a subset
of these positrons the position measured by the SRTD
can be compared to the result from the track measurement. 
The resolution of the track-SRTD matching
is measured to be 0.5~cm and systematic deviations are less than 0.1~cm 
(see Fig.~\ref{f:verteff}c and d). 
\end{itemize} 

The coordinates of the event vertex are determined from tracks reconstructed 
with the CTD and VXD. The $Z$ coordinate ($Z_{vertex}$) is determined on an event-by-event
basis. Since the transverse sizes of the beams 
are smaller than the resolutions for the $X$ and $Y$ coordinates of the 
vertex, the beam $X$ and $Y$ positions averaged over a single beam-fill
are used. For events which do not have a tracking vertex, the $Z$ 
coordinate is set to the nominal position of the interaction point ($Z = +6$ cm).

The probability of finding a vertex depends on the 
number of particles which traverse the tracking detectors and thus mainly 
on the hadronic angle \gammah . Figure~\ref{f:verteff}e, which shows the 
fraction of events having a reconstructed vertex as a function of \gammah ,
exhibits
the expected behaviour. For $\gammah > 70^\circ$ the vertex finding efficiency 
is greater than 95\perc. It decreases to around 50\perc\ for events
with $\gammah \sim 20^\circ$. The MC simulation reproduces the observed
behaviour reasonably well. Relative differences of the 
order of 5-10\perc\ are observed in 
the region below $\gammah = 30^\circ$. 
The drop in vertex finding efficiency in the region around $\gammah = 174^\circ$
results from the combination of the characteristics of the DIS final states
and the acceptance of the CTD in the rear direction. 
The effect is not fully reproduced by the MC simulation. 
All differences between data and MC simulation 
are taken into account in the determination
of the systematic uncertainties on the final results. 

 Figure~\ref{f:verteff}f shows the $Z$ resolution of the 
vertex reconstruction as a function of \gammah . For $\gammah~>~40^\circ$ the 
resolution is $\sigma^{track}_{_Z} = 0.2$~cm, while for events with \gammah\
close to zero~\footnote{The lines
of constant \gammah\ in the $(x,\qsd )$-plane are given
  in Fig.~\ref{f:KINPLANE}b for
$\gammah = 40^\circ,90^\circ$ and $135^\circ$.}
it becomes \linebreak 
$\sigma^{track}_{_Z}~=~2.5\pm~0.4$~cm. For events where no vertex is 
reconstructed, the effective resolution is determined by the length of the 
interaction region, which is $\sigma^{beam}_{_Z} = 12$~cm.
It is also possible to determine the 
$Z$ coordinate of the vertex from the measurements of the arrival times of
 energy deposits in the FCAL~\cite{b:ZEUS_F2}. 
In this measurement the resolution is $\sigma^{timing}_{_Z} =17$~cm,
and it is only used for studies of systematic errors.

For polar angles $45^\circ<\theta_e<135^\circ$ and positron transverse momenta
$p_{Te}>5\;\Gev$ the tracking
efficiency is greater than 98\perc. Thus when the positron angle, as determined
by the CAL, is smaller than $135^\circ$ a track is required to match the 
 positron identified in the CAL and the polar angle is taken from the track 
parameters. 
A successful match is obtained when the distance between the extrapolated
impact point of the track on the face of the CAL and the position determined by the CAL
is less than 5~cm. In this region the matching efficiency is above 97\perc\ 
and the combined efficiency for tracking and matching 
is greater than 95\perc .

The resolution of  the scattering angle measurement is 
$\sigma^{SRTD}_{\theta_e} = 2.0$~mrad for positrons reconstructed in the SRTD, 
$\sigma^{CAL}_{\theta_e} \approx 6.8~$mrad for positrons 
determined by the CAL and 
$\sigma^{track}_{\theta_e} \approx 3.4~$mrad 
for positrons with a matched track.
For events without a tracking vertex there is an additional error of
 $\sigma_{\theta_e}^{\overline{vertex}} \approx 80$~($\sin\theta_e$)~mrad.

\subsection{Positron energy determination}
The scattered positron loses energy in the passive material
in front of the CAL.  In the region relevant for this 
analysis this passive material constitutes about
1.5 radiation lengths except in areas around the rear
beampipe, $\theta \gsim 170^\circ$, and the solenoid support structure, 
$130^\circ \lsim \theta \lsim 145^\circ$, where it is up 
to 2.5 radiation lengths.
In the analysis, the measurement of the scattered positron energy
by the CAL is corrected for the energy loss in the passive material.

The correction for the CAL measurement of the scattered positrons
can be determined directly from the data using the following subsamples.

\begin{itemize}

 \item{At low $y$, the scattered positron energy is kinematically constrained
to be close to the
positron beam energy and to be primarily a function of the scattering angle.
For these events, called Kinematic Peak (KP) events, 
the mean positron energy is determined from the scattering angle 
($E_e^{\prime} \approx \frac{2E_e}{(1-\cos \theta_e)}$) to within
0.5\perc.  The KP events are selected by requiring $\yjb = \frac{\delta_h}{2 E_e} <0.04$,
and provide a calibration at $E^{\prime}_e \approx E_e = 27.5$ GeV for $\theta \gsim 135^\circ$.}

\item{For QED Compton events 
($ep \rightarrow e\gamma p$) observed in the main detector, energies of 
the positron and the photon can be predicted
precisely from the measurement of their scattering angles since the
transverse momentum of the scattered proton is small.
QED Compton events provide a calibration at 
$5 \lsim E_e^{\prime} \lsim 20$ GeV and
$\theta \gsim 160^\circ$.}

\item{In events from DIS $\rho^{\rm o}$ production,
$ep \rightarrow ep\rho^{\rm o}$~($\rho^{\rm o} \rightarrow \pi^+ \pi^-$), 
the angle of
the positron and the momenta of the 
$\pi^+$ and $\pi^-$, as measured with the CTD, give a precise determination
of the positron energy. 
DIS $\rho^{\rm o}$ events provide a calibration at $20 \lsim E^{\prime}_e \lsim 25$ GeV 
and $\theta \gsim 160^\circ$.}

\item{In the range  $30^\circ \lsim \theta \lsim 150^\circ$, 
momenta of the positrons
can be determined by the CTD.  Although the CAL energy resolution at
$E_e^{\prime} >10$ GeV is superior to the momentum resolution of the CTD, the
track momenta averaged over several events
give an independent check on the energy measurement of
the CAL.   }
\end{itemize}

In the fiducial volume of the SRTD, $\theta \gsim 167^\circ$, 
the correlation between the energy lost in the passive material in front
of the calorimeter and the energy deposited in the
SRTD is used to correct the calorimeter energy measurement (SRTD method). 
The corrections are determined using the QED Compton and the KP 
samples.  The DIS $\rho^{\rm o}$ events are used as checks on the
correction.  This procedure is described in detail elsewhere~\cite{b:LOWQ}.

In the RCAL outside the SRTD fiducial volume, $130^\circ \lsim \theta \lsim 
167^\circ$,
the observed energy shift of  KP positrons from the value expected
from kinematics, combined with  test beam results of the 
energy loss of electrons
in passive material, are used to determine the correction (KP method).  
This procedure is described in detail elsewhere~\cite{b:ZEUS_F2}.
Using the data from regions where 
both the KP method and the SRTD method can be applied,
the uncertainty in the KP method of correction is found to
be the same as that of the SRTD method.

The uncertainty in the energy determination in the RCAL after these corrections
is 2\perc\ at 10 GeV linearly decreasing to 1\perc\ at 27.5 GeV.

In the region covered by BCAL and FCAL, $\theta \lsim 130^\circ$, 
comparison with the measured
mean CTD momentum  
is used to correct the CAL energy measurement
(track method).  The energies corrected by the track method have
an uncertainty of 2\perc\ 
independent of the energy of the positron.  Where the KP method and the track method
corrections can both be applied, consistent results are found.

The resolution of the positron energy measurement can be
determined from \linebreak 
the data using QED Compton,  DIS $\rho^{\rm o}$, and 
KP events, and is found to be in the range \linebreak
$\sigma/E=(20-27)\%/\sqrt{E~{\rm(GeV)}}$,
depending on the trajectory of the positron through
the passive material in front of the CAL.

Studies identical to those  described above for the data were also performed
with MC event samples. The resulting corrections were applied to 
the DIS MC sample. 
Thus, any difference
in the CAL response between MC simulation and data is absorbed in
the corrections.
The resolution of the positron energy measurements in the MC simulation
is adjusted to reproduce the results found from the data.

The separate positron energy corrections for MC simulation and data
allow the determination of relative
calibrations for the CAL in MC simulation and data.
Before determining the hadronic energy,
adjustments, based on the difference in the magnitude, position 
dependence and energy
dependence of MC and data corrections,  
are made to bring the energy scale for 
data and MC simulation into agreement.

\subsection{Hadronic energy determination}

The hadronic energies, unlike the case of the positron energy, are not
corrected for energy loss in the passive material.
Instead,   the transverse momentum of the positron, $p_{Te}$, 
calculated using the positron energy
corrected as described in the last section, is compared to the
$p_{Th}$ of the hadrons in both the detector simulation and data.
From this comparison, uncertainties in the determination of the
hadronic energy are estimated.
The mean \ptrat\ as 
a function of \gammah\ agrees within 3\perc\  between
MC simulation and data
for the entire range of kinematics covered in this paper.
The \ptrat\ distributions in bins of \gammah\ are compared
(see Sect.~\ref{ANA}) for MC 
simulation and data, and are 
in good agreement.  An uncertainty of $\pm$3\perc\ is assigned
to the hadronic energy measurement, based on these comparisons.

\section{\bf Kinematic reconstruction}
\label{KINREC}

In deep inelastic scattering,
$e(k) + p(P) \rightarrow e (k^\prime) + X$,
the proton structure functions are expressed in terms 
of the negative of the four-momentum transfer squared, $Q^2$,
and Bjorken $x$.
In the absence of QED radiation,
\begin{equation}
\label{e:q2}
Q^2  =  - q^2 =  -(k-k^\prime)^2,
\end{equation}
\begin{equation}
\label{e:x} 
x    =  \frac{Q^2}{2P \cdot q} , 
\end{equation}
where $k$ and $P$  are the four-momenta of the 
incoming particles and $k^\prime$ is the four-momentum of the scattered 
lepton.
The fractional energy transferred to the proton
in its rest frame is $y = Q^2/(sx)$ where $s$ is the square
of the total center of mass energy of the lepton-proton collision
($s=90200$ ${\rm GeV}^2$).


The ZEUS detector measures both the scattered positron
and the hadronic system.  The four independent measured 
quantities $E_e^{\prime}$, $\theta_e$, $\delta_h$ and $p_{Th}$,
as described in the previous section, over-constrain the kinematic
variables $x$ and $Q^2$ (or equivalently, $y$ and $Q^2$).

In order to optimise the reconstruction of the kinematic variables,
both the resolution and 
robustness against possible systematic shifts (stability) of each
measured quantity must be considered.  

For the present analysis, a new method (PT) is used to reconstruct
the kinematic variables.  The PT method achieves both superior
resolution and stability in $x$ and $Q^2$ in the full kinematic
range covered, in comparison with reconstruction methods used in
our previous structure function measurements.

\subsection{Characteristics of standard reconstruction methods}

As discussed in Sect.~\ref{section:CALIB}, the positron variables, $E_e^{\prime}$ and
$\theta_e$ are measured with high precision, and the systematic
uncertainties are small.  The kinematic variables calculated from these
quantities are given by:
\begin{eqnarray}
  \ye &=& 1-\frac{E_e^{\prime}}{2E_e}(1-\cos\theta_e),  \label{YEL} \\
  Q^2_e &=& 2E_e E_e^{\prime}(1+\cos\theta_e).
\end{eqnarray}

This method of reconstruction, the electron (EL) method, 
gives good results at high $y$,
where $E_e^{\prime}$ is significantly different from the positron 
beam energy $E_e$, but at low $y$
($y \lsim 0.1$) both the resolution and stability of $\ye$ become
poor.

The Jacquet-Blondel (JB) method~\cite{b:JB} of kinematic 
reconstruction only uses information from the hadronic energy flow of
the event:
\begin{eqnarray}
   \yjb &=& \frac {\delta_h}{2E_e}, \label{YJB} \\
   \qjb &=& \frac {p_{Th}^2}{1-\yjb}.
\end{eqnarray}
$Q^2_e$ has better resolution than
$\qjb$ over the entire kinematic range while
at low $y$ ($y\lsim 0.04$), $\yjb$ has superior resolution in comparison
with $\ye$.  However, unlike
the case of the positron measurement, the energy loss of
the hadronic system in the passive material in
front of the CAL and the loss through the beam holes 
have to be determined using a simulation of the DIS final
states and of the detector effects.
This introduces substantial systematic
uncertainties.
 
The Double-Angle (DA) method~\cite{b:BKE} 
combines the information from the scattered lepton        
with that from the produced hadronic system:
\begin{eqnarray}
\qda &=& 4E_e^2 \frac{\sin \gammah(1+\cos \theta_e)}
{\sin \gammah + \sin \theta_e - 
\sin (\gammah + \theta_e)},  \label{QDA} \\
\yda &=& \frac {\sin \theta_e(1-\cos \gammah )}
{\sin \gammah + \sin \theta_e - 
\sin (\gammah + \theta_e)}. \label{YDA}
\end{eqnarray}
In this method the hadronic measurement enters the kinematic
reconstruction through the variable 
$\gammah$ (see eq.~\ref{COSG}), which depends on
the ratio of the measured quantities $\delta_h$ and $p_{Th}$. 
As a result, uncertainties in the hadronic energy measurement
tend to cancel leading
to a good stability of the reconstructed kinematic variables.
At the highest $y$, however, $\ye$ has better resolution than $\yda$.
At low $y$ ($\lsim 0.04$), and low $Q^2$, 
the noise in the CAL, which affects 
both $p_{Th}$ and $\delta_h$ has a large effect on $\qda$ and
$\yda$.  This region
was avoided in our
previous analyses, which used the DA method, by imposing
the cut $\yjb>0.04$.

\subsection{PT method}

The PT method provides an improved measurement of $y$ and $Q^2$ by
an efficacious combination of  
the information from the measurements of both the hadronic system and the 
positron.

This is done in two steps. 
 In the first step, the transverse momentum balance between
the positron and the hadron system is used event by event to
correct the hadronic energy
measurement of $y$ by using a functional form derived from the MC
simulation. 
In the second step, the advantages of the EL, JB
and DA methods are combined into a single reconstruction
method for the full kinematic range.

\subsubsection{The measurement of $y$ at low $y$}

In an ideal detector, transverse momentum conservation 
requires $p_{Th}=p_{Te}$ up to the negligible
transverse momentum carried by QED initial state radiation.  
Since the positron is measured in the calorimeter, energy carried by
QED final state radiation is mostly included in the measured positron
energy. 
Since $p_{Te}$
is a precisely  measured quantity (see above), 
$p_{Th}-p_{Te}$ gives, for each event, the difference between
the measured and the true hadronic transverse momentum.
The difference can result from hadronic energy loss through
the beam hole and/or in the passive material in front of CAL, as well
as from the finite energy resolution of CAL.
In first approximation the hadronic energy mismeasurement affects 
both $p_{Th}$ and $\delta_h$ in the same way.
Indeed, if the hadronic system consisted of the proton remnant carrying
no transverse momentum and a well collimated current jet, then an accurate
measurement of $y$ would be obtained by:
\begin{equation}
\ypr = \yjb/(\ptrat).
\label{SIMPLE}
\end{equation}  
In this idealised case, a precise measurement of $y$ is achieved
by determining the hadronic energy correction directly from the
data (on an event by event basis) rather than by a MC simulation.

In standard DIS events, in addition to a current jet, there
is hadronic energy flow in the region 
between the current jet and the direction of the 
proton remnant~\cite{b:HAD_PAPERS}.
As a result, the simple relationship given by eq.~\ref{SIMPLE} no longer
holds and is replaced by:
\begin{equation}
\ypr =
\yjb /{\cal C}  \label{CORF}
\end{equation} 
where
$\cal{C}$ is a correction function.  In the PT method, $\cal{C}$ 
is determined from
MC simulation of final states convoluted with the detector simulation,
as a function of the measured quantities $\ptrat$, $p_{Th}$ and $\gammah$  
giving $\cal{C} \it (\ptrat,p_{Th},\gammah)$.

\begin{figure}[p]
\begin{center}
\mbox{\epsfxsize=14cm\epsffile{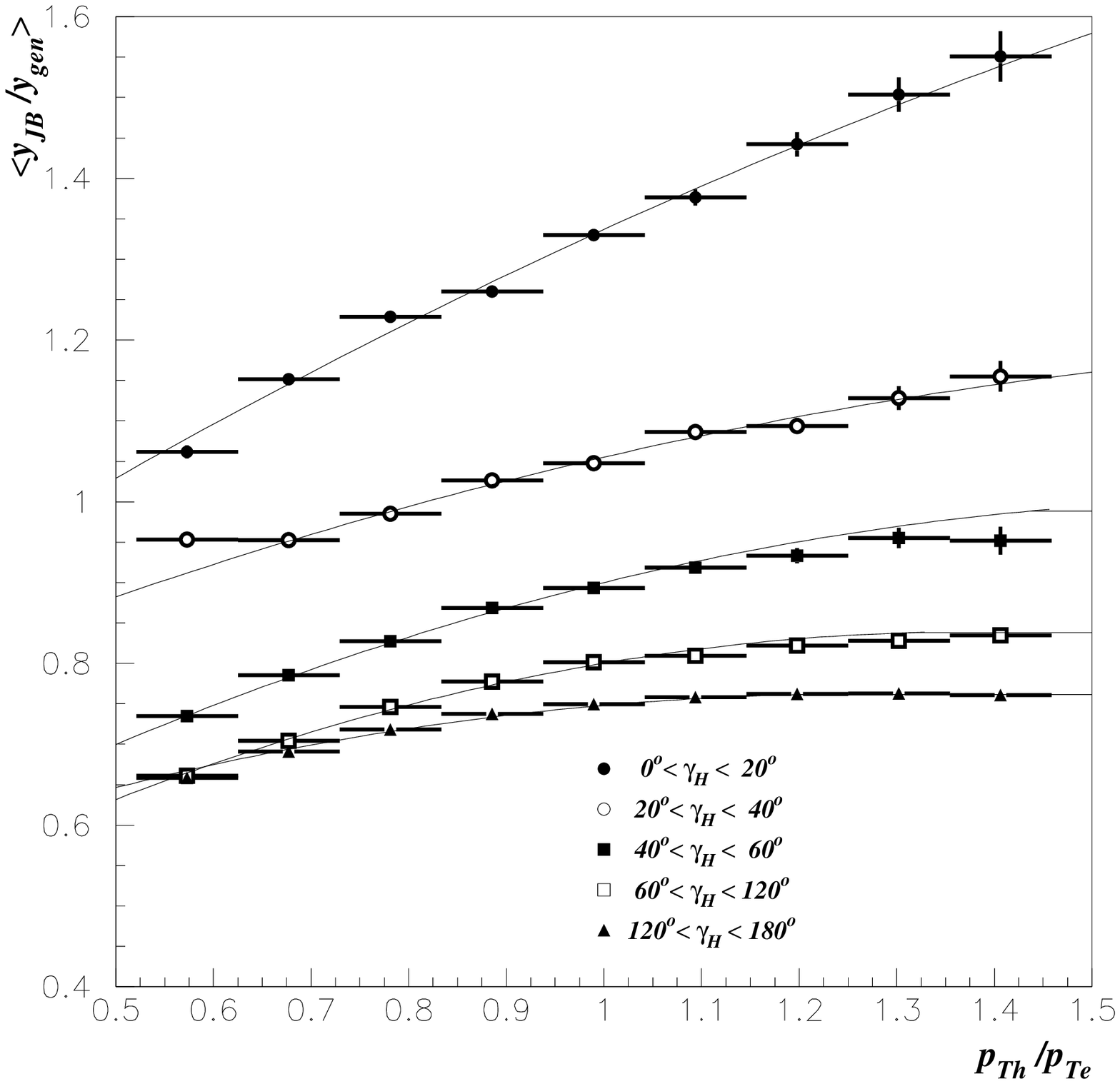}}
\end{center}
\caption[]
{\it The ratio $\yjbrat$ averaged over $p_{Th}$ as a function
of $\ptrat$ for several ranges of $\gammah$. The curves show
 the correction function ${\cal C}(\ptrat ,p_{Th} ,\gammah)$,
averaged over $p_{Th}$ as a function of \ptrat\ 
for several ranges in $\gammah$.  
}
\label{f:corf}
\end{figure}

The ratio 
$\yjbrat$ averaged over $p_{Th}$, $\langle\yjbrat\rangle$, 
as obtained from the MC simulation, 
is shown in Fig.~\ref{f:corf} as a function of $\ptrat$;
$\ygen$ is the
generated value of $y$.  
For small values of $\gammah$, $\langle\yjbrat\rangle$ rises
almost linearly with $\ptrat$.  This behaviour is as
expected from eq.~\ref{SIMPLE} and follows from the fact that
the hadronic energy deposit comes mainly from the current jet when 
$\gammah$ is small.  Note that the quantities $p_{Th}$ and
$\yjb$ are rather insensitive to the energy loss in the
forward beam hole.
The observed positive offset from unity of $\langle\yjbrat\rangle$ at $\ptrat = 1$ 
results from the CAL noise.~\footnote{
As an example of the effect of calorimeter noise on $\langle\yjbrat\rangle$ at low 
$\gammah$, consider the measurement of an event with true $y=0.01$ and
true $Q^2=10$ GeV$^2$.   
If one calorimeter cell in the RCAL produces a noise signal of 100~MeV, the
contribution to the measured $\yjb$ amounts roughly
to $\frac{200 {\rm MeV}}{55 {\rm GeV}} = 0.004$. Ignoring
the effects of energy loss and smearing,
$\yjb$ is reconstructed 40\% larger than the true $y$. 
The contribution of the same noise signal 
to the measured $p_{Th}$, on the other hand, is
small ($<0.1$ GeV) in comparison with the 
true $p_{T} = Q\sqrt{(1-y)} = 3.15$~GeV.}  

As $\gammah$ increases $\langle\yjbrat\rangle$ falls below
unity mainly due to energy loss in the passive material in the detector.
With increasing $\gammah$ an increasing
fraction of $p_{Th}$ is carried by particles produced in the region  
between the current jet
and the proton remnant.  The energy loss of these particles
enters directly into the measurement of $p_{Th}$ but to
a lesser extent into the measurement of $\yjb$ (see eqs.~\ref{DH} and~\ref{YJB}). 
This leads to a less linear dependence of $\langle\yjbrat\rangle$
on $\ptrat$.

As $p_{Th}$ increases
the energy flow between the current jet and the remnant becomes
relatively less in comparison to the energy flow in the current jet.
Thus, as a function of increasing $p_{Th}$, 
the $\ptrat$ dependence of $\yjbrat$
becomes more linear (not shown in figure).

The correction function $\cal{C} \it (\ptrat,p_{Th},\gammah)$
is parameterised using second order polynomials to describe 
the MC simulation of $\yjbrat$ in the
range $0.5<\ptrat<1.5$ which contains approximately 90\% of the events. 
The curves in Fig.~\ref{f:corf} show $\cal{C}$, averaged over $p_{Th}$, as a function of $\ptrat$ 
for several values of $\gammah$.

Application of the correction function, eq.~\ref{CORF},
 determines $y$ based on the 
{\it measured} values of $\ptrat$, $p_{Th}$ and $\gammah$, 
which reduces the dependence of the $y$ measurement on the MC simulation.
At the same time, by compensating for the deviations of
the hadronic energy measurement 
from the true values for each event individually, the resolution
of the $y$ measurement is improved.
It will be shown in Sect.~\ref{SYSUNC}, 
in the studies of systematic uncertainties, that the
structure function results do not depend strongly on the
details of the MC simulation of the DIS final states used in determining
the correction function. 

\subsubsection{Kinematic measurement in the full $y$ range}

When the current jet points in the backwards direction
($\gammah \gsim 90^\circ$), corresponding to the region of
high $y$, the hadronic system becomes less collimated,
and the particle loss through the rear beam hole is
not negligible.  These particles carry a small amount of transverse
momentum but make a large contribution to the decrease in $\delta_h$ and
therefore to the decrease in $\yjb$.  As a result
the ratio $\ptrat$ is less effective for determining the correction
to the hadronic measurement of $y$.
This can be seen in Fig.~\ref{f:corf}, where for $\gammah >120^\circ$, 
there is little  dependence of $\langle\yjbrat\rangle$ on $\ptrat$.

The measurement of $y$ from the positron (see eq.~\ref{YEL})
 at high $y$ does not 
suffer from these deficiencies and provides an accurate measurement.
Applying the $\Sigma$-correction, first introduced in ref.~\cite{b:SIGCO} 
and used by H1~\cite{b:H1_F2},
$\ypr$ and $\ye$ are combined:
\begin{equation}
\yprpr = \ypr \frac{1}{\ypr + 1 - \ye} .
\end{equation}

At this stage, the best estimate for the transverse momentum of the
hadronic system is $p_{Te}$ and for the hadronic sum $\sum_h(E_h-p_{Zh})$
is $2E_e^{\prime}\yprpr$.  This allows the calculation of the hadronic
angle, $\gammahc$, as follows:
\begin{equation}
\cos \gammahc = \frac { p_{Te}^2-4E_e^2{\ypps}}
{ p_{Te}^2+4E_e^2{\ypps}} .
\end{equation}
Since the Double-Angle method is the method least sensitive to energy scales,
the kinematic variables are calculated using eqs.~\ref{QDA} and~\ref{YDA} 
substituting $\gammahc$ for $\gammah$.

\begin{figure}[p]
\begin{center}
\mbox{\epsfxsize=14cm\epsffile{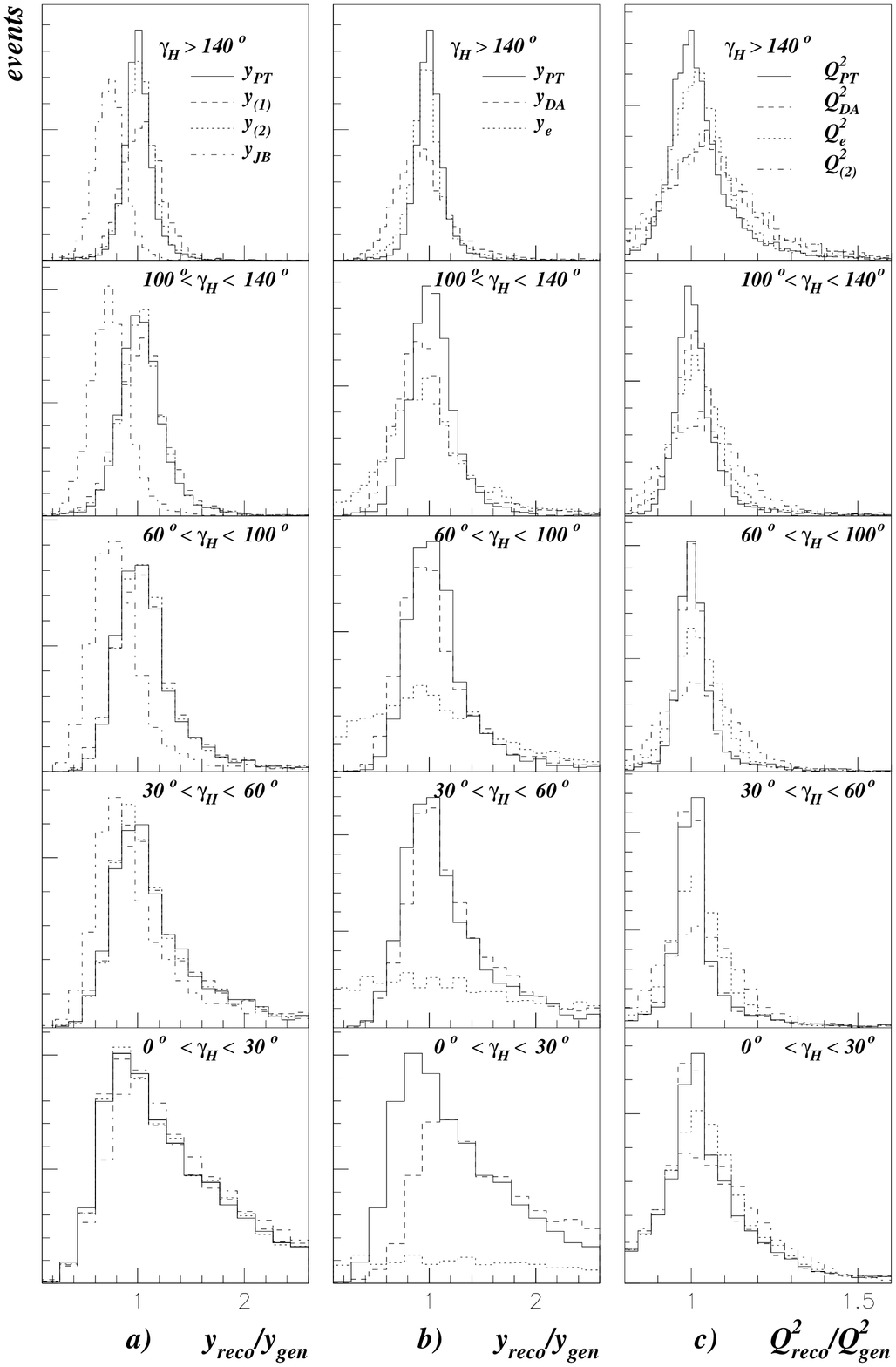}}

\end{center}
\caption[]
{\it a) The distributions of $\yjbrat$~(dashed-dotted), 
$\yprrat$~(dashed),
 $\yprprrat$~(dotted) and $\yptrat$~(solid) 
for several ranges in \gammah\ as determined from the  MC simulation.
b)  The distributions of 
$\yerat$~(dotted),
$\ydarat$~(dashed) and $\yptrat$~(solid)
for several ranges in \gammah\ as obtained from the MC simulation.
c) The corresponding distributions of $\qptrat$~(solid),
$\qdarat$~(dashed),
$\qerat$~(dotted) and
$\qprprrat$~(dashed-dotted). }
\label{f:RES}
\end{figure}

The kinematic variables 
determined in this way will be denoted by the subscript PT. 
Figure~\ref{f:RES}a shows the distributions
of $\yjbrat$, 
$\yprrat$, $\yprprrat$
and $\yptrat$ for several ranges in $\gammah$.
By construction, the correction from $\yjb$ to $\ypr$
centers the distribution around unity, while the further corrections 
yield improvements in resolution. 
Figure~\ref{f:RES}b shows the distributions of
$\yptrat$
together with $\ydarat$ and $\yerat$.
The corresponding distributions for the $Q^2$ reconstruction are shown
in Fig.~\ref{f:RES}c. 
 Both in $y$ and $Q^2$ the resolution of the 
PT method is superior to all other methods.
Also shown in Fig.~\ref{f:RES}c is the result for
 $ Q^2_{(2)} (=\frac{p_{Te}^2}{1-\yprpr})$. 
The resolution of this alternative method to reconstruct $Q^2$
is inferior to the resolution of the PT method.

\begin{figure}[p]
\begin{center}
\mbox{\epsfxsize=14cm\epsffile{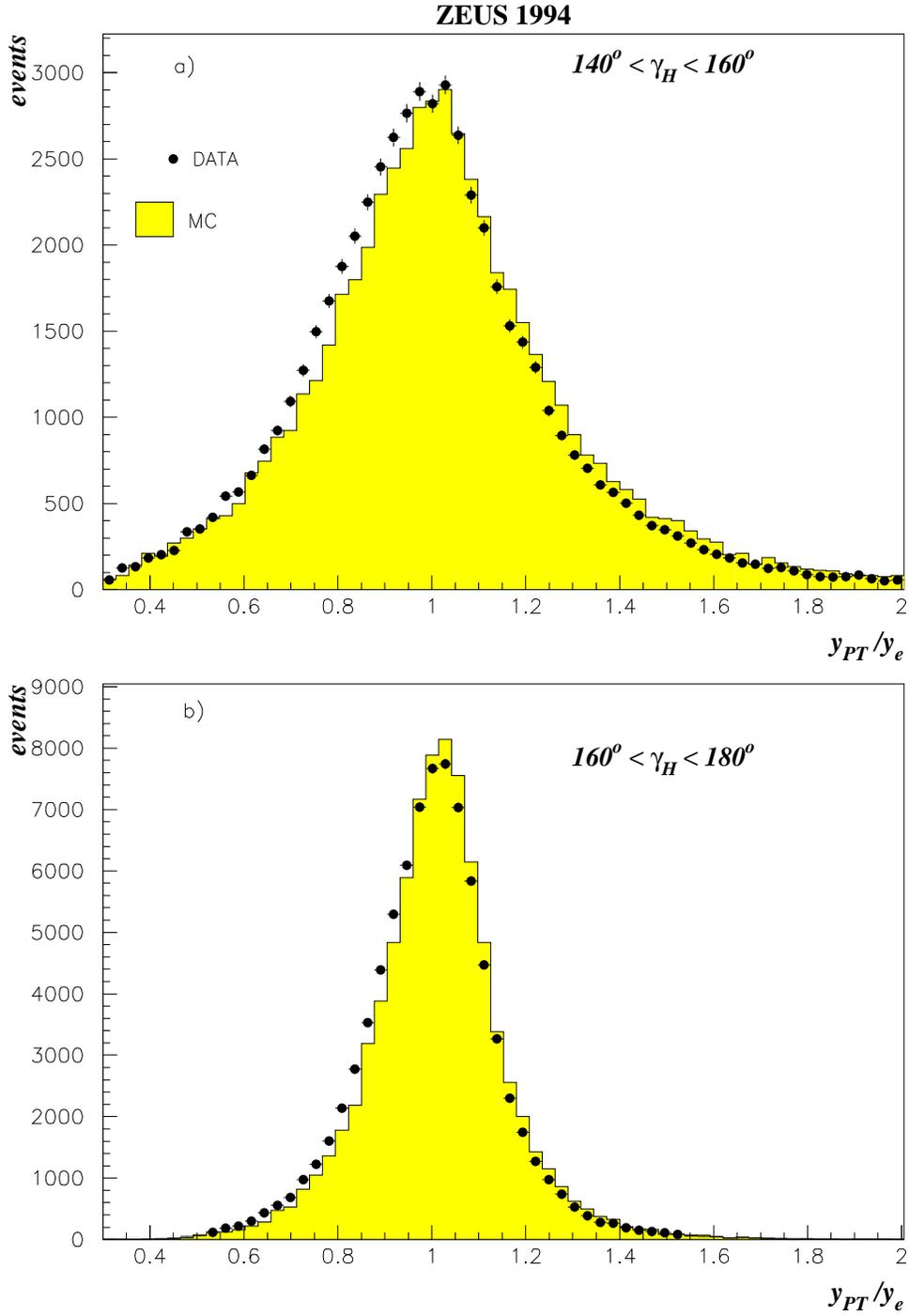}}
\end{center}
\caption[]
{\it The distributions of $\yptrat$ for a) $140^\circ< \gammah < 160^\circ$ 
and b) $160^\circ< \gammah < 180^\circ$. The points show the data and the shaded 
histograms show the results from the MC simulation.}
\label{f:YPTYE}
\end{figure}

For large $y$ one can
check the method  by comparing $\ypt$ with $\ye$.
Figure~\ref{f:YPTYE}a and b show the distributions of $\yptye$ for
the regions 
$140^\circ< \gammah < 160^\circ$ and $160^\circ< \gammah < 180^\circ$  
where $\ye$ is expected to give a reasonable estimate of $y$ . Also shown in 
Fig.~\ref{f:YPTYE} are the expectations from the MC simulation.
The r.m.s. widths 
($\sigma =9\%$ for $160^\circ< \gammah < 180^\circ$ and 
$\sigma =20\%$ for $140^\circ< \gammah < 160^\circ$) of the distributions
are in good agreement with the MC simulation 
and are dominated by the error on the 
measurement of $\ye$ . There are small shifts of about 2\perc\ between 
the data and MC simulation. These shifts are included in the study 
of the systematic uncertainties. 

\section{Event selection}
\label{ANA}
\begin{figure}[p]
\begin{center}
\mbox{\epsfxsize=14cm\epsffile{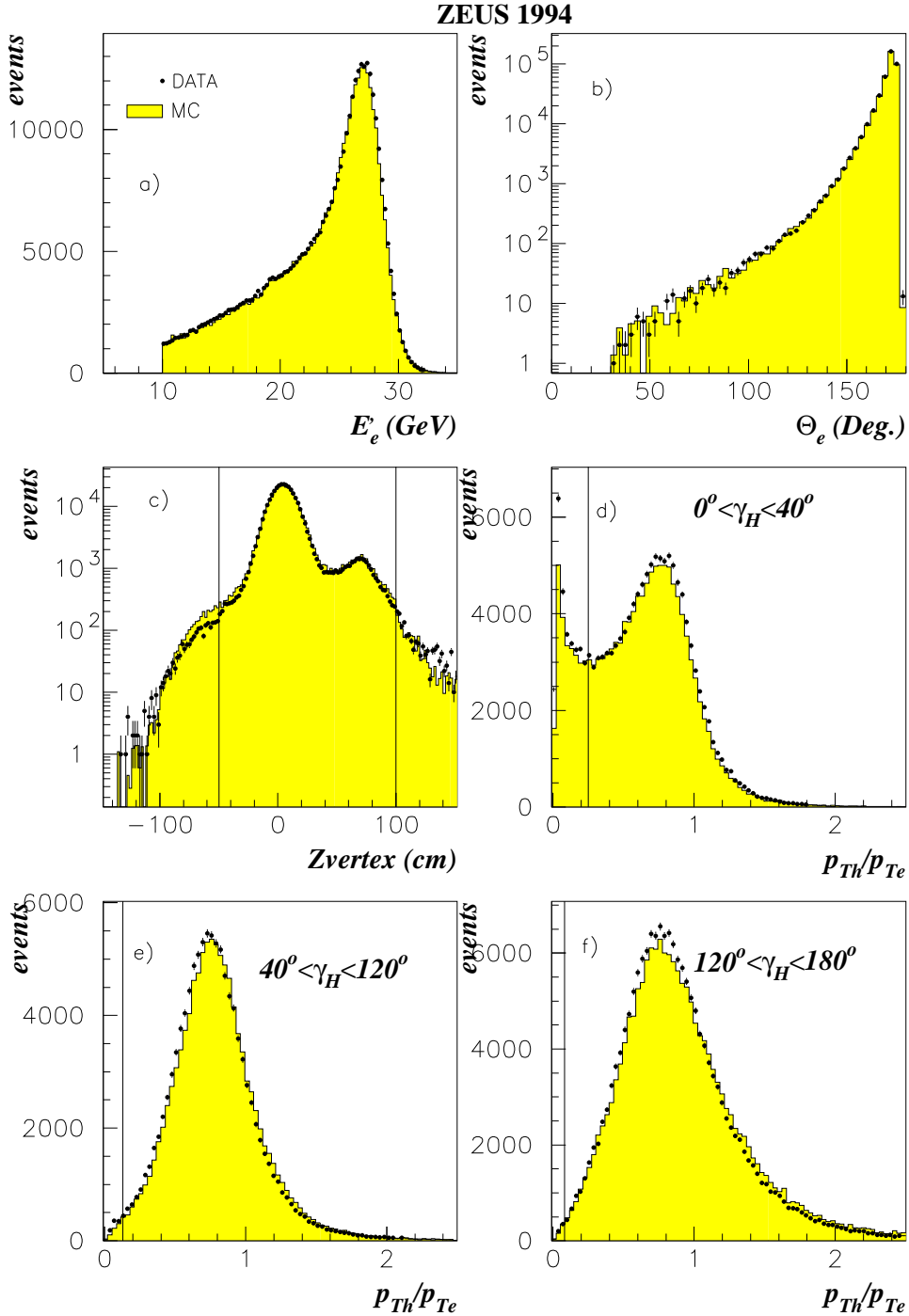}}
\end{center}
\caption[]
{\it Distributions of a) positron energy, b) positron angle, 
c) $Z_{vertex}$,
d)--f) \ptrat\ for different ranges of \gammah .
The vertical lines in c)--f) 
indicate the positions of the cuts used in the analysis. For d)-f) the cut values
are shown for the central value of $\gammah$ in each bin (see text).
The MC distributions are normalised to the integrated
luminosity of the data.}
\label{f:gendists}
\end{figure}

The following cuts were used to select NC DIS events:

\begin{itemize}
\item A positron candidate identified as described in Sect.~\ref{s:PIE}. 
\item $E^\prime_e > 10\;\Gev $, where $E^\prime_e$ is the corrected positron energy.
This cut ensures high and well understood 
positron finding efficiency and suppresses background from photoproduction.

\item $38\;\Gev < \delta < 65\;\Gev$, where $\delta = \sum_i (E_i-p_{Zi})$.
Here the sum runs 
over {\it all} calorimeter cells, i.e. including those belonging to the
identified positron. This cut removes events with large initial state 
radiation and further reduces the background from photoproduction.

\item A track match for $\theta < 135^\circ$.
 This condition suppresses
 events 
from  cosmic rays, halo-muons, photoproduction and
DIS events where an electromagnetic shower has been falsely identified
as the scattered positron. 

\item $\ye < 0.95$. This condition removes events where fake positrons  are found
in the FCAL.
\end{itemize}
A total of 680283 events pass the above cuts.
\begin{itemize}
\item Box cut.  Events with a scattered positron impact point in the RCAL inside
a box of 26~cm~$\times$~26~cm around the beampipe are rejected.
This ensures that 
the impact point is at least 2.5~cm away from the edge of the RCAL
and therefore guarantees 
full containment of the electromagnetic shower in the calorimeter.

\item $-50\;{\rm cm} < Z_{\it vertex} < 100 \;{\rm cm}$. 
This cut is performed
for events with a reconstructed tracking vertex and 
suppresses beam-gas background
events and the small fraction of the events where the vertex position
is incorrectly measured. Events without a tracking vertex are assigned 
the mean vertex $Z$ position and are accepted.
\end{itemize}
After these additional cuts a sample of 443421 events remains.
\begin{itemize}

\item $\ptrat > 0.3 - 0.001 \gammah$~(\gammah\ in degrees).
For the PT method to yield a reliable measurement of \x\ and \qsd\ 
the loss in hadronic transverse momentum
must be limited. There are  events where a substantial
part of the current jet remains in the forward beampipe. These events, which 
are produced at small \gammah\ , can be falsely reconstructed at large 
\gammah\ due to the CAL noise. This cut suppresses such events.

\end{itemize}
A total of 400627 events pass all the above selection cuts. 
For the accepted events 
Fig.~\ref{f:gendists}a,b show, respectively, the measured energy and
 scattering angle of the positron.
Figure~\ref{f:gendists}c shows the reconstructed
vertex distribution before applying the vertex cut
and Fig.~\ref{f:gendists}d-f show the 
distribution of \ptrat\ , where the \ptrat\
cut was removed,
for three ranges in \gammah\ .
The energy, angle, vertex and transverse momentum distributions are well reproduced 
by the MC simulation (also indicated in the figures). To obtain these 
distributions the MC events have been reweighted to the structure function
obtained from the QCD NLO fit described below.

\begin{figure}[p]
\begin{center}
\mbox{\epsfxsize=14cm\epsffile{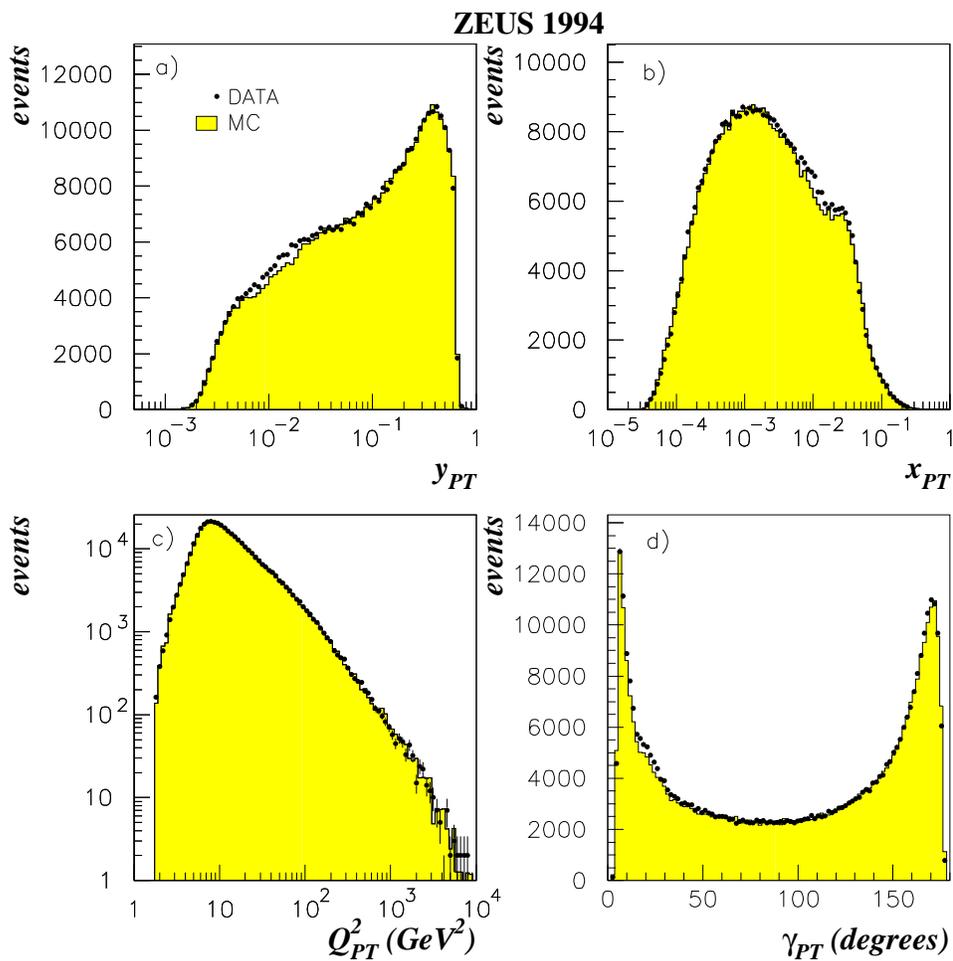}}
\end{center}
\caption[]
{\it The distributions of a) \ypt , b) \xpt , c) \qpt and d) \gammahc\ 
for data (points) compared with the distributions 
obtained from the MC simulation (shaded histograms).
The MC distributions are normalised to the integrated
luminosity of the data.}
\label{f:GENKIN}
\end{figure}

Figure~\ref{f:GENKIN} shows the distributions in \ypt , \xpt ,
\qpt\ and \gammahc\ together with the MC distributions
normalised to the integrated luminosity of the data. 
Good agreement is obtained 
except for the region around $y\sim 10^{-2}$ where the MC undershoots the data 
by up to $\sim$~5\perc . Equivalent deviations are observed in the \xpt\ and
\gammahc\ distributions.
This difference is not concentrated at a 
specific value of \qsd\ and has a negligible effect on the values of 
the extracted structure function.
 
\begin{figure}[p]
\setlength{\savelen}{1.0\unitlength}
\setlength{\unitlength}{1cm}
\begin{picture}(16,20)
\put(3.5,0.1){\mbox{\epsfxsize=11cm\epsffile{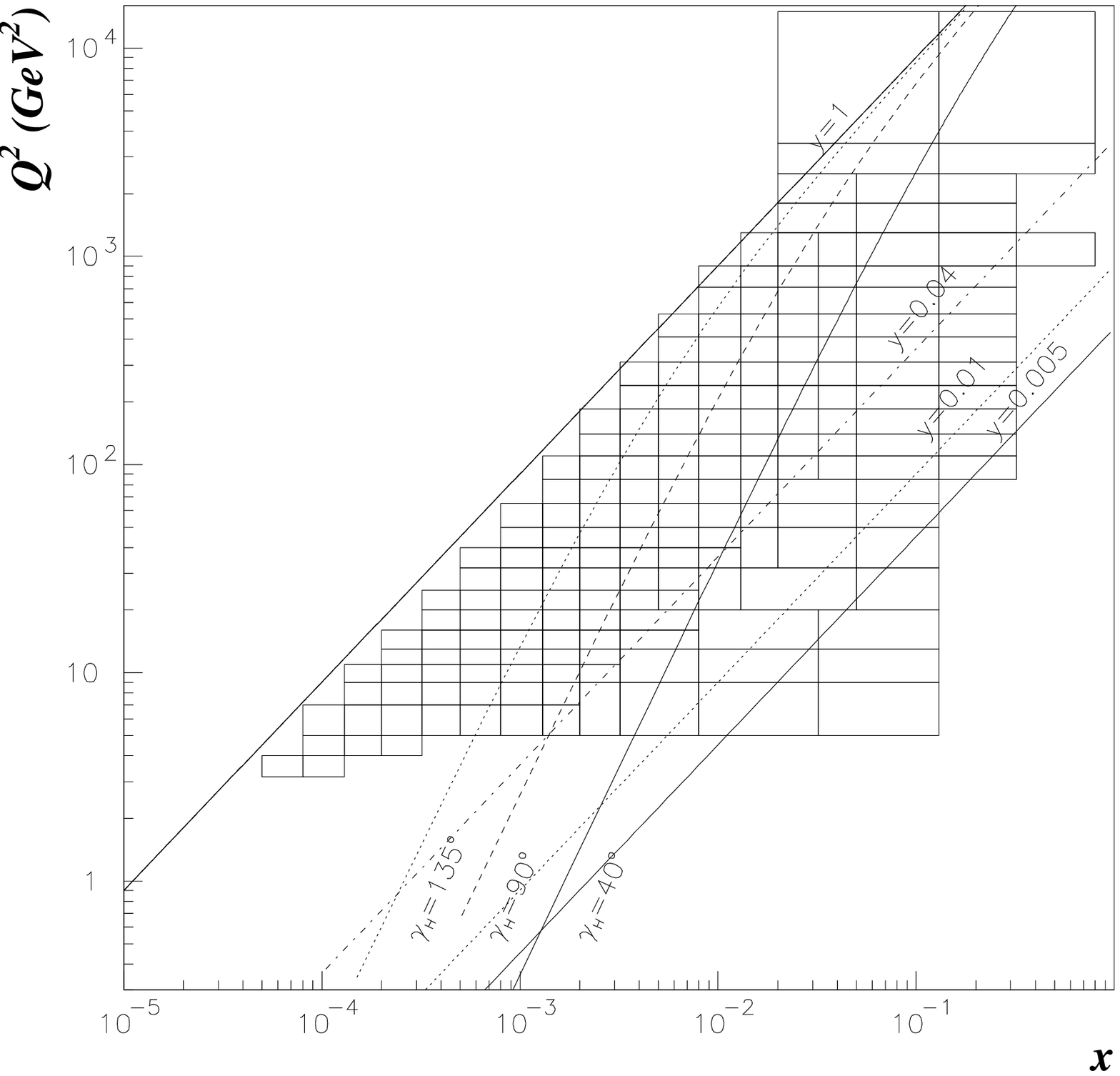}}}
\put(4.,11){\mbox{\epsfxsize=9.7cm\epsffile{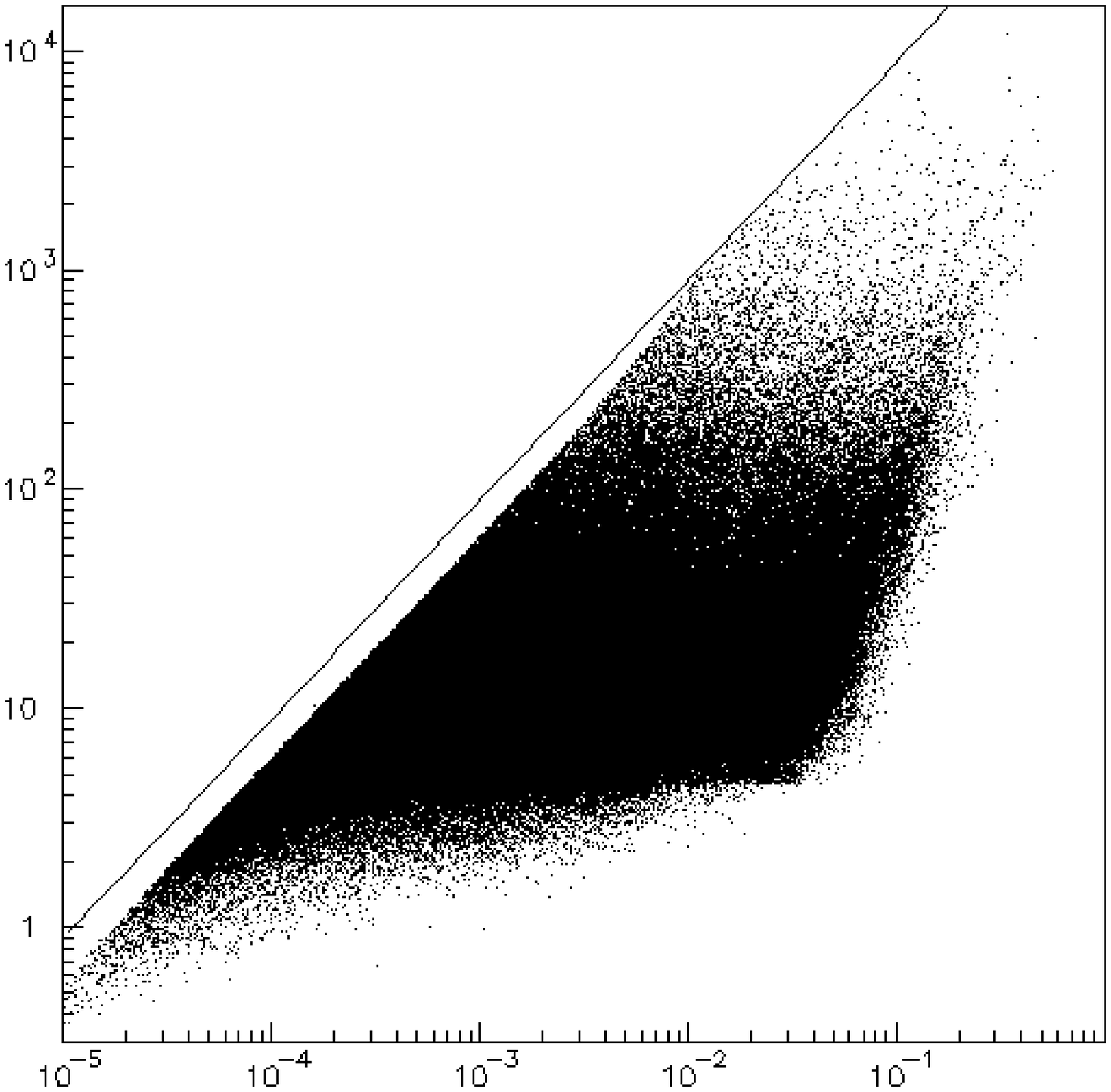}}}
\put(3.3,18){\mbox{\epsfxsize=0.7cm\epsfysize=2.1cm\epsffile{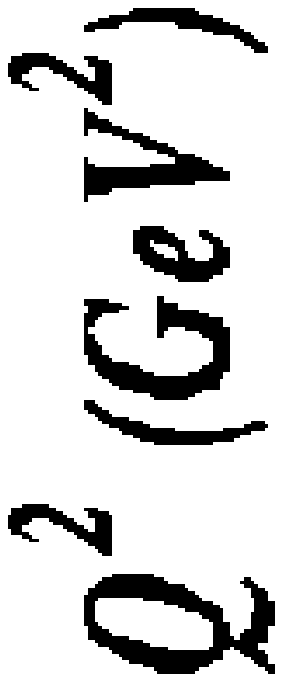}}}
\put(13,10.6){\mbox{\epsfxsize=0.32cm\epsfysize=0.38cm\epsffile{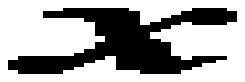}}}
\put(5,7){b)}
\put(5,17){a)}
\put(6.5,21){\large {\bf ZEUS 1994}}
\end{picture}
\setlength{\unitlength}{1.0\savelen}
\caption[]
{\it a)The distribution of the events in the (\x ,\qsd ) plane.
b) The (\x ,\qsd )-bins used in the structure function determination. Also 
indicated are lines of constant \y\ and of constant \gammah. The \gammah\
values for this figure are calculated directly from $x$ and $Q^2$.}
\label{f:KINPLANE}
\end{figure}

Figure~\ref{f:KINPLANE}a shows the distribution of events in the (\x ,\qsd )
plane.   
The (\x,\qsd) bins used for the determination of 
the structure function are shown in Fig.~\ref{f:KINPLANE}b. 
The bins have been chosen commensurate with the 
resolutions. At large \qsd\ and also at low \y\ larger bin sizes 
have been chosen
to obtain adequate statistics in each bin and to minimise bin-to-bin 
migrations as a result of the non-Gaussian tails. 
Furthermore,
large acceptance ($A>20\perc$) and purity ($P>30\perc$)~\footnote{Acceptance
is defined as the ratio of the number of events which are reconstructed to the number of 
generated events in a bin. Purity is defined as the fraction of reconstructed
events in a bin which originated from the same bin.} are required for each bin.
The lowest values of acceptance occur at the lowest \qsd\ values where the 
box cut becomes effective. The bins with lowest purity occur at the 
lowest \y\ values. 
In the majority of bins the acceptance is greater than 80\perc\ 
and the purity is
greater than 50\perc .

\begin{figure}[p]
\setlength{\savelen}{1.0\unitlength}
\setlength{\unitlength}{1cm}
\begin{picture}(16,20)
\put(4,0){\mbox{\epsfxsize=10cm\epsffile{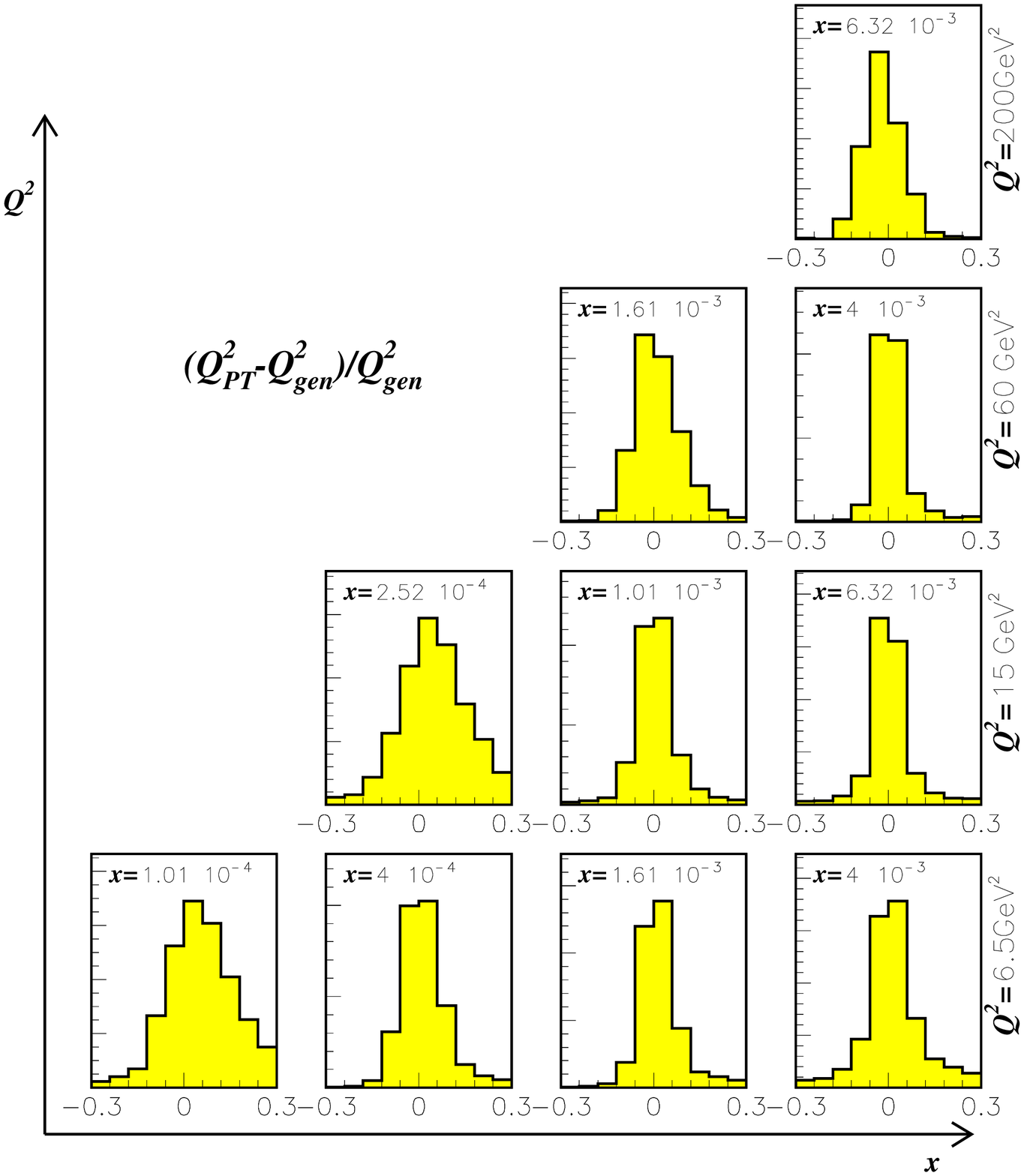}}}
\put(0,10){\mbox{\epsfxsize=10cm\epsffile{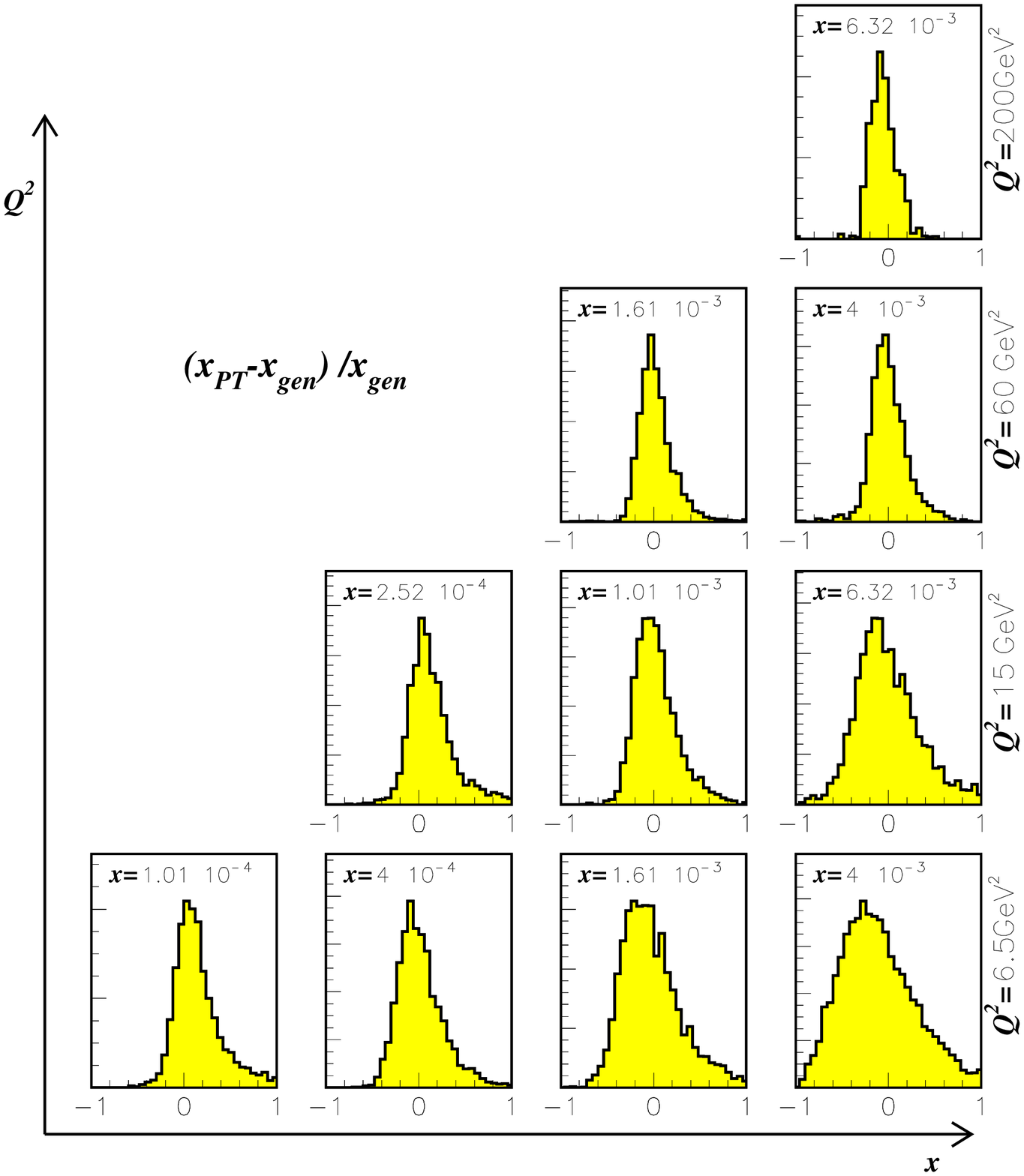}}}
\put(2,18.5){a)}
\put(6,8.5){b)}
\end{picture}
\setlength{\unitlength}{1.0\savelen}
\caption[]
{\it a)The distributions of $\frac{x_{PT}-x_{gen}}{x_{gen}}$ for different 
 (\x ,\qsd) bins, obtained by MC simulation.
b) The distributions of $\frac{Q^2_{PT}-Q^2_{gen}}{Q^2_{gen}}$ for 
the same (\x ,\qsd) bins.}
\label{f:resvsxq}
\end{figure}

Figure~\ref{f:resvsxq} shows the distributions in $(\frac{x_{PT} -x_{gen}}{x_{gen}})$ and 
$(\frac{Q^2_{PT} - Q^2_{gen}}{Q^2_{gen}})$  for a number of 
bins in  the (\x ,\qsd ) plane, where
$x_{gen}$ and $Q^2_{gen}$ are the generated values of $x$ and $Q^2$.
 The resolutions vary 
smoothly  with \x\ and \qsd.
The resolution in \x\ decreases from $\sigma_x\approx 25$\perc\ for low \y\
($y=\frac{Q^2}{xs}$) 
to $\sigma_x\approx 10$\perc\ for large \y . The resolution in \qsd\ is 
 $\sigma_{Q^2}\approx 10$\perc\ for large \y\ and improves to $\sigma_{Q^2}\approx 5$\perc\
for intermediate and low \y. At the lowest \y\ values
 tails appear due to the
vertex finding efficiency and resolution.

The numbers of events in each of the bins are given in Table~\ref{t:final}.

\begin{figure}[p]
\begin{center}
\mbox{\epsfxsize=14cm\epsffile{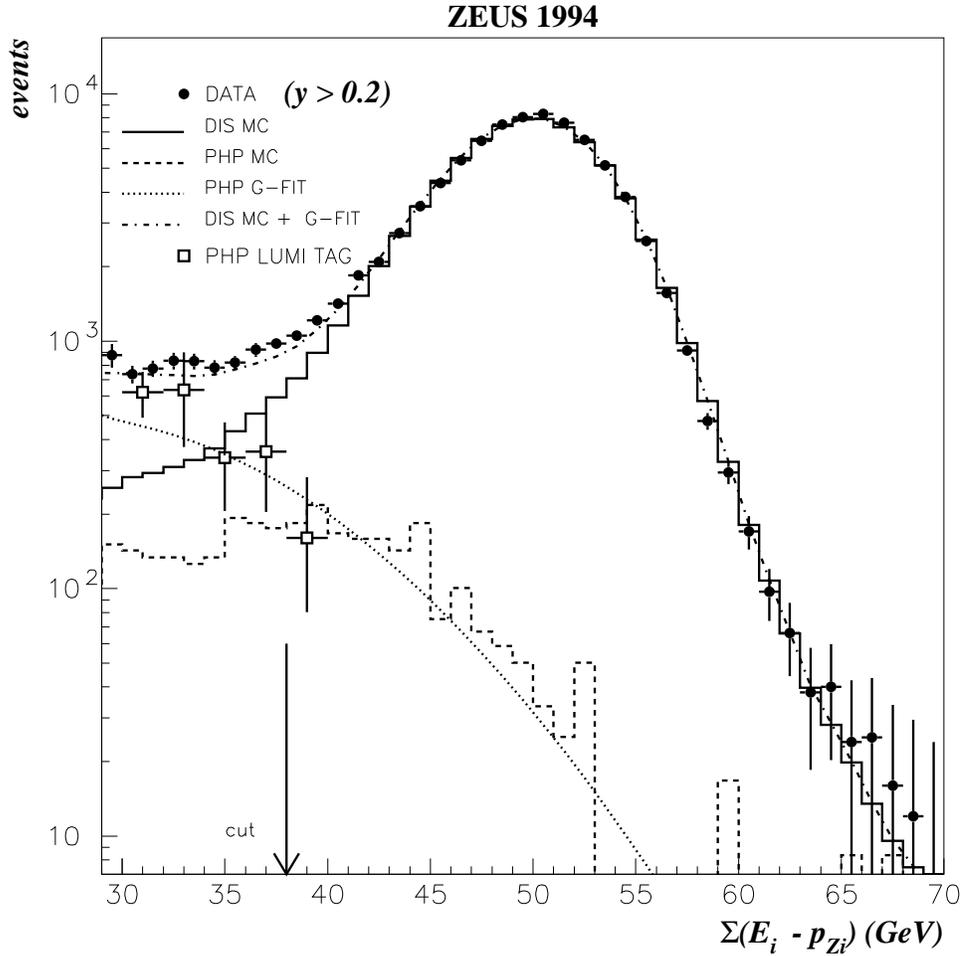}}
\end{center}
\caption[]
{\it The distribution of $\Sigma_i (E_i-p_{Zi})$ for data (solid dots), 
DIS MC simulation (solid histogram), photoproduction (PHP) 
MC simulation (dashed histogram) and 
LUMI tagged photoproduction data (open squares). The PHP distribution is
valid for $\Sigma_i (E_i-p_{Zi})>38\;\Gev$. The result of the 
fit to the data, described in the text, is shown as the dashed-dotted curve and
the photoproduction contribution to this fit is indicated by the dotted curve.
}
\label{f:php}
\end{figure}

The final sample contains a small number of background events which are not
due to deep inelastic neutral current scattering:
\begin{itemize}
\item  Non-$ep$ background. The level of background not associated with
$ep$ collisions is determined from the number of events observed in
unpaired or empty bunches. This background is subtracted statistically
taking into account the appropriate ratios of bunch currents and numbers of 
bunches. It amounts to less than 1\perc\ 
in all bins. For events in the bins of $\qsd > 1000\;\Gevsq$ all events 
were scanned visually and no non-$ep$ events were found.

\item  Photoproduction background. The events from the photoproduction 
MC code PYTHIA
were analysed in the same way as the data and the
number of events passing the selection cuts in each  bin 
was determined. This number is subtracted taking into account the 
ratio of equivalent luminosity of the data and the MC samples. 
As a check
the $\delta $ distributions for each of the bins were fitted 
to a shape extracted from  MC DIS events and a Gaussian shape, which
parameterises the photoproduction events
(the procedure is  described in~\cite{b:ZEUS_F2}). 
The $\delta$ distribution for $y>0.2$ is shown in Fig.~\ref{f:php}.
Also given are the MC expectations for DIS (solid histogram) and for 
photoproduction (dashed histogram). The flattening of the MC prediction
for photoproduction at $\delta <38\;\Gev$ is artificial and 
results from the cut of $W>190\;\Gev$ applied at the generator level. 
In the region $\delta <38\;\Gev$ the photoproduction
contribution can be measured directly through events tagged by the LUMI
positron tagger (the acceptance for these events tends to zero for 
$\delta >40\;\Gev$). The result of this measurement is shown
in Fig.~\ref{f:php}. Also shown  is the result
of the fit to the data (dashed-dotted curve) using the shape from the DIS MC 
distribution plus a Gaussian contribution (dotted curve) for the 
photoproduction background. The Gaussian contribution agrees with the 
MC photoproduction estimate for  $\delta >38\;\Gev$ and with
the measured photoproduction background.  For $\qsd > 1000\;\Gevsq$, the visual
scan found at most two events showing topologies consistent with
photoproduction events.
\end{itemize}

The total background contribution to each bin is given in Table~\ref{t:final}.
The maximum background fraction of about 6\perc\ occurs in bins of high \y .

\section{Proton structure function \Ft }
\label{section:rew}
In deep inelastic scattering the double differential cross section for 
inclusive $e^+p$ scattering is given in terms of the structure functions $F_i$:
\begin{equation}
\frac{d^2\sigma}{dx\;d\qsd} = \frac{2\pi \alpha^2}{xQ^4}
\left[Y_+ \Ft (x,\qsd )-y^2\FL (x,\qsd )-Y_{-}x\Fd (x,\qsd )\right]
(1+\delta_r(x,\qsd )) \label{TOTALFORM}
\end{equation}
where $Y_\pm = 1 \pm (1-y)^2$ and \x\ and \qsd\ are defined at the hadronic
vertex . In this equation \FL\ is the longitudinal structure 
function, \Fd\ is the parity violating 
term arising from the $Z^\circ$ exchange and 
$\delta_r$ is the electroweak radiative correction. 
Since all of these contributions to the cross section are expected to be small
in the kinematic region of the present measurement,
this can be rewritten as:
\begin{equation}
\frac{d^2\sigma}{dx\;d\qsd} = \frac{2\pi \alpha^2Y_+}{xQ^4}\Ft(x,\qsd) 
(1-\delta_{_{L}}-\delta_3)
(1+\delta_r).  \label{F2FORM}
\end{equation}

The \Ft\ structure function itself contains 
contributions from virtual photon and $Z^\circ$ exchange:
\begin{equation}
\Ft  = \Fem + \frac{\qsd}{(\qsd + M_Z^2)}\Fint
+ \frac{Q^4}{(\qsd + M_Z^2)^2}\Fwk = \Fem (1+\delta_{_{Z}})
\end{equation}
where $M_Z$ is the mass of the $Z^\circ$ and 
\Fem , \Fwk\ and \Fint\ are the contributions to \Ft\ due 
to photon exchange, $Z^\circ$ exchange
and $\gamma Z^\circ$ interference respectively.
We finally write:
\begin{equation}
\frac{d^2\sigma}{dx\;d\qsd} = \frac{2\pi \alpha^2Y_+}{xQ^4} \Fem
(1+\delta_{_{Z}})(1-\delta_{_{L}}-\delta_3)
(1+\delta_r).  
\end{equation}
The corrections, $\delta_{r,{_{Z}},3,{_{L}}}$,
are functions of \x\ and \qsd\ but are,  
to a good approximation, independent of \Ft, i.e. they are insensitive 
to the parton density distributions.

In this analysis we determine the structure function \Ft . 
In the Standard Model the difference between \Ft\ and \Fem\ is expected
to be less than
1\perc\ for values
of $\qsd <1000$ GeV$^2$. At higher \qsd\ the difference becomes
progressively larger. In this region therefore  we  present 
the values of \Fem\ as well as  \Ft. These are extracted from \Ft\  
through the correction $\delta{_{Z}}$ using the MRSA\cite{b:MRSA} parton 
distribution functions. 

An iterative procedure is used to extract the structure function
\Ft . Monte Carlo events are generated according to eq.~\ref{TOTALFORM} 
and so contain transverse and longitudinal photon and $Z^\circ$
contributions as well as radiative effects. In each step of the 
iteration the value of \Ft\  in a bin can be obtained from the ratio
of observed events in the data to the number of events observed in the 
MC simulation and the calculated values of $\delta_{r,{_{Z}},3,{_{L}}}$.
In the first step the acceptance correction is taken from 
the MC simulation which uses the parton distributions
given by MRSA. 
The result for \Ft\  from this first iteration is used for
a QCD fit using the DGLAP~\cite{b:GLAP} evolution equations
in next-to-leading order (QCD NLO fit) in a manner
very similar to that described in~\cite{b:ZEUS_glu}.
The evolution uses massless quarks of three flavours in the proton 
and the charm quark coefficient functions from references ~\cite{b:SMITH,b:GRS}
to ensure a smooth crossing of the charm threshold. 
The NMC~\cite{b:NMC} data for $Q^2 > 4~{\rm GeV^2}$ are used to 
constrain the QCD NLO fit at high $x$ ($\gsim 10^{-2}$). The fit 
also includes the low \qsd\ 1994 data from this experiment~\cite{b:LOWQ}.  

The parton distributions from the QCD NLO fit are used to 
 recalculate the 
cross section, including the new  $F_L$ contribution
as a function of \x\ and \qsd.
The MC 
events are reweighted using the calculated cross sections and the 
true hadronic \x\ and \qsd. This reweighted MC is then used to perform the 
acceptance correction bin by bin to the data,
leading to a new estimate of \Ft, specified at the mean values of $x$ and $Q^2$ in each bin.
The procedure is repeated until the \Ft\  values from two 
consecutive iterations change by less than 0.5\%.  The final result is reached in three iterations.
The functional form of the QCD NLO fit is used to quote the results for $F_2$  at
$x$ and $Q^2$ values which are chosen to be convenient  
for comparisons with other $F_2$ measurements.
Using alternative parametrisations for this correction has negligible effect
on the measured $F_2$.
It should be noted that the QCD NLO fit is used here only as a
parameterisation to obtain a
stable acceptance correction and not to perform a detailed QCD analysis.

The statistical errors of the \Ft\  values are calculated from the
number of events measured in the bins, including those
from the background subtraction, and the statistical error from the MC simulation.  
Since we are using bin by bin corrections, the correlations of statistical
errors between the \Ft\  measurements enter only via the finite statistics
of the MC sample.  The correlations are small given the relatively large
MC sample used in this analysis.
A correlation between the \Ft\  values of neighbouring bins
is present due to acceptance and smearing effects.
The sensitivity of the measured \Ft\  to these effects has been checked 
by comparing the \Ft\  obtained from the first iteration with the final
value.
The changes were within the statistical
errors of the final \Ft\ values.

The correction $\delta_{_{L}}$ is determined using the QCD
prescription for \FL~\cite{b:AM} and  the parton distributions from the
QCD NLO fit. It is significant only 
in the region of large \y\ where it reaches 12\perc (see Table~\ref{t:final}).
The $\delta_{_{Z}}$ and 
$\delta_3$ corrections are calculated from the MRSA parametrisation of the
parton distributions.
These corrections are negligible for values of \qsd\ below 1000~\Gevsq. 
For values of $Q^2$ above 1000~GeV$^2$ they strongly increase with
increasing $Q^2$ and have a slight dependence on \y .
 For the two highest \qsd\ bins the corrections are:

\newpage

\begin{itemize}
\item
$\qsd = 3000~\Gevsq$,
        \begin{itemize}
                \item[]{ $x=0.20$: $ \delta_{_{Z}}=+4\perc $ and $ \delta_3 = +5\perc$}
                \item[]{ $x=0.08$: $ \delta_{_{Z}}=+5\perc $ and $\delta_3 = +8\perc$}
        \end{itemize}
\item
$\qsd = 5000~\Gevsq$,
        \begin{itemize}
                \item[]{ $x=0.20$: $ \delta_{_{Z}}=+7.5\perc $ and $ \delta_3 = +11\perc$}
                \item[]{ $x=0.08$: $\delta_{_{Z}}=+8\perc $ and $ \delta_3 = +22\perc$.}
        \end{itemize}
\end{itemize}

The \Ft\ values are corrected for
higher order QED radiative effects not included in HERACLES.  These
corrections, including soft photon
exponentiation, were evaluated using the program HECTOR in the
leading log approximation~\cite{b:HECTOR}. They vary smoothly with
\qsd\ between 0.2\perc\ and 0.5\perc.


\subsection{Systematic uncertainties}
\label{SYSUNC}

\begin{figure}[p]
\begin{center}
\mbox{\epsfxsize=14cm\epsffile{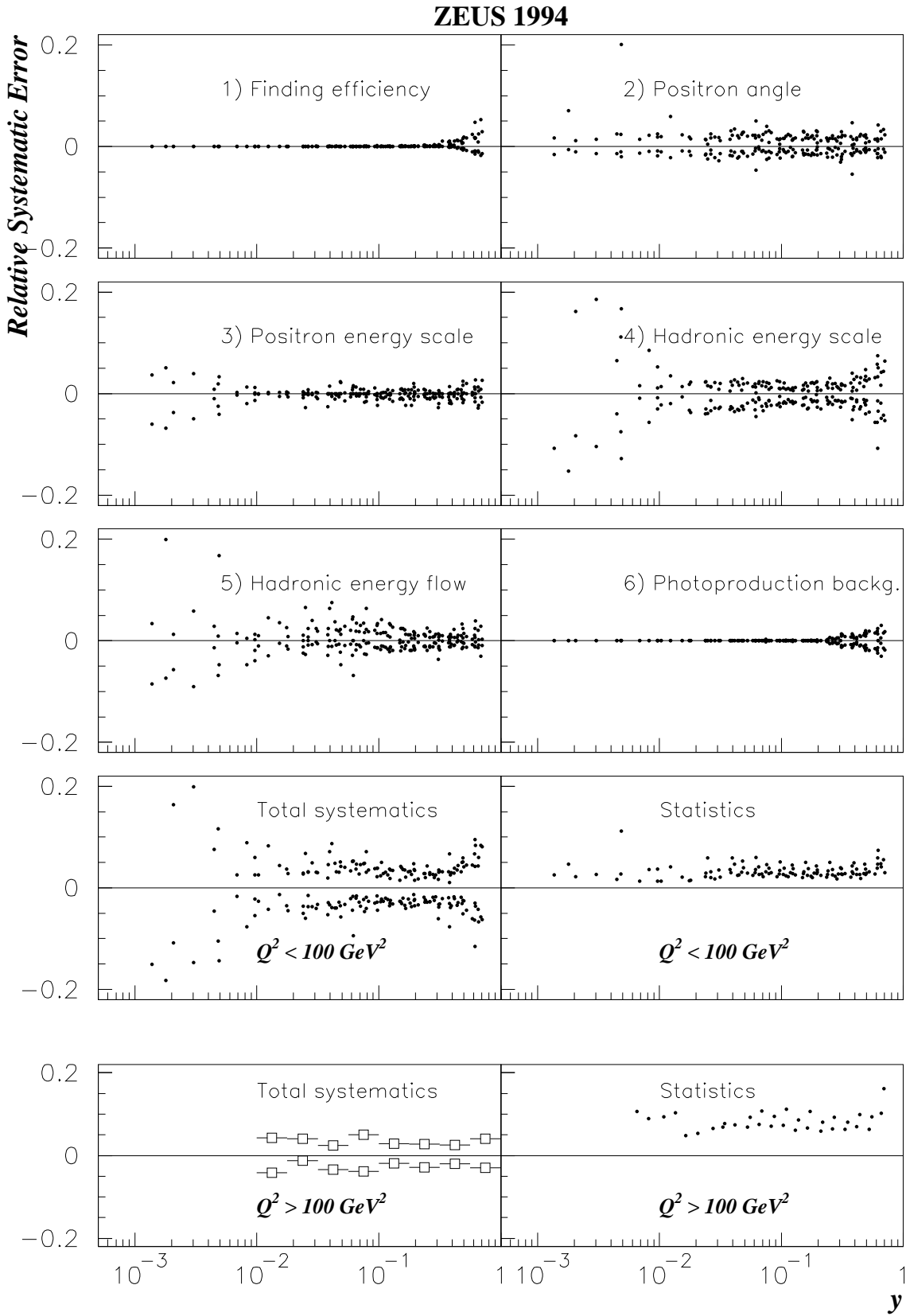}}
\end{center}
\caption[]
{\it Relative systematic errors as a 
function of \y\ for different categories
of systematic uncertainties for  bins with $Q^2<100\;{\rm\Gevsq}$.
The total systematic and statistical errors are also given.
The bottom two plots show total systematic and statistical errors for
bins with with $Q^2>100\;\Gevsq$. In categories 4) and 5), and the total systematic
error plot for $Q^2 < 100$ GeV$^2$, the four lowest $y$ points are off-scale 
 and are not shown.
}
\label{f:SYSER}
\end{figure}

Several factors contribute to the systematic uncertainties in the 
measurements of \Ft . In the following they are subdivided into 
six independent categories. For each category we summarise the 
checks that have been performed to estimate the size of the uncertainty.
Figure~\ref{f:SYSER} shows the size of the systematic uncertainties 
plotted as a function of 
\y\ for the six categories and also the total systematic error.
For reference (see appendix) each test applicable 
at $Q^2 < 100$ GeV$^2$ or $y<0.01$ is numbered; 
the number is given in brackets \{\}.
 
\begin{itemize}
\item[1] Positron finding and efficiency  
\begin{itemize}
\item[$\bullet$]
The positron identification efficiency is varied within the 
uncertainty found in the QED Compton study which is 
$\sim$3\% for positron energies $E^{\prime}_e = 10 $ GeV and negligible
for energies above 14 GeV \{1,2\}.
The effects on \Ft\ are negligible except in the lowest \x\ bins
where they reach $\sim$3\perc.

The parameters of SINISTRA94 are varied to check the effects from
overlaps of energy deposits from the hadronic final state
with those of the scattered positron \{1\}. This results in changes of \Ft\ 
of $\sim$4\perc, again in the lowest \x\ bins.

For $\theta_e<135^\circ$, where a track is required to match with the CAL positron
position,
a comparison of the number of rejected events in MC and data can accommodate 
a possible mismatch in efficiency of at most $\pm$2.5\perc.

In addition, an alternative electron
finder  employed in a previous analysis is used~\cite{b:ZEUS_F2}.
Consistent results are obtained over the full energy range.
\end{itemize}

\item[2]
Positron scattering angle
\begin{itemize}
\item[$\bullet$]
Changing the box cut from 26~cm~$\times$~26~cm to  28~cm~$\times$~28~cm
in both data and MC \{3,4\} 
has a small effect on the \Ft\ at low values of \qsd ($\leq 15$ GeV) and are negligible elsewhere.

The systematic uncertainty in the position of the scattered positron
is estimated from the difference between the extrapolated track
position at the face of the CAL
 and the position found from the CAL or the SRTD. In the region of the SRTD
these uncertainties are about 1~mm. They are 2~mm outside the SRTD. The largest
systematic error is obtained if the relative distance 
between the two halves of the RCAL is changed by $\pm$2~mm \{5,6\}.

For positions determined by the RCAL we additionally 
changed the distance of the measured impact point with
respect to the beam-line by $\pm$2~mm \{7,8\}.

For positron angles measured by the tracking detector,
the scattering angle $\theta_{e}$ is changed by the estimated 
systematic uncertainty of the track angle of $\pm0.2^\circ$.
  
These changes were made in the data 
while leaving the MC simulation unchanged.
This had an effect of about 1\% on the measured $F_2$.

\item[$\bullet$]
The absolute vertex reconstruction efficiency in the MC simulation
is decreased by 1\% overall \{9\} or increased by 3\% for 
$\gammah < 40^\circ$ \{10\}. 
These changes are motivated by  the study of the  efficiency as a function
of the hadronic angle (see Fig.~\ref{f:verteff}e).  
Also, the cut on the $Z_{vertex}$ position is tightened by
demanding --28~cm~$<Z_{vertex}<$~40cm to estimate a possible
uncertainty from the luminosity of the events in the satellite
bunch \{11\}. 
Effects on the measured $F_2$ are typically 1-2\%.

As an additional check, we required that the 
vertex determined from the CAL timing was --40~cm~$<Z^{timing}_{vertex}<$~30~cm 
 for the events that had no tracking vertex.
The effects on the measured \Ft\ are similar to those described 
above and are not included in
the final systematic uncertainties.
\end{itemize}

\item[3] Positron energy scale 
\begin{itemize}
\item[$\bullet$]
 The systematic uncertainties in  \Ft\  due
to the  uncertainties in the absolute
calorimeter energy calibration for scattered positrons
is estimated  by changing the energy scale in the RCAL in 
the MC simulation by 2\% at 10 GeV  linearly decreasing to
 1\% at 27.5 GeV. 
For BCAL the energy scale is modified by $\pm 2\perc$. The magnitude of these
shifts represents our present understanding of the positron energy measurement
 (see Sect.~\ref{section:CALIB}) \{12,13\}. 
Systematic variations in \Ft\ of 1-2\perc\ are
observed.
\end{itemize}
\item[4]
Hadronic energy measurement
\begin{itemize}
\item[$\bullet$]{
The energy scale in the MC simulation is changed by
$\pm$3\% while leaving the positron energy scale unchanged \{14,15\}.
The value of 3\perc\ is based on detailed comparisons of the distributions of
the quantity \ptrat\ from data and MC 
(see Fig.~\ref{f:gendists}d-f) and on the deviation
of the mean of $\yptye$ from unity for high \y\
 (see Fig.~\ref{f:YPTYE}). 
The effects on \Ft\ are typically 4\%  in the bins at very high and very low $y$, while the effects
are about 1\% at moderate $y$.
In addition the energy scales of the different calorimeter
parts are changed independently by 3\% in turn \{16-19\},
which resulted in similar effects on $F_2$. }
\item[$\bullet$]{
The noise cut on isolated EMC(HAC) cells of 100(150) MeV was changed in the MC 
simulation by
 $\pm$10 MeV which reflects the uncertainty in the MC simulation of
low energy deposits in the CAL and the 
modeling of the uranium radioactivity \{20,21\}.
Variations in \Ft\ of up to 25\% are observed in the bins at very low \y.
Elsewhere the variations in \Ft\ are within 2\%.}
\end{itemize}

\newpage

\item[5]{ Hadronic energy flow
\begin{itemize}
\item[$\bullet$]
The correction function, ${\cal C}(\ptrat ,p_{Th} ,\gammah)$, depends on 
the hadronic energy flow. The correction function was also determined
using a MC simulation which uses the LEPTO
MEPS model for the description of the hadronic 
final state, instead of CDMBGF. This model 
also provides a reasonable
description of the hadronic energy flow in DIS events~\cite{b:HAD_PAPERS}.
The new correction function was applied to the data.  Acceptance corrections were 
then made with the sample of CDMBGF MC events with the original
correction function applied \{22\}.  
The changes in \Ft\  are typically 2\perc\  in all bins.
\item[$\bullet$]
The hadronic energy flow for diffractive events is different
to that for non-diffractive events. To investigate the sensitivity of the 
PT method to the size of the diffractive component
we reweighted  the diffractive scattering cross section in the MC simulation
such that it agrees with the measurements of \cite{b:gunter} while leaving the
correction function ${\cal C}$ unchanged \{23\}. 
As a separate check, to allow for 
possible additional diffractive-like events we increased the
diffractive cross section by
 an additional 100\perc \{24\}.
Effects at the 2\perc\ level in \Ft\  are observed in each check.

\item[$\bullet$]
The fraction of events removed by the \ptrat\ cut is
sensitive to the amount of $p_{Th}$ lost in the forward beampipe and so
also to the details of the fragmentation. We removed the \ptrat\ cut,
 while
extending the correction curves to include the region of very low 
\ptrat \{25\}. This resulted in changes of 3\% percent
at low \y .
\item[$\bullet$]
We determined the correction functions without allowing for 
a dependence on $p_{Th}$ \{26\},
resulting in changes of less than 2\perc\ in \Ft\ .
Reducing, by a factor of two, the size of $\gammah$ intervals in which the correction
function is determined resulted in changes of typically 1-2\perc \{27\}.
\end{itemize} }
\item[6] Background subtraction 
\begin{itemize}
\item[$\bullet$]
As described above, 
the photoproduction background as given by 
the PYTHIA MC simulation 
agrees with the determination from the fitting procedure 
described in section~\ref{ANA} (see Fig.~\ref{f:php}), which in turn agrees
with the photoproduction contribution extracted from events tagged by the 
LUMI positron detector. Uncertainties in the LUMI positron detection 
efficiency, the assumed shape in the fitting procedure and the positron
misidentification in the PYTHIA MC simulation lead to an overall uncertainty
in the photoproduction background of $\pm$50\perc \{28,29\}.
This results in an uncertainty in \Ft\ which is at most 3\perc\ for high \y\
and negligible elsewhere.
\end{itemize}
\end{itemize}
The total systematic uncertainty for all bins 
is determined by adding in quadrature separately the positive
and negative deviations from the above mentioned categories.

At high $Q^2$ ($> 100 $ GeV$^2$), 
a significant contribution to the apparent systematic deviations
is due to the limited number of events in the data. 
From the lower $Q^2$ region we know that
the systematic uncertainties depend mainly on \y.
Therefore, all bins with $Q^2>100$ GeV$^2$ and $\y > 0.01$ are combined, for
the systematic error determination,
in fixed \y\ intervals. 
The uncertainty in the track matching efficiency
contributes only at $Q^2 > $ 350 GeV$^2$ and is taken into account
separately.

Figure~\ref{f:SYSER} shows the positive and negative systematic errors for
each bin with $Q^2 <$ 100 GeV$^2$ as a function of \y .
For the majority of the bins the total systematic uncertainty
is  below 5\perc .
As \y\ increases above $y=0.5$ the uncertainty 
grows to around 10\perc.
For $y<0.01$ the uncertainty increases mainly because of 
the CAL noise.

The combined results for the bins with $Q^2 >$ 100 GeV$^2$ are also shown.

The systematic errors do not include the uncertainty  
in the measurement of the integrated luminosity ($\pm 1.5\perc$), 
the overall trigger efficiency ($\pm 1.0\perc$)
or the uncertainty 
due to higher order electroweak radiative corrections ($\pm 0.5\perc$). 
These effects lead to a combined 
uncertainty of  2\perc\ on the overall normalisation of \Ft.

In addition to the above studies a complete analysis was done using
the electron (EL) method of kinematic reconstruction. In the region
of high $y$, where this method is reliable, the results were 
compatible with this analysis.
 
\section{Results}
The values of \Ft\  are given in Tables~\ref{t:final} and 
\ref{t:final2} together with their 
systematic and statistical uncertainties. 
In the appendix we give the $F_2$ values with
 $Q^2 < 100~{\rm GeV^2}$ or $y<0.01$ obtained from  
each systematic check used in the calculation of the systematic uncertainties.
Also shown in Tables~\ref{t:final} and 
\ref{t:final2} are the number of events
and the estimated number of background events in each bin as well as the 
correction due to the longitudinal structure function, $\delta_{_{L}}$.
The \Ft\  values are displayed versus \x\ for fixed values of \qsd\ in 
Fig.~11 together with the low \qsd\ data 
from our experiment~\cite{b:LOWQ} and 
the high \x\ results from NMC~\cite{b:NMC}.
The rise of \Ft\ for $x\rightarrow 0$ is measured with much improved 
precision. 
Also shown is the result of the QCD NLO fit used 
for the acceptance determination.
The data from this analysis at large \x\ 
reach the \x\ range covered by the fixed target
experiments. In the 
overlap region the agreement is good.  
The results agree well with
the recently published results of the H1 collaboration~\cite{b:H11994} 
in the kinematic region where the data sets overlap.

Figure~\ref{f:f2vsq} shows the \Ft\  values as a function of \qsd\ for
fixed \x . Scaling violations are observed, 
which decrease as \x\ increases.
The \qsd -range  
of the ZEUS data 
has increased substantially with respect to 
our previous data. The data now span
more than a decade in \qsd\ for \x\ values 
as low as $x=4\cdot 10^{-4}$ and almost three decades at $x\approx 0.1$.

In Fig.~13 the data are shown
together with the  NLO
predictions of GRV94~\cite{b:GRV94} and the NLO parameterisations
MRSA$^\prime$~\cite{b:MRSA}
 and CTEQ3~\cite{b:CTEQ}. 
The data of the fixed target
experiments~\cite{b:NMC,b:E665,b:BCDMS,b:SLAC} are also shown.
At large \qsd\ all parameterisations represent the data 
well. For $\qsd \lsim 45\;\Gevsq$ CTEQ3 undershoots the measured \Ft\ values
in the region $10^{-4} \lsim x \lsim 10^{-2}$. MRSA$^\prime$ overshoots the 
data below $x=10^{-3}$ at low \qsd\ ($\qsd \lsim 8.5\;\Gevsq$), whereas GRV94
overshoots the data at $\qsd \lsim 18\;\Gevsq$ and $x<10^{-3}$. Our QCD NLO
fit reproduces the data over the full kinematic range indicating that 
NLO DGLAP evolution can give a consistent description of the data.
The details of this fit will be included in a forthcoming paper.

\section{Conclusions}

We have presented new measurements of the proton structure function \Ft\ 
from an analysis of inelastic positron proton neutral current scattering
using data obtained with the ZEUS detector at HERA during 1994.
A new method for determining the event kinematics has allowed us
to measure  \Ft\ over a substantially larger phase space than before
by this experiment 
and with systematic uncertainties reduced to below 5\perc\ in most 
of the ($\x,\qsd$) space. The data cover \qsd\ values between 3.2 and 15000~\Gevsq\ 
and \x\ values between $5\cdot 10^{-5}$ and $0.8$.

At large \x\ values
 and \qsd\ values up to 70~\Gevsq, where the new data reach the 
\x -range covered by fixed-target experiments, good agreement with these 
experiments is observed. The data show the rise of \Ft\ towards small \x\ 
with much improved precision. Strong scaling violations are observed for
$x<0.02$. The measured \x -\qsd\ behaviour of \Ft\ can be described 
by QCD using NLO DGLAP evolution in the full kinematic range.

\section{Acknowledgements}
The strong support and encouragement of the DESY Directorate has
been invaluable. 
The experiment was made possible by the inventiveness and the diligent
efforts of the HERA machine group.  The design, construction and 
installation of the ZEUS detector have been made possible by the
ingenuity and dedicated efforts of many people from inside DESY and
from the home institutes who are not listed as authors. Their 
contributions are acknowledged with great appreciation. 
Useful conversations with J. Bl\"{u}mlein, S. Riemersma, 
H. Spiesberger and J.~Smith are also gratefully acknowledged.

\newpage

\topcaption{\it The measured $F_2$ values.  The values of $x$ 
and $Q^2$ at which $F_2$ are determined
are shown outside the brackets, while the bin boundaries are shown inside
the brackets.  The number of events in each bin as well as the
estimated number of background events are given.  The corrections for
$F_L$, $\delta_L$ (see eq.~\ref{F2FORM}), are also given.
}
\label{t:final}
\tablefirsthead{\head}
\tablehead{\cpt \head}
\tabletail{\hline}
\tablelasttail{\\ \hline}
\begin{center}
\begin{supertabular}{|c|c|c|c|c|}
\mvqbin{     3.5}{    3.2}{    4.0}
\mva{ 6.3}{5.0}{8.0}{5}{ 2196}{  101}
\mvd{ 0.875}{0.052}{0.083}{0.101}{ 7.8}
\mva{ 1.0}{0.8}{1.3}{4}{ 3721}{  128}
\mvd{ 0.939}{0.040}{0.063}{0.072}{ 2.4}
\mvqbin{     4.5}{    4.0}{    5.0}
\mva{ 1.0}{0.8}{1.3}{4}{ 5131}{  211}
\mvd{ 1.030}{0.028}{0.061}{0.053}{ 4.5}
\mva{ 1.6}{1.3}{2.0}{4}{ 4671}{  103}
\mvd{ 1.038}{0.028}{0.037}{0.034}{ 1.5}
\mva{ 2.5}{2.0}{3.2}{4}{ 2967}{   39}
\mvd{ 0.932}{0.029}{0.031}{0.033}{ 0.5}
\mvqbin{     6.5}{    5.0}{    7.0}
\mva{ 1.0}{0.8}{1.3}{4}{ 3192}{  179}
\mvd{ 1.221}{0.036}{0.098}{0.076}{11.8}
\mva{ 1.6}{1.3}{2.0}{4}{ 8529}{  245}
\mvd{ 1.138}{0.021}{0.051}{0.034}{ 3.6}
\mva{ 2.5}{2.0}{3.2}{4}{ 9101}{  150}
\mvd{ 1.038}{0.018}{0.021}{0.026}{ 1.2}
\mva{ 4.0}{3.2}{5.0}{4}{ 6911}{   50}
\mvd{ 0.951}{0.019}{0.031}{0.028}{ 0.4}
\mva{ 6.3}{5.0}{8.0}{4}{ 5909}{   60}
\mvd{ 0.841}{0.018}{0.026}{0.017}{ 0.2}
\mva{ 1.0}{0.8}{1.3}{3}{ 5278}{   10}
\mvd{ 0.776}{0.017}{0.036}{0.028}{ 0.1}
\mva{ 1.6}{1.3}{2.0}{3}{ 4655}{   20}
\mvd{ 0.753}{0.019}{0.024}{0.030}{ 0.0}
\mvqbin{     6.5}{    5.0}{    9.0}
\mva{ 2.5}{2.0}{3.2}{3}{ 9522}{   90}
\mvd{ 0.645}{0.012}{0.032}{0.024}{ 0.0}
\mva{ 4.0}{3.2}{8.0}{3}{15723}{   40}
\mvd{ 0.629}{0.009}{0.018}{0.028}{ 0.0}
\mva{ 1.6}{0.8}{3.2}{2}{16255}{   20}
\mvd{ 0.495}{0.008}{0.037}{0.023}{ 0.0}
\mva{ 0.4}{0.3}{1.3}{1}{ 1736}{    0}
\mvd{ 0.403}{0.019}{0.149}{0.073}{ 0.0}
\mvqbin{     8.5}{    7.0}{    9.0}
\mva{ 1.6}{1.3}{2.0}{4}{ 3666}{  109}
\mvd{ 1.337}{0.037}{0.060}{0.081}{ 7.1}
\mva{ 2.5}{2.0}{3.2}{4}{ 6502}{  132}
\mvd{ 1.151}{0.024}{0.037}{0.023}{ 2.3}
\mva{ 4.0}{3.2}{5.0}{4}{ 6000}{   77}
\mvd{ 0.993}{0.021}{0.028}{0.019}{ 0.8}
\mva{ 6.3}{5.0}{8.0}{4}{ 5983}{   40}
\mvd{ 0.920}{0.019}{0.014}{0.025}{ 0.3}
\mva{ 1.0}{0.8}{1.3}{3}{ 5827}{   50}
\mvd{ 0.831}{0.018}{0.031}{0.025}{ 0.1}
\mva{ 1.6}{1.3}{2.0}{3}{ 4657}{   30}
\mvd{ 0.704}{0.017}{0.023}{0.015}{ 0.0}
\mvqbin{    10.0}{    9.0}{   11.0}
\mva{ 1.6}{1.3}{2.0}{4}{  870}{   35}
\mvd{ 1.359}{0.075}{0.114}{0.078}{10.9}
\mva{ 2.5}{2.0}{3.2}{4}{ 3904}{  106}
\mvd{ 1.231}{0.032}{0.043}{0.043}{ 3.4}
\mva{ 4.0}{3.2}{5.0}{4}{ 4415}{   57}
\mvd{ 1.136}{0.026}{0.030}{0.022}{ 1.1}
\mva{ 6.3}{5.0}{8.0}{4}{ 4479}{   28}
\mvd{ 0.968}{0.022}{0.030}{0.021}{ 0.4}
\mva{ 1.0}{0.8}{1.3}{3}{ 4370}{   48}
\mvd{ 0.857}{0.021}{0.024}{0.023}{ 0.1}
\mva{ 1.6}{1.3}{2.0}{3}{ 3737}{   10}
\mvd{ 0.792}{0.021}{0.021}{0.022}{ 0.1}
\mva{ 2.5}{2.0}{3.2}{3}{ 3971}{   50}
\mvd{ 0.715}{0.018}{0.034}{0.028}{ 0.0}
\mvqbin{    10.0}{    9.0}{   13.0}
\mva{ 6.3}{3.2}{8.0}{3}{12284}{   40}
\mvd{ 0.593}{0.008}{0.022}{0.021}{ 0.0}
\mva{ 1.6}{0.8}{3.2}{2}{15549}{   30}
\mvd{ 0.517}{0.007}{0.013}{0.009}{ 0.0}
\mva{ 0.8}{0.3}{1.3}{1}{ 3685}{    0}
\mvd{ 0.395}{0.010}{0.098}{0.059}{ 0.0}
\mvqbin{    12.0}{   11.0}{   13.0}
\mva{ 2.5}{2.0}{3.2}{4}{ 2311}{   94}
\mvd{ 1.300}{0.042}{0.059}{0.057}{ 5.4}
\mva{ 4.0}{3.2}{5.0}{4}{ 3052}{   50}
\mvd{ 1.128}{0.031}{0.036}{0.024}{ 1.7}
\mva{ 6.3}{5.0}{8.0}{4}{ 3339}{   20}
\mvd{ 1.102}{0.027}{0.020}{0.037}{ 0.6}
\mva{ 1.0}{0.8}{1.3}{3}{ 3242}{   40}
\mvd{ 0.903}{0.023}{0.031}{0.029}{ 0.2}
\mva{ 1.6}{1.3}{2.0}{3}{ 2668}{   70}
\mvd{ 0.789}{0.023}{0.028}{0.021}{ 0.1}
\mva{ 2.5}{2.0}{3.2}{3}{ 2849}{   10}
\mvd{ 0.746}{0.021}{0.030}{0.024}{ 0.0}
\mvqbin{    15.0}{   13.0}{   16.0}
\mva{ 2.5}{2.0}{3.2}{4}{ 1463}{   84}
\mvd{ 1.559}{0.065}{0.093}{0.105}{ 9.5}
\mva{ 4.0}{3.2}{5.0}{4}{ 3247}{   62}
\mvd{ 1.338}{0.035}{0.051}{0.029}{ 2.9}
\mva{ 6.3}{5.0}{8.0}{4}{ 3421}{   65}
\mvd{ 1.092}{0.027}{0.029}{0.030}{ 1.0}
\mva{ 1.0}{0.8}{1.3}{3}{ 3487}{   20}
\mvd{ 0.992}{0.025}{0.035}{0.034}{ 0.3}
\mva{ 1.6}{1.3}{2.0}{3}{ 2978}{   20}
\mvd{ 0.908}{0.024}{0.021}{0.025}{ 0.1}
\mva{ 2.5}{2.0}{3.2}{3}{ 3022}{    0}
\mvd{ 0.793}{0.021}{0.037}{0.012}{ 0.0}
\mva{ 6.3}{3.2}{8.0}{3}{ 5535}{   10}
\mvd{ 0.662}{0.014}{0.020}{0.025}{ 0.0}
\mvqbin{    15.0}{   13.0}{   20.0}
\mva{ 1.6}{0.8}{3.2}{2}{14175}{   50}
\mvd{ 0.553}{0.007}{0.014}{0.015}{ 0.0}
\mva{ 0.8}{0.3}{1.3}{1}{ 5160}{    0}
\mvd{ 0.370}{0.008}{0.061}{0.040}{ 0.0}
\mvqbin{    18.0}{   16.0}{   20.0}
\mva{ 4.0}{3.2}{5.0}{4}{ 2672}{   85}
\mvd{ 1.438}{0.041}{0.059}{0.037}{ 4.5}
\mva{ 6.3}{5.0}{8.0}{4}{ 3006}{   87}
\mvd{ 1.166}{0.031}{0.038}{0.037}{ 1.5}
\mva{ 1.0}{0.8}{1.3}{3}{ 3118}{   39}
\mvd{ 1.015}{0.028}{0.037}{0.025}{ 0.5}
\mva{ 1.6}{1.3}{2.0}{3}{ 2682}{   29}
\mvd{ 0.932}{0.026}{0.017}{0.028}{ 0.2}
\mva{ 2.5}{2.0}{3.2}{3}{ 2737}{   10}
\mvd{ 0.879}{0.024}{0.027}{0.017}{ 0.1}
\mva{ 6.3}{3.2}{8.0}{3}{ 4755}{    0}
\mvd{ 0.629}{0.013}{0.019}{0.021}{ 0.0}
\mvqbin{    22.0}{   20.0}{   25.0}
\mva{ 4.0}{3.2}{5.0}{4}{ 1363}{   50}
\mvd{ 1.507}{0.058}{0.101}{0.095}{ 7.5}
\mva{ 6.3}{5.0}{8.0}{4}{ 2568}{   60}
\mvd{ 1.324}{0.037}{0.033}{0.051}{ 2.3}
\mva{ 1.0}{0.8}{1.3}{3}{ 2498}{    8}
\mvd{ 1.081}{0.028}{0.030}{0.031}{ 0.8}
\mva{ 1.6}{1.3}{2.0}{3}{ 2136}{   20}
\mvd{ 0.930}{0.026}{0.039}{0.025}{ 0.3}
\mva{ 2.5}{2.0}{3.2}{3}{ 2240}{    0}
\mvd{ 0.906}{0.025}{0.020}{0.026}{ 0.1}
\mva{ 4.0}{3.2}{5.0}{3}{ 1892}{    0}
\mvd{ 0.749}{0.022}{0.039}{0.033}{ 0.0}
\mva{ 6.3}{5.0}{8.0}{3}{ 2003}{    0}
\mvd{ 0.701}{0.021}{0.032}{0.020}{ 0.0}
\mvqbin{    22.0}{   20.0}{   32.0}
\mva{ 1.0}{0.8}{1.3}{2}{ 3613}{    0}
\mvd{ 0.601}{0.013}{0.018}{0.030}{ 0.0}
\mva{ 2.5}{1.3}{5.0}{2}{ 9701}{    0}
\mvd{ 0.490}{0.007}{0.012}{0.011}{ 0.0}
\mva{ 0.8}{0.5}{1.3}{1}{ 2325}{    0}
\mvd{ 0.338}{0.009}{0.067}{0.050}{ 0.0}
\mvqbin{    27.0}{   25.0}{   32.0}
\mva{ 6.3}{5.0}{8.0}{4}{ 2314}{   60}
\mvd{ 1.460}{0.041}{0.047}{0.027}{ 3.8}
\mva{ 1.0}{0.8}{1.3}{3}{ 2356}{   35}
\mvd{ 1.194}{0.032}{0.030}{0.021}{ 1.2}
\mva{ 1.6}{1.3}{2.0}{3}{ 2060}{    0}
\mvd{ 1.140}{0.033}{0.030}{0.042}{ 0.4}
\mva{ 2.5}{2.0}{3.2}{3}{ 1854}{   11}
\mvd{ 0.928}{0.028}{0.024}{0.021}{ 0.2}
\mva{ 4.0}{3.2}{5.0}{3}{ 1695}{    8}
\mvd{ 0.765}{0.024}{0.037}{0.018}{ 0.1}
\mva{ 6.3}{5.0}{8.0}{3}{ 1709}{    2}
\mvd{ 0.674}{0.021}{0.021}{0.019}{ 0.0}
\mvqbin{    35.0}{   32.0}{   40.0}
\mva{ 6.3}{5.0}{8.0}{4}{  966}{   35}
\mvd{ 1.565}{0.073}{0.066}{0.046}{ 7.3}
\mva{ 1.0}{0.8}{1.3}{3}{ 1756}{   25}
\mvd{ 1.370}{0.044}{0.014}{0.043}{ 2.2}
\mva{ 1.6}{1.3}{2.0}{3}{ 1543}{    8}
\mvd{ 1.149}{0.038}{0.023}{0.028}{ 0.7}
\mva{ 2.5}{2.0}{3.2}{3}{ 1631}{    0}
\mvd{ 1.018}{0.032}{0.040}{0.032}{ 0.3}
\mva{ 4.0}{3.2}{5.0}{3}{ 1423}{    0}
\mvd{ 0.887}{0.030}{0.034}{0.014}{ 0.1}
\mva{ 6.3}{5.0}{8.0}{3}{ 1306}{    0}
\mvd{ 0.749}{0.027}{0.040}{0.013}{ 0.0}
\mva{ 1.0}{0.8}{1.3}{2}{ 1307}{   10}
\mvd{ 0.681}{0.025}{0.013}{0.036}{ 0.0}
\mvqbin{    35.0}{   32.0}{   50.0}
\mva{ 1.6}{1.3}{2.0}{2}{ 2026}{    0}
\mvd{ 0.604}{0.017}{0.017}{0.022}{ 0.0}
\mva{ 2.5}{2.0}{5.0}{2}{ 3845}{    0}
\mvd{ 0.570}{0.012}{0.025}{0.007}{ 0.0}
\mva{ 0.8}{0.5}{1.3}{1}{ 2323}{    0}
\mvd{ 0.461}{0.013}{0.054}{0.049}{ 0.0}
\mvqbin{    45.0}{   40.0}{   50.0}
\mva{ 1.0}{0.8}{1.3}{3}{ 1431}{   25}
\mvd{ 1.441}{0.053}{0.074}{0.035}{ 3.9}
\mva{ 1.6}{1.3}{2.0}{3}{ 1293}{   33}
\mvd{ 1.178}{0.045}{0.036}{0.045}{ 1.3}
\mva{ 2.5}{2.0}{3.2}{3}{ 1390}{   11}
\mvd{ 1.071}{0.037}{0.016}{0.025}{ 0.4}
\mva{ 4.0}{3.2}{5.0}{3}{ 1172}{   10}
\mvd{ 0.894}{0.034}{0.031}{0.022}{ 0.2}
\mva{ 6.3}{5.0}{8.0}{3}{ 1129}{    0}
\mvd{ 0.804}{0.031}{0.057}{0.033}{ 0.1}
\mva{ 1.0}{0.8}{1.3}{2}{ 1108}{    0}
\mvd{ 0.745}{0.029}{0.031}{0.043}{ 0.0}
\mvqbin{    60.0}{   50.0}{   65.0}
\mva{ 1.0}{0.8}{1.3}{3}{  790}{    8}
\mvd{ 1.553}{0.073}{0.067}{0.066}{ 8.0}
\mva{ 1.6}{1.3}{2.0}{3}{ 1250}{    8}
\mvd{ 1.339}{0.050}{0.039}{0.024}{ 2.5}
\mva{ 2.5}{2.0}{3.2}{3}{ 1294}{    8}
\mvd{ 1.100}{0.040}{0.033}{0.030}{ 0.8}
\mva{ 4.0}{3.2}{5.0}{3}{ 1166}{    0}
\mvd{ 1.012}{0.039}{0.022}{0.041}{ 0.3}
\mva{ 6.3}{5.0}{8.0}{3}{ 1118}{    0}
\mvd{ 0.890}{0.035}{0.044}{0.020}{ 0.1}
\mva{ 1.0}{0.8}{1.3}{2}{  952}{    0}
\mvd{ 0.672}{0.028}{0.025}{0.014}{ 0.0}
\mva{ 1.6}{1.3}{2.0}{2}{  760}{    0}
\mvd{ 0.614}{0.029}{0.053}{0.024}{ 0.0}
\mva{ 2.5}{2.0}{5.0}{2}{ 1719}{    0}
\mvd{ 0.627}{0.020}{0.020}{0.010}{ 0.0}
\mva{ 0.8}{0.5}{1.3}{1}{ 1391}{    0}
\mvd{ 0.521}{0.019}{0.046}{0.040}{ 0.0}
\mvqbin{    70.0}{   65.0}{   85.0}
\mva{ 1.6}{1.3}{2.0}{3}{  862}{   25}
\mvd{ 1.318}{0.062}{0.066}{0.041}{ 3.6}
\mva{ 2.5}{2.0}{3.2}{3}{ 1107}{   20}
\mvd{ 1.228}{0.049}{0.021}{0.075}{ 1.2}
\mva{ 4.0}{3.2}{5.0}{3}{  885}{    0}
\mvd{ 0.917}{0.039}{0.043}{0.013}{ 0.4}
\mva{ 6.3}{5.0}{8.0}{3}{  852}{   10}
\mvd{ 0.887}{0.040}{0.033}{0.031}{ 0.1}
\mva{ 1.0}{0.8}{1.3}{2}{  816}{    0}
\mvd{ 0.774}{0.035}{0.037}{0.036}{ 0.0}
\mva{ 1.6}{1.3}{2.0}{2}{  616}{    0}
\mvd{ 0.697}{0.037}{0.036}{0.024}{ 0.0}
\mva{ 2.5}{2.0}{5.0}{2}{ 1390}{    0}
\mvd{ 0.607}{0.021}{0.019}{0.025}{ 0.0}
\mva{ 0.8}{0.5}{1.3}{1}{ 1244}{    0}
\mvd{ 0.484}{0.018}{0.029}{0.026}{ 0.0}
\mvqbin{    90.0}{   85.0}{  110.0}
\mva{ 1.6}{1.3}{2.0}{3}{  299}{    0}
\mvd{ 1.363}{0.101}{0.114}{0.050}{ 6.6}
\mva{ 2.5}{2.0}{3.2}{3}{  746}{    8}
\mvd{ 1.275}{0.063}{0.038}{0.045}{ 2.1}
\mva{ 4.0}{3.2}{5.0}{3}{  702}{    8}
\mvd{ 1.068}{0.054}{0.046}{0.031}{ 0.7}
\mva{ 6.3}{5.0}{8.0}{3}{  689}{    0}
\mvd{ 0.974}{0.049}{0.035}{0.020}{ 0.2}
\mva{ 1.0}{0.8}{1.3}{2}{  667}{    0}
\mvd{ 0.845}{0.043}{0.033}{0.037}{ 0.1}
\mva{ 1.6}{1.3}{2.0}{2}{  456}{    0}
\mvd{ 0.619}{0.037}{0.033}{0.058}{ 0.0}
\mva{ 2.5}{2.0}{3.2}{2}{  468}{    0}
\mvd{ 0.556}{0.033}{0.039}{0.019}{ 0.0}
\mva{ 4.0}{3.2}{5.0}{2}{  473}{    0}
\mvd{ 0.526}{0.031}{0.035}{0.032}{ 0.0}
\mva{ 0.8}{0.5}{1.3}{1}{  987}{    0}
\mvd{ 0.453}{0.019}{0.037}{0.019}{ 0.0}
\mva{ 2.0}{1.3}{3.2}{1}{  122}{    0}
\mvd{ 0.255}{0.029}{0.080}{0.037}{ 0.0}
\mvqbin{   120.0}{  110.0}{  140.0}
\mva{ 2.5}{2.0}{3.2}{3}{  477}{    8}
\mvd{ 1.467}{0.093}{0.038}{0.030}{ 4.1}
\mva{ 4.0}{3.2}{5.0}{3}{  519}{   17}
\mvd{ 1.176}{0.073}{0.030}{0.024}{ 1.3}
\mva{ 6.3}{5.0}{8.0}{3}{  520}{    0}
\mvd{ 1.112}{0.066}{0.031}{0.032}{ 0.4}
\mva{ 1.0}{0.8}{1.3}{2}{  466}{    0}
\mvd{ 0.836}{0.051}{0.024}{0.016}{ 0.1}
\mva{ 1.6}{1.3}{2.0}{2}{  321}{    0}
\mvd{ 0.610}{0.043}{0.031}{0.023}{ 0.0}
\mva{ 2.5}{2.0}{3.2}{2}{  366}{    0}
\mvd{ 0.653}{0.045}{0.016}{0.022}{ 0.0}
\mva{ 4.0}{3.2}{5.0}{2}{  345}{    0}
\mvd{ 0.519}{0.036}{0.012}{0.017}{ 0.0}
\mva{ 0.8}{0.5}{1.3}{1}{  716}{    0}
\mvd{ 0.451}{0.022}{0.019}{0.019}{ 0.0}
\mva{ 2.0}{1.3}{3.2}{1}{  125}{    0}
\mvd{ 0.207}{0.022}{0.073}{0.031}{ 0.0}
\mvqbin{   150.0}{  140.0}{  185.0}
\mva{ 2.5}{2.0}{3.2}{3}{  159}{    0}
\mvd{ 1.275}{0.130}{0.051}{0.037}{ 7.1}
\mva{ 4.0}{3.2}{5.0}{3}{  376}{    8}
\mvd{ 1.013}{0.070}{0.026}{0.020}{ 2.1}
\mva{ 6.3}{5.0}{8.0}{3}{  418}{    0}
\mvd{ 0.991}{0.063}{0.028}{0.028}{ 0.7}
\mva{ 1.0}{0.8}{1.3}{2}{  382}{    0}
\mvd{ 0.835}{0.056}{0.024}{0.016}{ 0.2}
\mva{ 1.6}{1.3}{2.0}{2}{  316}{    0}
\mvd{ 0.685}{0.050}{0.020}{0.013}{ 0.1}
\mva{ 2.5}{2.0}{3.2}{2}{  314}{    0}
\mvd{ 0.719}{0.054}{0.036}{0.027}{ 0.0}
\mva{ 4.0}{3.2}{5.0}{2}{  297}{    0}
\mvd{ 0.522}{0.039}{0.012}{0.018}{ 0.0}
\mva{ 0.8}{0.5}{1.3}{1}{  573}{    0}
\mvd{ 0.420}{0.022}{0.017}{0.005}{ 0.0}
\mva{ 2.0}{1.3}{3.2}{1}{  204}{    0}
\mvd{ 0.288}{0.026}{0.033}{0.045}{ 0.0}
\mvqbin{   200.0}{  185.0}{  240.0}
\mva{ 4.0}{3.2}{5.0}{3}{  186}{    0}
\mvd{ 1.035}{0.096}{0.026}{0.021}{ 4.2}
\mva{ 6.3}{5.0}{8.0}{3}{  267}{    0}
\mvd{ 1.041}{0.084}{0.027}{0.021}{ 1.3}
\mva{ 1.0}{0.8}{1.3}{2}{  258}{    0}
\mvd{ 0.832}{0.067}{0.023}{0.024}{ 0.4}
\mva{ 1.6}{1.3}{2.0}{2}{  230}{    0}
\mvd{ 0.720}{0.062}{0.021}{0.014}{ 0.1}
\mva{ 2.5}{2.0}{3.2}{2}{  189}{    0}
\mvd{ 0.631}{0.060}{0.032}{0.024}{ 0.0}
\mva{ 4.0}{3.2}{5.0}{2}{  194}{    0}
\mvd{ 0.541}{0.050}{0.013}{0.018}{ 0.0}
\mva{ 0.8}{0.5}{1.3}{1}{  372}{    0}
\mvd{ 0.410}{0.027}{0.016}{0.005}{ 0.0}
\mva{ 2.0}{1.3}{3.2}{1}{  178}{    0}
\mvd{ 0.271}{0.025}{0.012}{0.011}{ 0.0}
\mvqbin{   250.0}{  240.0}{  310.0}
\mva{ 4.0}{3.2}{5.0}{3}{   67}{    0}
\mvd{ 1.353}{0.219}{0.054}{0.040}{ 7.2}
\mva{ 6.3}{5.0}{8.0}{3}{  184}{    0}
\mvd{ 1.176}{0.116}{0.030}{0.024}{ 2.1}
\mva{ 1.0}{0.8}{1.3}{2}{  200}{    0}
\mvd{ 0.916}{0.085}{0.026}{0.026}{ 0.6}
\mva{ 1.6}{1.3}{2.0}{2}{  147}{    0}
\mvd{ 0.699}{0.074}{0.020}{0.013}{ 0.2}
\mva{ 2.5}{2.0}{3.2}{2}{  127}{    0}
\mvd{ 0.544}{0.061}{0.016}{0.010}{ 0.1}
\mva{ 4.0}{3.2}{5.0}{2}{  148}{    0}
\mvd{ 0.583}{0.063}{0.029}{0.022}{ 0.0}
\mva{ 0.8}{0.5}{1.3}{1}{  275}{    0}
\mvd{ 0.435}{0.034}{0.010}{0.015}{ 0.0}
\mva{ 2.0}{1.3}{3.2}{1}{  148}{    0}
\mvd{ 0.267}{0.027}{0.011}{0.011}{ 0.0}
\mvqbin{   350.0}{  310.0}{  410.0}
\mva{ 6.3}{5.0}{8.0}{3}{  129}{    0}
\mvd{ 1.107}{0.127}{0.044}{0.032}{ 4.8}
\mva{ 1.0}{0.8}{1.3}{2}{  144}{    0}
\mvd{ 0.814}{0.086}{0.021}{0.016}{ 1.3}
\mva{ 1.6}{1.3}{2.0}{2}{  120}{    0}
\mvd{ 0.661}{0.076}{0.019}{0.019}{ 0.4}
\mva{ 2.5}{2.0}{3.2}{2}{   95}{    0}
\mvd{ 0.446}{0.055}{0.013}{0.009}{ 0.1}
\mva{ 4.0}{3.2}{5.0}{2}{  103}{    0}
\mvd{ 0.672}{0.089}{0.034}{0.025}{ 0.0}
\mva{ 0.8}{0.5}{1.3}{1}{  199}{    0}
\mvd{ 0.414}{0.037}{0.010}{0.014}{ 0.0}
\mva{ 2.0}{1.3}{3.2}{1}{  130}{    0}
\mvd{ 0.262}{0.029}{0.011}{0.003}{ 0.0}
\mvqbin{   450.0}{  410.0}{  530.0}
\mva{ 6.3}{5.0}{8.0}{3}{   51}{    0}
\mvd{ 1.722}{0.348}{0.082}{0.066}{ 8.7}
\mva{ 1.0}{0.8}{1.3}{2}{   89}{    0}
\mvd{ 0.929}{0.128}{0.033}{0.030}{ 2.4}
\mva{ 1.6}{1.3}{2.0}{2}{   84}{    0}
\mvd{ 0.864}{0.125}{0.032}{0.033}{ 0.7}
\mva{ 2.5}{2.0}{3.2}{2}{   89}{    0}
\mvd{ 0.688}{0.095}{0.026}{0.026}{ 0.2}
\mva{ 4.0}{3.2}{5.0}{2}{   52}{    0}
\mvd{ 0.435}{0.075}{0.017}{0.014}{ 0.1}
\mva{ 0.8}{0.5}{1.3}{1}{  148}{    0}
\mvd{ 0.455}{0.048}{0.026}{0.021}{ 0.0}
\mva{ 2.0}{1.3}{3.2}{1}{  115}{    0}
\mvd{ 0.348}{0.043}{0.016}{0.010}{ 0.0}
\mvqbin{   650.0}{  530.0}{  710.0}
\mva{ 1.0}{0.8}{1.3}{2}{   76}{    0}
\mvd{ 1.067}{0.163}{0.051}{0.041}{ 5.7}
\mva{ 1.6}{1.3}{2.0}{2}{   71}{    0}
\mvd{ 0.897}{0.142}{0.032}{0.029}{ 1.6}
\mva{ 2.5}{2.0}{3.2}{2}{   78}{    0}
\mvd{ 0.728}{0.109}{0.027}{0.028}{ 0.5}
\mva{ 4.0}{3.2}{5.0}{2}{   43}{   10}
\mvd{ 0.471}{0.102}{0.018}{0.018}{ 0.1}
\mva{ 0.8}{0.5}{1.3}{1}{  104}{    0}
\mvd{ 0.445}{0.056}{0.025}{0.020}{ 0.0}
\mva{ 2.0}{1.3}{3.2}{1}{   96}{    0}
\mvd{ 0.344}{0.046}{0.012}{0.014}{ 0.0}
\mvqbin{   800.0}{  710.0}{  900.0}
\mva{ 1.0}{0.8}{1.3}{2}{   22}{    0}
\mvd{ 0.890}{0.241}{0.042}{0.034}{ 9.3}
\mva{ 1.6}{1.3}{2.0}{2}{   40}{    0}
\mvd{ 0.819}{0.169}{0.029}{0.026}{ 2.6}
\mva{ 2.5}{2.0}{3.2}{2}{   48}{    0}
\mvd{ 0.657}{0.122}{0.023}{0.021}{ 0.8}
\mva{ 4.0}{3.2}{5.0}{2}{   32}{    0}
\mvd{ 0.512}{0.115}{0.019}{0.019}{ 0.2}
\mva{ 0.8}{0.5}{1.3}{1}{   54}{    0}
\mvd{ 0.293}{0.048}{0.011}{0.009}{ 0.0}
\mva{ 2.0}{1.3}{3.2}{1}{   45}{    0}
\mvdl{ 0.227}{0.041}{0.008}{0.010}{ 0.0}
\end{supertabular}
\end{center}

\begin{table}
\begin{center}
\caption{  \it
The measured $F_2$  and $F_2^{em}$ values for 
$Q^2 > 1000$ ${\rm GeV^2}$ (see text).  
The values of $x$ 
and $Q^2$ at which $F_2$  and $F_2^{em}$ are determined
are shown outside the brackets, while the bin boundaries are shown inside
the brackets.  The number of events in each bin as well as the
estimated number of background events are given.  The corrections
for $F_L$, $\delta_L$ (see eq.~\ref{F2FORM}), are also given.
}
\label{t:final2}
\begin{tabular}{|c|c|c|c|c|c|} 
\headhq
\mvqbinhq{  1200.0}{  900.0}{ 1300.0}
\mva{ 1.6}{1.3}{2.0}{2}{   34}{    0}
\pkd{ 0.645}{0.634}{0.135}{0.030}{0.024}{ 6.8}
\mva{ 2.5}{2.0}{3.2}{2}{   41}{    0}
\pkd{ 0.652}{0.641}{0.128}{0.023}{0.021}{ 1.9}
\mva{ 4.0}{3.2}{5.0}{2}{   32}{    0}
\pkd{ 0.586}{0.577}{0.132}{0.021}{0.019}{ 0.5}
\mva{ 0.8}{0.5}{1.3}{1}{   54}{    0}
\pkd{ 0.321}{0.316}{0.052}{0.012}{0.010}{ 0.1}
\mva{ 2.0}{1.3}{3.2}{1}{   48}{    0}
\pkd{ 0.308}{0.304}{0.057}{0.017}{0.014}{ 0.0}
\mva{ 5.1}{3.2}{8.0}{1}{   11}{    0}
\pkd{ 0.056}{0.055}{0.020}{0.003}{0.002}{ 0.0}
\mvqbinhq{  1500.0}{ 1300.0}{ 1800.0}
\mva{ 2.5}{2.0}{5.0}{2}{   52}{    0}
\pkd{ 0.614}{0.601}{0.105}{0.028}{0.023}{ 3.2}
\mva{ 0.8}{0.5}{1.3}{1}{   46}{    0}
\pkd{ 0.605}{0.593}{0.119}{0.022}{0.022}{ 0.1}
\mva{ 2.0}{1.3}{3.2}{1}{   29}{    0}
\pkd{ 0.216}{0.212}{0.048}{0.012}{0.010}{ 0.0}
\mvqbinhq{  2000.0}{ 1800.0}{ 2500.0}
\mva{ 4.0}{2.0}{5.0}{2}{   28}{    0}
\pkd{ 0.704}{0.682}{0.169}{0.024}{0.022}{ 1.7}
\mva{ 0.8}{0.5}{1.3}{1}{   19}{    0}
\pkd{ 0.362}{0.351}{0.099}{0.013}{0.013}{ 0.2}
\mva{ 2.0}{1.3}{3.2}{1}{   25}{    0}
\pkd{ 0.412}{0.401}{0.110}{0.015}{0.013}{ 0.0}
\mvqbinhq{  3000.0}{ 2500.0}{ 3500.0}
\mva{ 0.8}{0.2}{1.3}{1}{   15}{    0}
\pkd{ 0.238}{0.227}{0.067}{0.008}{0.007}{ 0.5}
\mva{ 2.0}{1.3}{8.0}{1}{   26}{    0}
\pkd{ 0.479}{0.458}{0.127}{0.017}{0.014}{ 0.0}
\mvqbinhq{  5000.0}{ 3500.0}{15000.0}
\mva{ 0.8}{0.2}{1.3}{1}{   12}{    0}
\pkd{ 0.613}{0.558}{0.209}{0.026}{0.022}{ 1.7}
\mva{ 2.0}{1.3}{8.0}{1}{   17}{    0}
\pkd{ 0.211}{0.194}{0.056}{0.007}{0.007}{ 0.1}
\hline
\end{tabular}
\end{center}
\end{table}
\clearpage
\newpage

\clearpage
\newpage

\clearpage

\clearpage
\clearpage

\clearpage

\newcounter{subfig}
\setcounter{subfig}{1}
\renewcommand{\thefigure}{\arabic{figure}\alph{subfig}}
\clearpage
\begin{figure}[p]
\begin{center}
{\mbox{\epsfxsize=14.5cm\epsffile{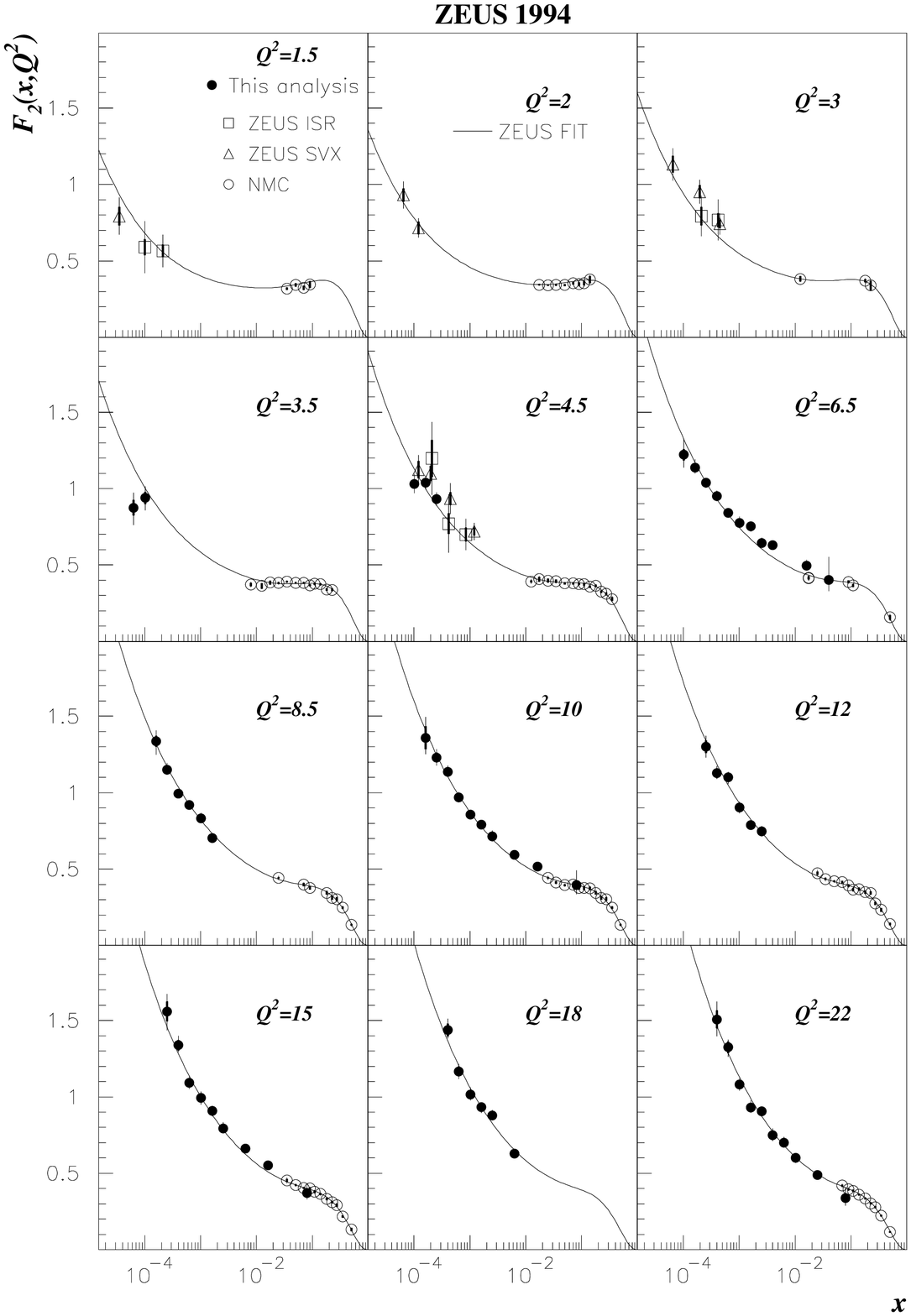}}}
\vspace*{-0.65cm}
\caption[]
{\it Structure function \Ft\  as a function of \x\ for 
fixed values of \qsd\ {\rm(}$1.5\;\Gevsq<\qsd<22\;\Gevsq$).The solid
dots correspond to the  data
from this analysis. The ``ISR'' (open squares) 
and ``SVX'' (open triangles) data show the ZEUS results obtained
previously. The inner thick error bars represent the
statistical error, the full error bars correspond to the statistical
and systematic errors added in quadrature.
The results from NMC are also shown (open circles).
The solid lines indicate the QCD NLO 
fit to the data used for the acceptance correction. 
The $Q^2$ values are indicated in units of ${\rm GeV^2}$.
}
\label{f:f2vsx}
\end{center}
\end{figure}
\addtocounter{figure}{-1}
\addtocounter{subfig}{1}
\clearpage
\begin{figure}[p]
\begin{center}
{\mbox{\epsfxsize=14.5cm\epsffile{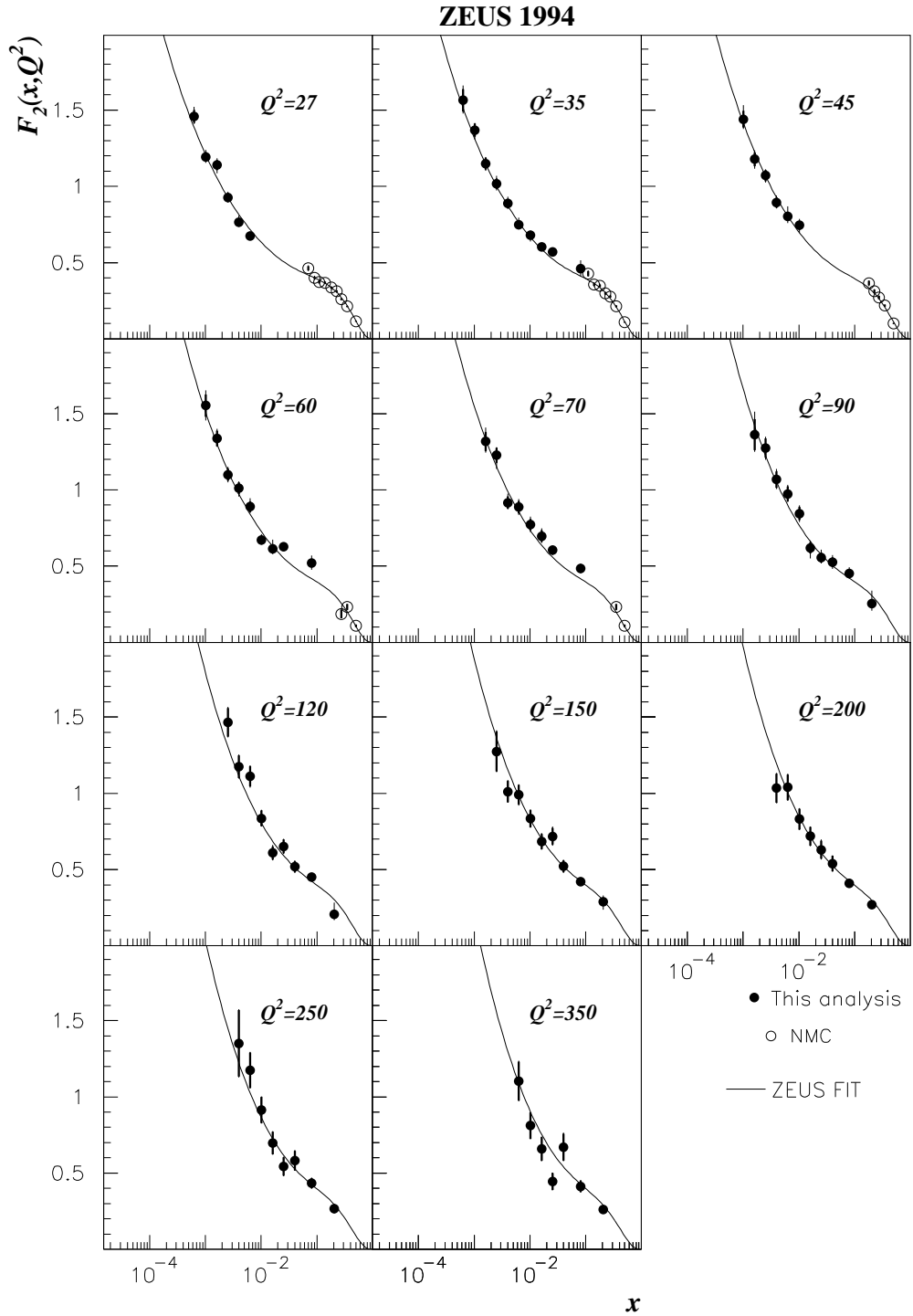}}}
\caption[\thefigure a]
{\it Structure function \Ft\  as a function of \x\ for 
fixed values of \qsd\ {\rm(}$27\;\Gevsq<\qsd<350\;\Gevsq$\rm{)}. 
The $Q^2$ values are indicated in units of ${\rm GeV^2}$.
}
\end{center}
\end{figure}

\addtocounter{figure}{-1}
\addtocounter{subfig}{1}
\clearpage
\begin{figure}[p]
\begin{center}
{\mbox{\epsfxsize=14.5cm\epsffile{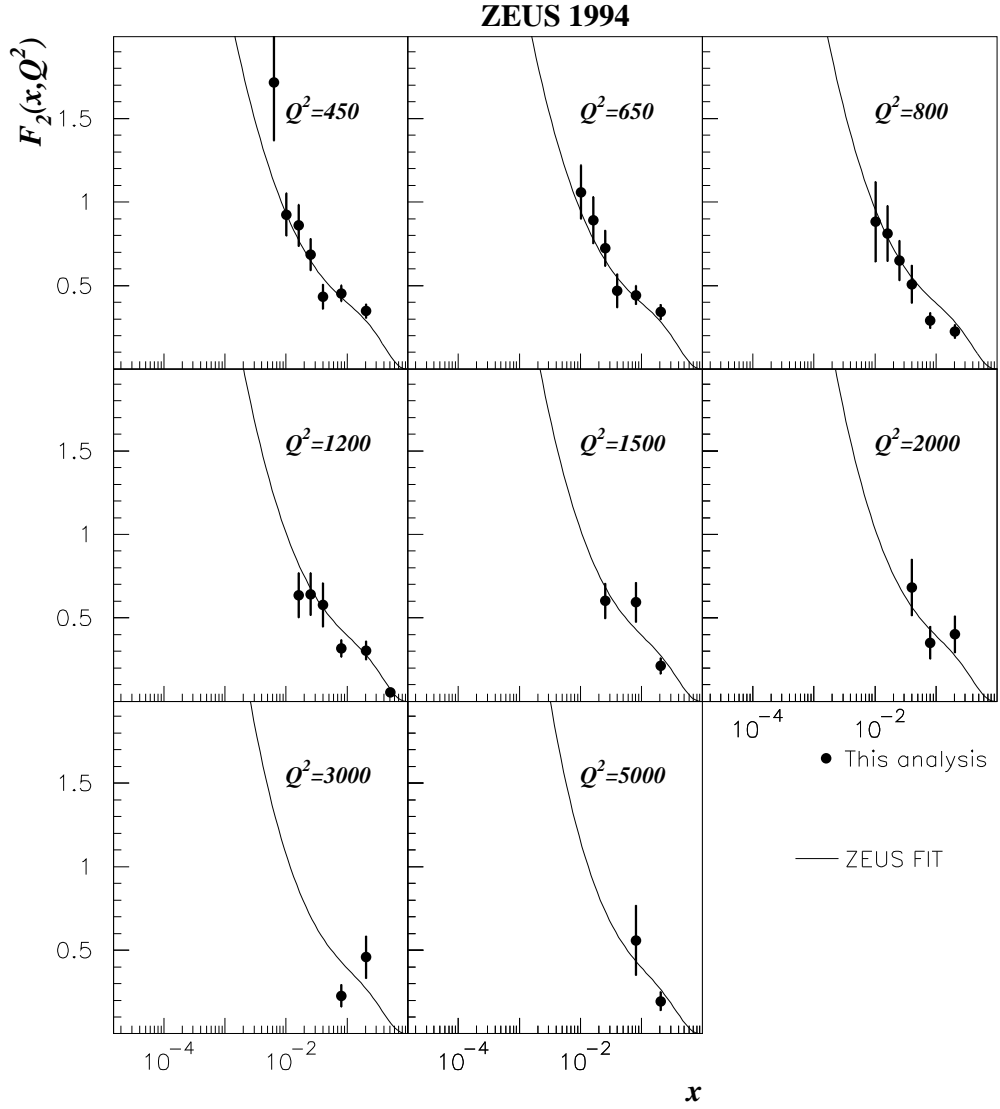}}}
\caption[]
{\it Structure function \Ft\ (\Fem\ for $Q^2>1000\;\Gevsq$) as a function of \x\ for 
fixed values of \qsd\ {\rm(}$450\;\Gevsq<\qsd<5000\;\Gevsq$\rm{)}.
The $Q^2$ values are indicated in units of ${\rm GeV^2}$.
}
\end{center}
\end{figure}

\setcounter{subfig}{1}
\renewcommand{\thefigure}{\arabic{figure}}

\clearpage
\begin{figure}[p]
\begin{center}
{\mbox{\epsfxsize=15cm\epsffile{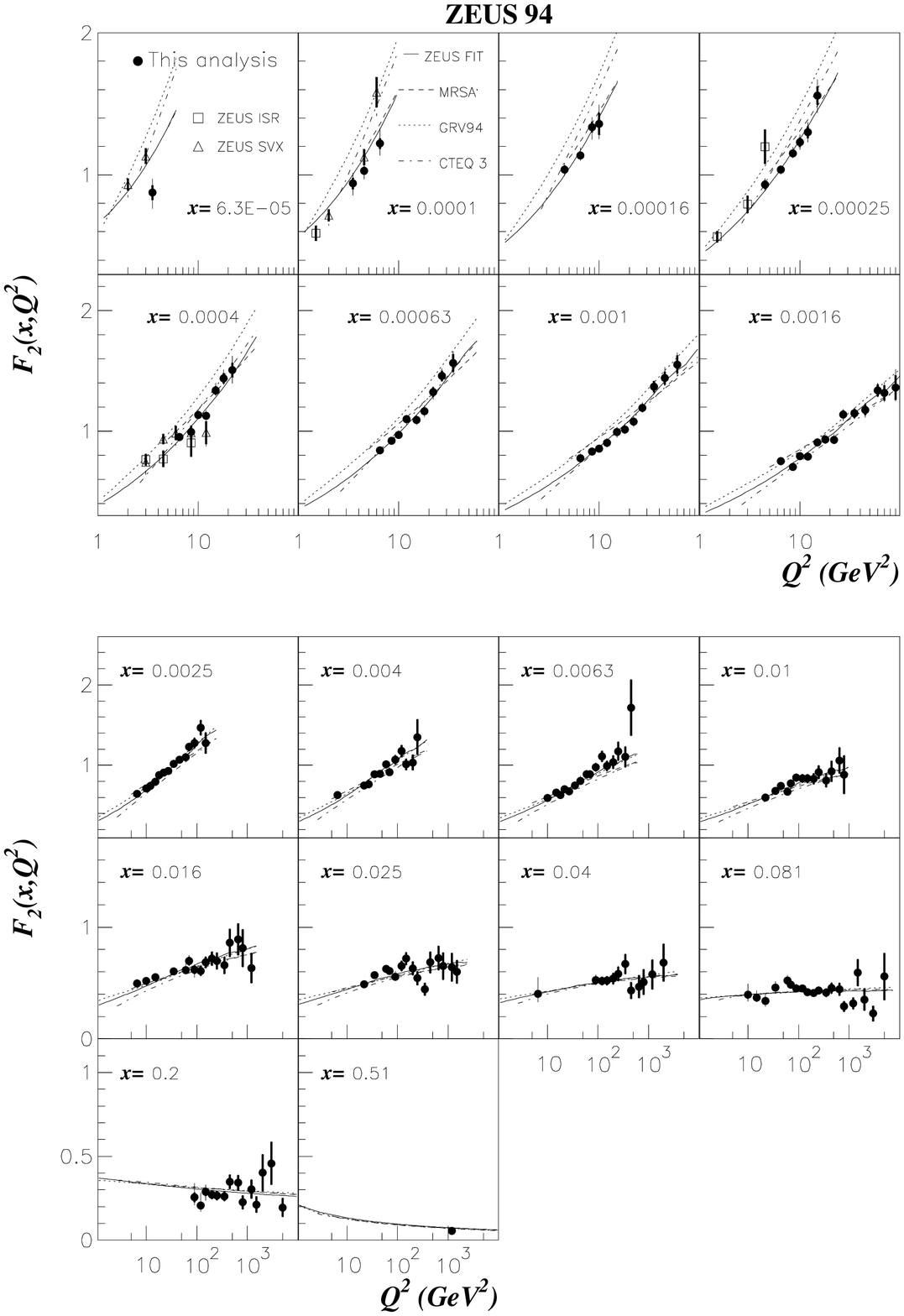}}}
\caption[]
{\it Structure function \Fem\  as a function of \qsd\ for 
fixed values of \x. The data from this analysis correspond to the 
solid dots. The ``ISR'' (open squares) 
and ``SVX'' (open triangles) data show the ZEUS results obtained
previously.
The solid line indicates the QCD NLO 
fit to the data used for acceptance correction. Also indicated are the
 MRSA$^\prime$, GRV94 and CTEQ3 parameterisations.}
\label{f:f2vsq}
\end{center}
\end{figure}

\setcounter{subfig}{1}
\renewcommand{\thefigure}{\arabic{figure}\alph{subfig}}

\clearpage
\begin{figure}[p]
\begin{center}
{\mbox{\epsfxsize=14.5cm\epsffile{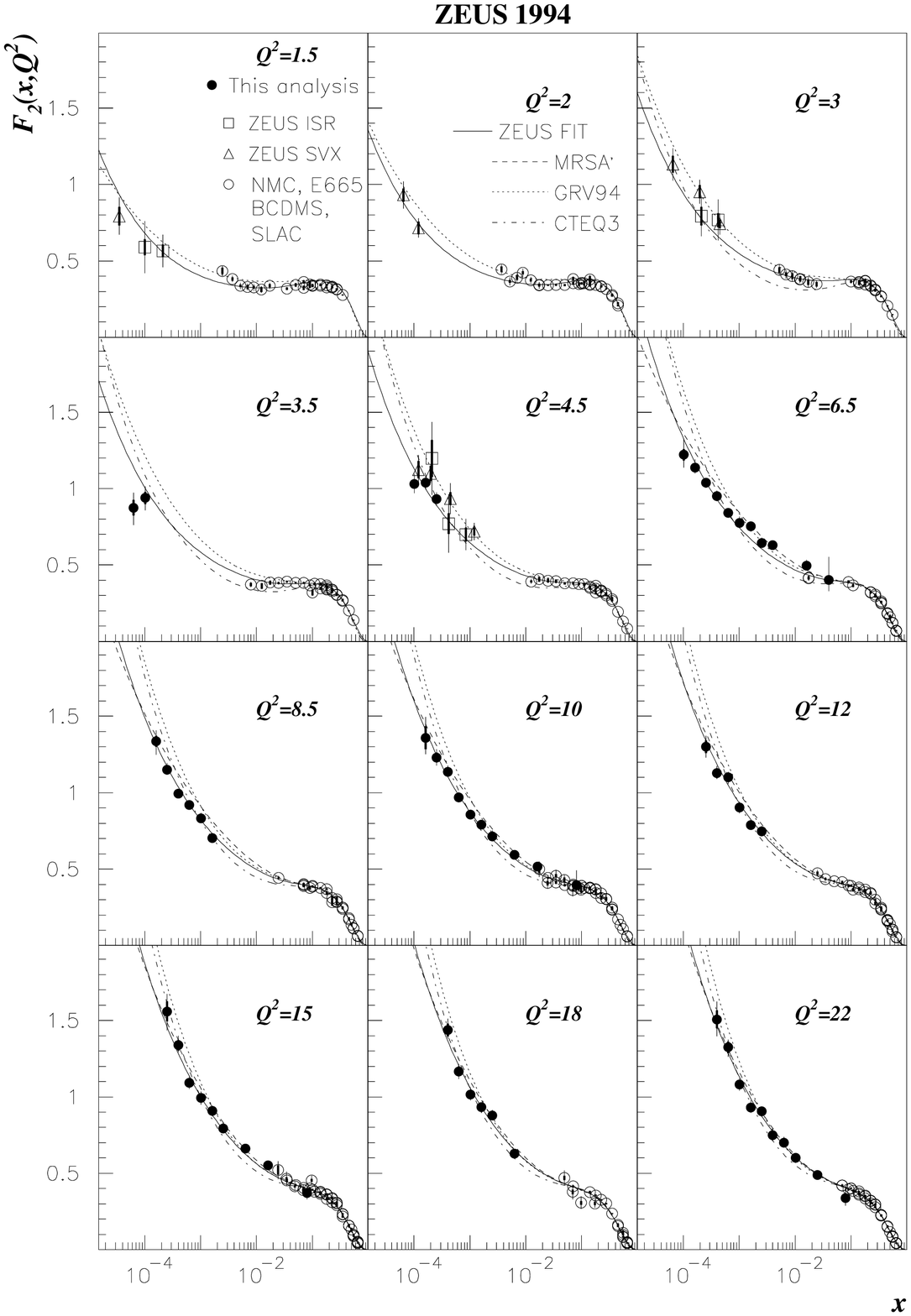}}}
\vspace*{-0.65cm}
\caption[]
{\it Structure function \Ft\  as a function of \x\ for 
fixed values of \qsd.
The solid dots
correspond to the  data
from this analysis. The ``ISR'' (open squares) 
and ``SVX'' (open triangles) data show the ZEUS results obtained
previously. 
The inner thick error bars represent the
statistical error, the full error bars correspond to the statistical
and systematic errors added in quadrature.
The results from NMC, E665, BCDMS and SLAC are also shown (open circles).
The solid line indicates the QCD NLO 
fit to the data used for acceptance correction.
Also indicated are the
 MRSA$^\prime$, GRV94 and CTEQ3 parameterisations.
The $Q^2$ values are given in ${\rm GeV^2}$.
}
\label{f:f2vsx2}
\end{center}
\end{figure}
\addtocounter{figure}{-1}
\addtocounter{subfig}{1}

\clearpage
\begin{figure}[p]
\begin{center}
{\mbox{\epsfxsize=14.5cm\epsffile{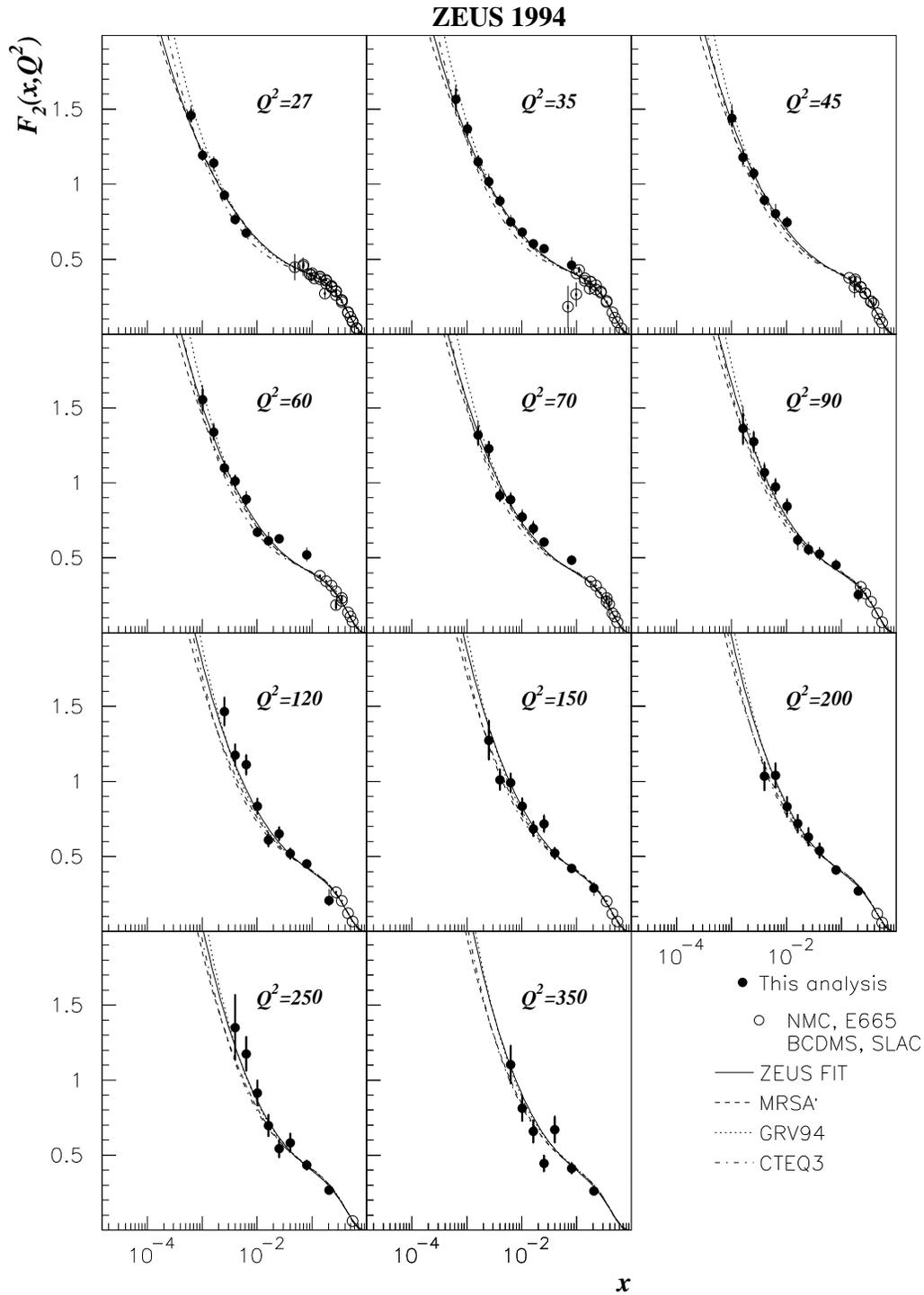}}}
\caption[]
{\it Structure function \Ft\  as a function of \x\ for 
fixed values of \qsd .
The solid line indicates the QCD NLO 
fit to the data used for acceptance correction.
Also indicated are the
 MRSA$^\prime$, GRV94 and CTEQ3 parameterisations.
The $Q^2$ values are given in ${\rm GeV^2}$.
}
\end{center}
\end{figure}

\addtocounter{figure}{-1}
\addtocounter{subfig}{1}

\clearpage
\begin{figure}[p]
\begin{center}
{\mbox{\epsfxsize=14.5cm\epsffile{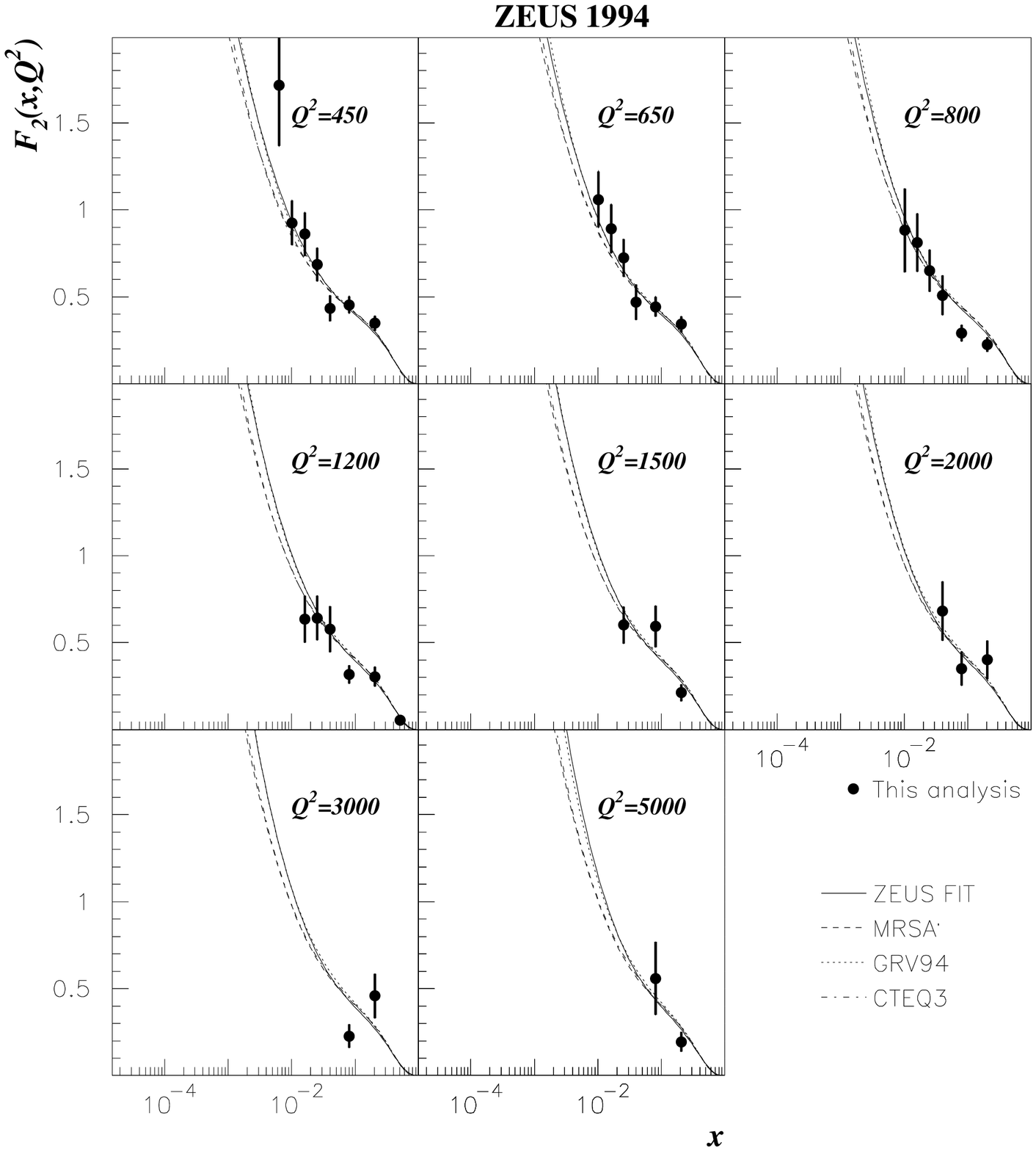}}}
\caption[]
{\it Structure function \Ft\ (\Fem\ for $Q^2>1000\;\Gevsq$)  as a function of \x\ for 
fixed values of \qsd .
The solid curves indicates the QCD NLO 
fit to the data used for acceptance correction.
Also indicated are the
 MRSA$^\prime$, GRV94 and CTEQ3 parameterisations.
The $Q^2$ values are given in ${\rm GeV^2}$.
}
\end{center}
\end{figure}

\setcounter{subfig}{1}
\renewcommand{\thefigure}{\arabic{figure}}

\clearpage
\newpage

\appendix
\section{Systematic uncertainties and their correlations}

In Sect.~\ref{SYSUNC}, the determination of the systematic uncertainties
of the $F_2$ values was discussed.  The resulting systematic
errors are given in Tables~\ref{t:final} and~\ref{t:final2}.  Bin to bin correlations exist
for the systematic errors.  Given the small
statistical and systematic uncertainties of the present measurement
these correlations should be taken into account
when performing fits.

\subsection{$F_2$ values resulting from systematic checks}
\label{ap:a1}

Tables~\ref{t:alla} and~\ref{t:allb} give the $F_2$ values obtained for each of the
29 different checks of systematic uncertainties described in 
Sect.~\ref{SYSUNC}.\footnote{
This Table and a short description of its recommended usage 
is available in electronic form from the ZEUS WWW page
whose current address is http://zow00.desy.de:8000/.  It can also
be obtained by contacting the authors.}
As noted above, the systematic uncertainties for the $F_2$ measurements
at $Q^2 > 100$ GeV$^2$ and  $y=\frac{Q^2}{xs} > 0.01$ are small
compared to the statistical errors.  Therefore, these
$F_2$ values are given only for the data with  
$Q^2 < 100$ GeV$^2$ or  $y < 0.01$.

In the calculation of the total systematic uncertainty given in
Tables~\ref{t:final} and~\ref{t:final2}, the deviations from the central value for each
check were added in 
quadrature, except for the two checks concerning
diffractively produced events (n$^o$ 23, 24) whose contributions
were calculated from the maximum offset, positive or negative, 
produced by the checks of n$^o$ 23 and 24.

Fits to the $F_2$ data, which take the systematic error correlations into
account, may be performed in the following way.
First the $F_2$ values with their statistical errors as
given in Tables~\ref{t:final} and~\ref{t:final2} are fitted.  This yields the standard
values of the fit parameters ($a_i$) and their statistical
errors ($\sigma_i(stat)$).  The fit is then 
repeated for each of the 29 sets of $F_2$ values listed in
Tables~\ref{t:alla} and~\ref{t:allb} resulting in 29 different sets of fit parameters
$a^f_i$, $f=1-29$.  An estimate of the systematic errors of the
fit parameters may be obtained by adding in quadrature, for each fit parameter,
the differences from the central parameter values,
$\sigma_i(syst) = \sqrt{\sum_f (a^f_i-a_i)^2}$.

Note that the overall normalisation error of 2\% (see Sect.~\ref{SYSUNC})
must be taken into account separately.

\subsection{Compact parametrisation of uncertainty correlations}

Although we recommend the use of Tables~\ref{t:alla} and~\ref{t:allb}, in determining
the effect of correlated uncertainties in a fit to $F_2$, the
large number of systematic checks involved may make this a
cumbersome process.  Therefore, a somewhat more compact representation
of the correlated systematic uncertainties is provided below.

We have investigated the dependence on  $y$ and $Q^2$ 
of the relative deviations, $\Delta(F_2)/F_2$, of the 
$F_2$ values listed in Tables~\ref{t:alla} and~\ref{t:allb}
from the central values of $F_2$.
These deviations can be grouped in six different classes of systematic 
behaviour, one of which is a function of $Q^2$ only and the others are  
functions of $y$ only.
\begin{eqnarray}
 A:\:&\frac{\Delta(F_2)}{F_2}\;[\%] & = \: a_A - b_A\cdot\log_{10} Q^2  \nonumber \\
 B:\:&\frac{\Delta(F_2)}{F_2}\;[\%] & = \: {a_B}/{|\log_{10} y|^2}  \nonumber \\
 C:\:&\frac{\Delta(F_2)}{F_2}\;[\%] & = \: {a_C}/{|\log_{10} y|^2} - b_C\cdot|\log_{10} y|^7  \nonumber \\
 D:\:&\frac{\Delta(F_2)}{F_2}\;[\%] & = \: {a_D}/{|\log_{10} y|^2} + b_D\cdot|\log_{10} y|^7 - c_D \nonumber \\
 E:\:&\frac{\Delta(F_2)}{F_2}\;[\%] & = \: a_E\cdot|\log_{10} y|^2 - b_E\cdot|\log_{10} y|^4 \nonumber \\
 F:\:&\frac{\Delta(F_2)}{F_2}\;[\%] & = \:  -a_F + b_F\cdot|\log_{10} y| \nonumber
\end{eqnarray}
The deviations produced by the systematic checks fall into one 
of the classes listed above (positive class) or into  
a class (negative class) where the signs of the coefficients are
reversed (these classes are denoted by, e.g. --A). 
The parameters are determined
by fitting the deviations produced by each check separately to the 
appropriate functional form. For each class the parameters obtained from 
each fit are then added in quadrature.

The results for the 
six classes of systematic errors are given in 
Table~\ref{t:app} and their dependence on $Q^2$ and $y$ are 
displayed graphically in Fig.~\ref{f:app}.\footnote{
A FORTRAN program of these functions is available 
through the ZEUS WWW home page whose current address 
is http://zow00.desy.de:8000/.  
It can also be obtained by contacting the authors.}
Note that these functions only represent the systematic
errors within the kinematic range of the present $F_2$ measurements. 

These functional forms may be used for systematic studies 
to reproduce the offsets of the values of 
$F_2$ for $Q^2< 100$~${\rm GeV^2}$ or $y < 0.01$
given in this paper. 
Outside this kinematic region,  
the systematic errors given in Table~\ref{t:final} can be treated as 
uncorrelated errors. 

The systematic deviations of the $F_2$ values as described by 
the 6+6 classes determined can
be added to the central values of $F_2$ to produce 12 sets
of $F_2$ data.  These 12 sets may be used, in the same way
as described in Sect.~\ref{ap:a1}  for the 29 sets, to determine
the effect of correlated uncertainties on fits to $F_2$.

These functions have been tested using the QCD NLO
fit described in Sect.~\ref{section:rew}.  The systematic uncertainties 
of the fit parameters are
consistent with the more
complete treatment described in Sect.~\ref{ap:a1}. 
It should be noted, however, that the simplification of the systematic
correlations results in some loss
of information, and that the use of the full treatment described
in Sect.~\ref{ap:a1} is recommended if the most reliable estimates
of the effect of correlated uncertainties are required.

\newcounter{subtab}
\setcounter{subtab}{1}
\newcounter{aptab}
\setcounter{aptab}{1}
\setcounter{table}{0}
\renewcommand{\thetable}{\Alph{aptab}\arabic{table}\alph{subtab}}

\setcounter{figure}{0}
\renewcommand{\thefigure}{\Alph{aptab}\arabic{figure}}

\clearpage
\newpage
\tabcolsep 1.2pt
\begin{table}[P]
\begin{sideways}\begin{minipage}[b]{\textheight}
\tiny
\begin{center}
\vspace{1.5cm} 
\begin{tabular}{|l|r|l|l|l|l|l|l|l|l|l|l|l|l|l|l|l|l|l|l|l|l|l|l|l|l|l|l|l|l|l|l|} \cline{1-2}  
\multicolumn{1}{|c|}{ }& \multicolumn{1}{c|}{$Q^2$}& \multicolumn{30}{c}{ } \\      
\multicolumn{1}{|c|}{$x$} & $({\rm GeV^2})$ &
\multicolumn{1}{c}{ \begin{rotate}{45} 0: $F_2$ standard \end{rotate}}&
\multicolumn{1}{c}{\begin{rotate}{45} 1: Finding efficiency + \end{rotate}}&
\multicolumn{1}{c}{\begin{rotate}{45} 2: Finding efficiency - \end{rotate}}&
\multicolumn{1}{c}{\begin{rotate}{45} 3: BOX X 28cm  \end{rotate}}&
\multicolumn{1}{c}{\begin{rotate}{45} 4: BOX Y 28cm  \end{rotate}} &
\multicolumn{1}{c}{\begin{rotate}{45} 5: Wide gap \end{rotate}}&
\multicolumn{1}{c}{\begin{rotate}{45} 6: Narrow gap \end{rotate}}&
\multicolumn{1}{c}{\begin{rotate}{45} 7: R+0.2cm \end{rotate}}&
\multicolumn{1}{c}{\begin{rotate}{45} 8: R-0.2cm  \end{rotate}}&
\multicolumn{1}{c}{\begin{rotate}{45} 9: Vertex efficiency -1\%  \end{rotate}}&
\multicolumn{1}{c}{\begin{rotate}{45} 10: Vertex efficiency low $\gamma_H$ +3\%  \end{rotate}}&
\multicolumn{1}{c}{\begin{rotate}{45} 11: Vertex cut (-28,40) \end{rotate}}&
\multicolumn{1}{c}{\begin{rotate}{45} 12: Positron Energy scale +  \end{rotate}}&
\multicolumn{1}{c}{\begin{rotate}{45} 13: Positron Energy scale -  \end{rotate}}&
\multicolumn{1}{c}{\begin{rotate}{45} 14: Hadron scale -3\%  \end{rotate}}&
\multicolumn{1}{c}{\begin{rotate}{45} 15: Hadron scale +3\% \end{rotate}}&
\multicolumn{1}{c}{\begin{rotate}{45} 16: Hadron scale F/B/RCAL +3/0/-3 \% \end{rotate}} &
\multicolumn{1}{c}{\begin{rotate}{45} 17: Hadron scale F/B/RCAL -3/0/+3 \% \end{rotate}} &
\multicolumn{1}{c}{\begin{rotate}{45} 18: Hadron scale F/B/RCAL 0/+3/0 \% \end{rotate}} &
\multicolumn{1}{c}{\begin{rotate}{45} 19: Hadron scale F/B/RCAL 0/-3/0 \% \end{rotate}} &
\multicolumn{1}{c}{\begin{rotate}{45} 20: Noise + \end{rotate}} &
\multicolumn{1}{c}{\begin{rotate}{45} 21: Noise - \end{rotate}} &
\multicolumn{1}{c}{\begin{rotate}{45} 22: MEPS \end{rotate}}&
\multicolumn{1}{c}{\begin{rotate}{45} 23: Diffractive~\cite{b:gunter} (take maximum 23,24) \end{rotate}}&
\multicolumn{1}{c}{\begin{rotate}{45} 24: Diffractive x 2 \end{rotate}}&
\multicolumn{1}{c}{\begin{rotate}{45} 25: Relax $\ptrat$ cut  \end{rotate}} &
\multicolumn{1}{c}{\begin{rotate}{45} 26: 1 $P_{Th}$ bin \end{rotate}}&
\multicolumn{1}{c}{\begin{rotate}{45} 27: More $\gamma$ bins \end{rotate}}&
\multicolumn{1}{c}{\begin{rotate}{45} 28: PHP background x 1.5 \end{rotate}}&
\multicolumn{1}{c}{\begin{rotate}{45} 29: PHP background x 0.5 \end{rotate}} \\ \hline

6.3 $\times 10^{-5}$&   3.5&0.875&0.917&0.862&0.866&0.875&0.884&0.861&0.875&0.875&0.870&0.875&0.871&0.882&0.861&0.923&0.805&0.917&0.817&0.856&0.883&0.864&0.887&0.870&0.896&0.887&0.876&0.880&0.864&0.854&0.896 \\
1.0 $\times 10^{-4}$&   3.5&0.939&0.949&0.937&0.980&0.889&0.954&0.929&0.939&0.939&0.946&0.939&0.939&0.952&0.928&0.973&0.905&0.954&0.910&0.935&0.947&0.923&0.943&0.936&0.946&0.946&0.938&0.935&0.945&0.926&0.952 \\
1.0 $\times 10^{-4}$&   4.5&1.030&1.046&1.023&1.032&1.034&1.036&1.019&1.030&1.030&1.027&1.030&1.031&1.055&1.011&1.069&0.998&1.056&1.006&1.019&1.030&1.024&1.038&1.028&1.036&1.034&1.028&1.016&1.031&1.010&1.050 \\
1.6 $\times 10^{-4}$&   4.5&1.038&1.042&1.037&1.056&1.015&1.046&1.024&1.038&1.038&1.041&1.038&1.060&1.044&1.037&1.034&1.034&1.028&1.046&1.032&1.047&1.029&1.049&1.036&1.029&1.041&1.036&1.034&1.031&1.028&1.047 \\
2.5 $\times 10^{-4}$&   4.5&0.932&0.933&0.932&0.930&0.918&0.952&0.928&0.932&0.932&0.935&0.932&0.951&0.935&0.934&0.929&0.924&0.914&0.937&0.922&0.933&0.924&0.931&0.940&0.939&0.935&0.930&0.943&0.918&0.931&0.934 \\
1.0 $\times 10^{-4}$&   6.5&1.221&1.257&1.204&1.220&1.227&1.235&1.208&1.221&1.222&1.221&1.221&1.246&1.254&1.201&1.282&1.173&1.266&1.179&1.213&1.237&1.214&1.226&1.215&1.212&1.229&1.223&1.222&1.216&1.199&1.243 \\
1.6 $\times 10^{-4}$&   6.5&1.138&1.151&1.135&1.135&1.133&1.158&1.125&1.138&1.138&1.139&1.138&1.157&1.148&1.137&1.164&1.114&1.158&1.125&1.130&1.145&1.133&1.147&1.138&1.135&1.136&1.138&1.149&1.142&1.128&1.147 \\
2.5 $\times 10^{-4}$&   6.5&1.038&1.042&1.038&1.031&1.042&1.046&1.022&1.038&1.038&1.035&1.038&1.044&1.030&1.032&1.045&1.032&1.032&1.043&1.030&1.046&1.028&1.045&1.046&1.033&1.035&1.039&1.037&1.037&1.032&1.044 \\
4.0 $\times 10^{-4}$&   6.5&0.951&0.951&0.951&0.954&0.963&0.962&0.939&0.951&0.951&0.953&0.951&0.962&0.941&0.951&0.961&0.948&0.946&0.960&0.941&0.964&0.941&0.959&0.953&0.957&0.958&0.951&0.959&0.933&0.951&0.951 \\
6.3 $\times 10^{-4}$&   6.5&0.841&0.841&0.841&0.837&0.847&0.857&0.834&0.841&0.841&0.841&0.841&0.851&0.835&0.843&0.842&0.839&0.832&0.848&0.838&0.845&0.837&0.843&0.844&0.852&0.848&0.842&0.830&0.848&0.841&0.841 \\
1.0 $\times 10^{-3}$&   6.5&0.776&0.776&0.776&0.779&0.786&0.786&0.770&0.776&0.776&0.779&0.776&0.787&0.783&0.776&0.777&0.777&0.766&0.782&0.779&0.772&0.752&0.793&0.797&0.783&0.782&0.782&0.766&0.780&0.776&0.776 \\
1.6 $\times 10^{-3}$&   6.5&0.753&0.753&0.753&0.745&0.755&0.760&0.740&0.753&0.753&0.755&0.753&0.762&0.750&0.752&0.747&0.752&0.752&0.757&0.755&0.746&0.729&0.764&0.767&0.755&0.760&0.756&0.750&0.747&0.753&0.753 \\
2.5 $\times 10^{-3}$&   6.5&0.645&0.645&0.645&0.634&0.652&0.652&0.636&0.645&0.645&0.645&0.645&0.651&0.644&0.645&0.643&0.650&0.643&0.651&0.650&0.642&0.627&0.655&0.655&0.653&0.660&0.650&0.640&0.662&0.645&0.645 \\
4.0 $\times 10^{-3}$&   6.5&0.629&0.629&0.629&0.626&0.631&0.637&0.619&0.629&0.629&0.629&0.628&0.633&0.625&0.630&0.630&0.629&0.627&0.631&0.631&0.628&0.605&0.643&0.619&0.629&0.624&0.634&0.629&0.627&0.629&0.629 \\
1.6 $\times 10^{-2}$&   6.5&0.495&0.495&0.495&0.505&0.498&0.501&0.488&0.495&0.495&0.494&0.495&0.494&0.490&0.499&0.489&0.496&0.491&0.495&0.498&0.489&0.526&0.477&0.489&0.494&0.490&0.506&0.495&0.503&0.495&0.495 \\
4.0 $\times 10^{-2}$&   6.5&0.403&0.403&0.403&0.431&0.407&0.409&0.401&0.403&0.403&0.403&0.405&0.404&0.376&0.424&0.393&0.416&0.414&0.394&0.405&0.400&0.522&0.344&0.375&0.405&0.392&0.484&0.407&0.402&0.403&0.403 \\
1.6 $\times 10^{-4}$&   8.5&1.337&1.368&1.323&1.336&1.336&1.347&1.327&1.337&1.334&1.339&1.337&1.352&1.340&1.333&1.370&1.280&1.363&1.293&1.325&1.343&1.314&1.337&1.327&1.343&1.342&1.337&1.329&1.338&1.316&1.357 \\
2.5 $\times 10^{-4}$&   8.5&1.151&1.161&1.150&1.151&1.152&1.168&1.135&1.150&1.151&1.151&1.151&1.166&1.145&1.148&1.171&1.148&1.159&1.157&1.145&1.155&1.143&1.159&1.152&1.147&1.146&1.149&1.150&1.160&1.143&1.159 \\
4.0 $\times 10^{-4}$&   8.5&0.993&0.996&0.993&0.995&0.989&1.009&0.981&0.993&0.993&0.990&0.993&0.995&0.991&1.001&0.996&0.991&0.987&0.997&0.991&1.006&0.990&1.001&1.002&1.000&0.994&0.995&0.996&0.981&0.991&0.994 \\
6.3 $\times 10^{-4}$&   8.5&0.920&0.921&0.920&0.914&0.927&0.930&0.916&0.920&0.920&0.922&0.920&0.922&0.915&0.912&0.919&0.918&0.907&0.924&0.910&0.924&0.907&0.922&0.920&0.921&0.921&0.919&0.919&0.916&0.920&0.920 \\
1.0 $\times 10^{-3}$&   8.5&0.831&0.832&0.831&0.837&0.830&0.842&0.818&0.831&0.831&0.832&0.831&0.833&0.829&0.840&0.835&0.828&0.823&0.842&0.830&0.839&0.812&0.847&0.844&0.841&0.842&0.833&0.834&0.833&0.831&0.831 \\
1.6 $\times 10^{-3}$&   8.5&0.704&0.705&0.704&0.708&0.703&0.713&0.694&0.704&0.704&0.705&0.704&0.716&0.700&0.703&0.708&0.702&0.700&0.707&0.708&0.701&0.698&0.713&0.714&0.708&0.713&0.707&0.699&0.711&0.704&0.704 \\
1.6 $\times 10^{-4}$&  10.0&1.359&1.432&1.336&1.359&1.359&1.379&1.366&1.353&1.398&1.364&1.359&1.371&1.371&1.349&1.412&1.307&1.381&1.337&1.353&1.375&1.355&1.362&1.343&1.375&1.349&1.356&1.395&1.323&1.339&1.380 \\
2.5 $\times 10^{-4}$&  10.0&1.231&1.246&1.227&1.230&1.231&1.238&1.218&1.231&1.228&1.229&1.231&1.254&1.240&1.219&1.247&1.203&1.247&1.209&1.226&1.239&1.221&1.238&1.238&1.230&1.226&1.232&1.230&1.233&1.218&1.243 \\
4.0 $\times 10^{-4}$&  10.0&1.136&1.138&1.136&1.136&1.142&1.135&1.135&1.135&1.137&1.137&1.136&1.153&1.137&1.129&1.155&1.132&1.140&1.139&1.126&1.146&1.127&1.142&1.139&1.145&1.139&1.136&1.137&1.123&1.133&1.140 \\
6.3 $\times 10^{-4}$&  10.0&0.968&0.969&0.968&0.967&0.967&0.981&0.951&0.968&0.969&0.967&0.968&0.980&0.964&0.976&0.972&0.971&0.964&0.981&0.960&0.978&0.962&0.980&0.974&0.979&0.976&0.966&0.969&0.968&0.967&0.969 \\
1.0 $\times 10^{-3}$&  10.0&0.857&0.857&0.857&0.861&0.867&0.870&0.847&0.857&0.857&0.854&0.857&0.861&0.849&0.859&0.853&0.852&0.845&0.864&0.852&0.855&0.849&0.856&0.866&0.866&0.862&0.860&0.848&0.852&0.856&0.857 \\
1.6 $\times 10^{-3}$&  10.0&0.792&0.793&0.792&0.797&0.795&0.799&0.785&0.792&0.792&0.790&0.792&0.796&0.789&0.792&0.797&0.790&0.790&0.800&0.792&0.791&0.773&0.801&0.801&0.798&0.803&0.794&0.793&0.785&0.792&0.792 \\
2.5 $\times 10^{-3}$&  10.0&0.715&0.715&0.715&0.716&0.721&0.728&0.698&0.715&0.715&0.714&0.714&0.714&0.708&0.710&0.717&0.715&0.711&0.722&0.718&0.715&0.693&0.728&0.738&0.722&0.724&0.719&0.716&0.721&0.715&0.715 \\
6.3 $\times 10^{-3}$&  10.0&0.593&0.593&0.593&0.596&0.597&0.601&0.587&0.592&0.593&0.592&0.593&0.597&0.592&0.594&0.591&0.593&0.589&0.594&0.597&0.589&0.574&0.602&0.586&0.596&0.597&0.595&0.593&0.607&0.593&0.593 \\
1.6 $\times 10^{-2}$&  10.0&0.517&0.518&0.517&0.518&0.520&0.524&0.511&0.517&0.517&0.517&0.517&0.518&0.516&0.519&0.519&0.520&0.515&0.523&0.520&0.514&0.522&0.515&0.517&0.518&0.515&0.517&0.525&0.517&0.517&0.517 \\
8.1 $\times 10^{-2}$&  10.0&0.395&0.395&0.395&0.397&0.395&0.401&0.389&0.395&0.395&0.395&0.396&0.396&0.371&0.409&0.387&0.398&0.395&0.389&0.397&0.394&0.490&0.354&0.363&0.394&0.384&0.410&0.392&0.392&0.395&0.395 \\
2.5 $\times 10^{-4}$&  12.0&1.300&1.322&1.290&1.300&1.301&1.305&1.294&1.295&1.307&1.299&1.300&1.310&1.320&1.317&1.334&1.261&1.316&1.277&1.292&1.307&1.297&1.303&1.305&1.306&1.309&1.298&1.305&1.283&1.276&1.325 \\
4.0 $\times 10^{-4}$&  12.0&1.128&1.140&1.128&1.127&1.130&1.150&1.117&1.130&1.130&1.129&1.128&1.138&1.127&1.121&1.143&1.114&1.128&1.132&1.125&1.137&1.119&1.134&1.134&1.134&1.139&1.127&1.130&1.129&1.119&1.138 \\
6.3 $\times 10^{-4}$&  12.0&1.102&1.104&1.102&1.100&1.102&1.113&1.088&1.101&1.102&1.099&1.102&1.107&1.096&1.110&1.101&1.089&1.086&1.108&1.080&1.107&1.091&1.104&1.104&1.113&1.103&1.102&1.100&1.099&1.102&1.102 \\
1.0 $\times 10^{-3}$&  12.0&0.903&0.903&0.903&0.904&0.904&0.912&0.892&0.902&0.903&0.905&0.903&0.915&0.889&0.908&0.900&0.908&0.893&0.909&0.906&0.903&0.892&0.910&0.923&0.913&0.905&0.885&0.905&0.910&0.903&0.903 \\
1.6 $\times 10^{-3}$&  12.0&0.789&0.789&0.789&0.793&0.793&0.808&0.780&0.788&0.789&0.788&0.789&0.800&0.791&0.792&0.794&0.791&0.783&0.794&0.788&0.795&0.772&0.792&0.796&0.798&0.795&0.792&0.786&0.798&0.789&0.789 \\
2.5 $\times 10^{-3}$&  12.0&0.746&0.746&0.746&0.747&0.753&0.755&0.731&0.746&0.746&0.745&0.746&0.756&0.735&0.747&0.745&0.743&0.743&0.755&0.750&0.744&0.732&0.758&0.756&0.757&0.764&0.745&0.746&0.739&0.746&0.746 \\
2.5 $\times 10^{-4}$&  15.0&1.559&1.597&1.541&1.559&1.559&1.556&1.550&1.526&1.574&1.562&1.559&1.578&1.567&1.523&1.612&1.507&1.594&1.510&1.547&1.567&1.535&1.575&1.556&1.564&1.544&1.554&1.564&1.559&1.512&1.606 \\
4.0 $\times 10^{-4}$&  15.0&1.338&1.352&1.337&1.337&1.338&1.349&1.329&1.331&1.345&1.339&1.338&1.351&1.341&1.336&1.366&1.320&1.360&1.333&1.333&1.356&1.329&1.358&1.333&1.347&1.334&1.341&1.340&1.329&1.329&1.347 \\
6.3 $\times 10^{-4}$&  15.0&1.092&1.093&1.092&1.092&1.092&1.099&1.083&1.092&1.091&1.090&1.092&1.108&1.081&1.093&1.094&1.084&1.084&1.096&1.079&1.101&1.078&1.098&1.104&1.108&1.100&1.092&1.081&1.095&1.087&1.098 \\
1.0 $\times 10^{-3}$&  15.0&0.992&0.993&0.992&0.993&0.992&1.002&0.986&0.990&0.992&0.992&0.992&1.013&0.996&0.999&0.995&0.976&0.981&0.998&0.976&1.002&0.978&1.001&1.012&1.001&0.998&0.994&0.995&0.976&0.992&0.992 \\
1.6 $\times 10^{-3}$&  15.0&0.908&0.908&0.908&0.906&0.911&0.914&0.903&0.908&0.909&0.905&0.908&0.920&0.907&0.902&0.911&0.910&0.902&0.914&0.907&0.907&0.892&0.914&0.918&0.913&0.910&0.913&0.909&0.892&0.908&0.908 \\
2.5 $\times 10^{-3}$&  15.0&0.793&0.793&0.793&0.794&0.795&0.802&0.790&0.792&0.794&0.791&0.793&0.806&0.790&0.791&0.791&0.803&0.790&0.798&0.807&0.790&0.786&0.801&0.811&0.802&0.801&0.796&0.787&0.812&0.793&0.793 \\
6.3 $\times 10^{-3}$&  15.0&0.662&0.662&0.662&0.663&0.662&0.671&0.652&0.661&0.662&0.663&0.662&0.675&0.660&0.667&0.662&0.661&0.659&0.664&0.665&0.658&0.640&0.670&0.657&0.664&0.666&0.662&0.664&0.658&0.662&0.662 \\
1.6 $\times 10^{-2}$&  15.0&0.553&0.553&0.553&0.553&0.553&0.558&0.548&0.552&0.553&0.552&0.552&0.562&0.552&0.553&0.552&0.552&0.548&0.557&0.555&0.549&0.543&0.559&0.551&0.559&0.550&0.548&0.552&0.551&0.553&0.553 \\
8.1 $\times 10^{-2}$&  15.0&0.370&0.370&0.370&0.371&0.370&0.375&0.366&0.370&0.370&0.370&0.371&0.370&0.357&0.379&0.365&0.378&0.376&0.367&0.373&0.368&0.429&0.341&0.349&0.368&0.368&0.376&0.373&0.372&0.370&0.370 \\
4.0 $\times 10^{-4}$&  18.0&1.438&1.465&1.432&1.438&1.438&1.447&1.436&1.429&1.445&1.433&1.438&1.460&1.440&1.421&1.469&1.425&1.451&1.428&1.430&1.449&1.435&1.449&1.448&1.440&1.438&1.430&1.437&1.456&1.417&1.459 \\
6.3 $\times 10^{-4}$&  18.0&1.166&1.170&1.166&1.167&1.167&1.161&1.152&1.159&1.174&1.165&1.166&1.191&1.168&1.160&1.181&1.154&1.170&1.170&1.153&1.179&1.159&1.171&1.178&1.170&1.168&1.169&1.167&1.145&1.153&1.180 \\
1.0 $\times 10^{-3}$&  18.0&1.015&1.016&1.015&1.015&1.015&1.036&1.009&1.017&1.014&1.014&1.015&1.012&1.002&1.028&1.017&1.025&1.005&1.034&1.009&1.022&1.004&1.018&1.020&1.022&1.017&1.001&1.022&1.014&1.013&1.016 \\
1.6 $\times 10^{-3}$&  18.0&0.932&0.932&0.932&0.932&0.932&0.941&0.914&0.933&0.932&0.936&0.932&0.941&0.929&0.933&0.935&0.929&0.920&0.936&0.928&0.940&0.925&0.937&0.933&0.938&0.936&0.926&0.920&0.936&0.931&0.934 \\
2.5 $\times 10^{-3}$&  18.0&0.879&0.880&0.879&0.880&0.879&0.884&0.872&0.875&0.880&0.876&0.879&0.883&0.873&0.878&0.881&0.889&0.872&0.882&0.888&0.874&0.870&0.888&0.895&0.886&0.889&0.881&0.881&0.874&0.879&0.879 \\
6.3 $\times 10^{-3}$&  18.0&0.629&0.629&0.629&0.630&0.630&0.633&0.621&0.631&0.629&0.628&0.629&0.635&0.623&0.630&0.629&0.626&0.627&0.634&0.631&0.627&0.613&0.641&0.626&0.633&0.634&0.630&0.628&0.640&0.629&0.629 \\
\hline
\end{tabular}
\end{center}
\vspace{-0.3cm}
\caption{\label{t:alla} \it  The extracted $F_2$ values for each systematic
check at values of $Q^2< 100$ {\rm GeV}$^2$ $or$ $y=\frac{Q^2}{xs} < 0.01$.
}
\end{minipage}\end{sideways}
\end{table}

\addtocounter{table}{-1}
\addtocounter{subtab}{1}

\newpage
\begin{table}[P]
\begin{sideways}\begin{minipage}[b]{\textheight}
\tiny
\begin{center}
\vspace{-0.5cm}
\begin{tabular}{|l|r|l|l|l|l|l|l|l|l|l|l|l|l|l|l|l|l|l|l|l|l|l|l|l|l|l|l|l|l|l|l|} \hline
\multicolumn{1}{|c|}{ }& \multicolumn{1}{c|}{$Q^2$}& \multicolumn{30}{c|}{\em Systematic check} \\  \cline{3-32}
\multicolumn{1}{|c|}{$x$} & $({\rm GeV^2})$ & \multicolumn{1}{c|}{\em 0}&\multicolumn{1}{c|}{\em 1}&\multicolumn{1}{c|}{\em 2}&\multicolumn{1}{c|}{\em 3}&\multicolumn{1}{c|}{\em 4}&\multicolumn{1}{c|}{\em 5}&\multicolumn{1}{c|}{\em 6}&\multicolumn{1}{c|}{\em 7}&\multicolumn{1}{c|}{\em 8}&\multicolumn{1}{c|}{\em 9}&\multicolumn{1}{c|}{\em 10}&\multicolumn{1}{c|}{\em 11}&\multicolumn{1}{c|}{\em 12}&\multicolumn{1}{c|}{\em 13}&\multicolumn{1}{c|}{\em 14}&\multicolumn{1}{c|}{\em 15}&\multicolumn{1}{c|}{\em 16}&\multicolumn{1}{c|}{\em 17}&\multicolumn{1}{c|}{\em 18}&\multicolumn{1}{c|}{\em 19}&\multicolumn{1}{c|}{\em 20}&\multicolumn{1}{c|}{\em 21}&\multicolumn{1}{c|}{\em 22}&\multicolumn{1}{c|}{\em 23}&\multicolumn{1}{c|}{\em 24}&\multicolumn{1}{c|}{\em 25}&\multicolumn{1}{c|}{\em 26}&\multicolumn{1}{c|}{\em 27}&\multicolumn{1}{c|}{\em 28}&\multicolumn{1}{c|}{\em 29} \\ \hline
4.0 $\times 10^{-4}$&  22.0&1.507&1.534&1.493&1.507&1.507&1.519&1.500&1.506&1.488&1.513&1.507&1.526&1.526&1.515&1.572&1.436&1.562&1.462&1.497&1.522&1.496&1.516&1.505&1.493&1.505&1.514&1.525&1.498&1.479&1.536 \\
6.3 $\times 10^{-4}$&  22.0&1.324&1.332&1.324&1.324&1.324&1.335&1.320&1.325&1.330&1.327&1.324&1.338&1.304&1.326&1.331&1.292&1.324&1.307&1.307&1.333&1.315&1.336&1.334&1.330&1.323&1.323&1.309&1.337&1.311&1.338 \\
1.0 $\times 10^{-3}$&  22.0&1.081&1.083&1.081&1.081&1.081&1.083&1.064&1.065&1.085&1.079&1.081&1.094&1.080&1.073&1.093&1.080&1.076&1.095&1.072&1.094&1.069&1.086&1.089&1.088&1.089&1.079&1.085&1.069&1.079&1.083 \\
1.6 $\times 10^{-3}$&  22.0&0.930&0.930&0.930&0.930&0.930&0.939&0.923&0.920&0.938&0.930&0.930&0.954&0.921&0.938&0.930&0.928&0.922&0.938&0.927&0.937&0.930&0.940&0.944&0.936&0.928&0.911&0.931&0.945&0.930&0.930 \\
2.5 $\times 10^{-3}$&  22.0&0.906&0.906&0.906&0.906&0.906&0.912&0.897&0.904&0.912&0.907&0.906&0.916&0.898&0.899&0.897&0.906&0.900&0.913&0.903&0.900&0.896&0.913&0.914&0.913&0.913&0.906&0.893&0.915&0.906&0.906 \\
4.0 $\times 10^{-3}$&  22.0&0.749&0.750&0.749&0.750&0.749&0.756&0.743&0.747&0.754&0.750&0.749&0.763&0.752&0.754&0.760&0.741&0.739&0.757&0.747&0.751&0.729&0.755&0.777&0.752&0.755&0.754&0.763&0.728&0.749&0.749 \\
6.3 $\times 10^{-3}$&  22.0&0.701&0.701&0.701&0.701&0.701&0.705&0.696&0.697&0.703&0.701&0.701&0.722&0.700&0.701&0.691&0.709&0.701&0.700&0.707&0.692&0.690&0.714&0.696&0.703&0.707&0.703&0.695&0.717&0.701&0.701 \\
1.0 $\times 10^{-2}$&  22.0&0.601&0.601&0.601&0.601&0.601&0.609&0.591&0.598&0.601&0.601&0.601&0.605&0.596&0.603&0.604&0.600&0.597&0.606&0.606&0.600&0.577&0.612&0.589&0.603&0.600&0.592&0.601&0.607&0.601&0.601 \\
2.5 $\times 10^{-2}$&  22.0&0.490&0.490&0.490&0.490&0.490&0.494&0.486&0.488&0.494&0.490&0.488&0.497&0.489&0.491&0.491&0.491&0.487&0.494&0.492&0.488&0.487&0.492&0.489&0.491&0.487&0.481&0.496&0.487&0.490&0.490 \\
8.1 $\times 10^{-2}$&  22.0&0.338&0.338&0.338&0.338&0.338&0.340&0.334&0.336&0.341&0.338&0.342&0.337&0.321&0.352&0.328&0.346&0.343&0.331&0.340&0.334&0.400&0.305&0.308&0.337&0.336&0.350&0.334&0.355&0.338&0.338 \\
6.3 $\times 10^{-4}$&  27.0&1.460&1.475&1.454&1.460&1.460&1.468&1.460&1.484&1.452&1.455&1.460&1.476&1.453&1.470&1.477&1.450&1.467&1.455&1.456&1.463&1.447&1.461&1.473&1.466&1.462&1.456&1.461&1.461&1.443&1.476 \\
1.0 $\times 10^{-3}$&  27.0&1.194&1.196&1.194&1.194&1.194&1.200&1.189&1.206&1.204&1.194&1.194&1.208&1.185&1.193&1.199&1.184&1.195&1.193&1.183&1.201&1.186&1.203&1.208&1.196&1.199&1.194&1.188&1.194&1.187&1.200 \\
1.6 $\times 10^{-3}$&  27.0&1.140&1.141&1.140&1.140&1.140&1.147&1.125&1.131&1.133&1.139&1.140&1.155&1.113&1.148&1.145&1.154&1.124&1.155&1.143&1.145&1.142&1.140&1.142&1.145&1.144&1.141&1.117&1.138&1.140&1.140 \\
2.5 $\times 10^{-3}$&  27.0&0.928&0.929&0.928&0.928&0.928&0.934&0.928&0.926&0.935&0.929&0.928&0.945&0.926&0.935&0.921&0.925&0.921&0.935&0.925&0.926&0.917&0.931&0.935&0.931&0.934&0.919&0.929&0.915&0.928&0.928 \\
4.0 $\times 10^{-3}$&  27.0&0.765&0.765&0.765&0.765&0.765&0.772&0.761&0.758&0.776&0.764&0.765&0.783&0.765&0.772&0.767&0.767&0.765&0.770&0.773&0.766&0.750&0.769&0.787&0.774&0.778&0.770&0.768&0.771&0.763&0.767 \\
6.3 $\times 10^{-3}$&  27.0&0.674&0.674&0.674&0.674&0.674&0.676&0.672&0.668&0.680&0.675&0.674&0.689&0.672&0.670&0.667&0.673&0.669&0.673&0.674&0.665&0.663&0.679&0.673&0.680&0.687&0.673&0.674&0.670&0.674&0.674 \\
6.3 $\times 10^{-4}$&  35.0&1.565&1.589&1.551&1.565&1.565&1.583&1.556&1.566&1.545&1.564&1.565&1.577&1.553&1.561&1.607&1.545&1.590&1.557&1.556&1.574&1.564&1.579&1.572&1.551&1.556&1.574&1.561&1.553&1.544&1.586 \\
1.0 $\times 10^{-3}$&  35.0&1.370&1.374&1.369&1.370&1.370&1.371&1.355&1.366&1.351&1.368&1.370&1.373&1.346&1.359&1.371&1.367&1.361&1.367&1.365&1.376&1.361&1.368&1.370&1.374&1.371&1.370&1.370&1.353&1.360&1.380 \\
1.6 $\times 10^{-3}$&  35.0&1.149&1.150&1.149&1.149&1.149&1.150&1.153&1.156&1.145&1.149&1.149&1.158&1.152&1.156&1.137&1.157&1.136&1.157&1.136&1.150&1.141&1.156&1.154&1.157&1.154&1.145&1.152&1.135&1.146&1.152 \\
2.5 $\times 10^{-3}$&  35.0&1.018&1.020&1.018&1.018&1.018&1.027&1.005&1.028&1.008&1.018&1.018&1.043&1.009&1.027&1.016&1.006&1.009&1.021&1.006&1.025&1.007&1.023&1.019&1.023&1.017&1.018&1.005&1.043&1.018&1.018 \\
4.0 $\times 10^{-3}$&  35.0&0.887&0.889&0.887&0.887&0.887&0.891&0.885&0.885&0.886&0.888&0.887&0.886&0.883&0.885&0.896&0.896&0.887&0.898&0.895&0.892&0.883&0.895&0.912&0.896&0.892&0.890&0.884&0.876&0.887&0.887 \\
6.3 $\times 10^{-3}$&  35.0&0.749&0.749&0.749&0.749&0.749&0.757&0.750&0.751&0.758&0.746&0.749&0.747&0.754&0.755&0.748&0.750&0.748&0.747&0.756&0.743&0.737&0.755&0.755&0.757&0.765&0.757&0.770&0.768&0.749&0.749 \\
1.0 $\times 10^{-2}$&  35.0&0.681&0.682&0.681&0.681&0.681&0.682&0.674&0.672&0.688&0.681&0.681&0.689&0.664&0.679&0.670&0.684&0.680&0.682&0.682&0.673&0.666&0.683&0.662&0.684&0.685&0.680&0.675&0.689&0.681&0.681 \\
1.6 $\times 10^{-2}$&  35.0&0.604&0.604&0.604&0.604&0.604&0.609&0.599&0.597&0.604&0.602&0.604&0.610&0.606&0.602&0.609&0.598&0.593&0.612&0.607&0.602&0.591&0.614&0.600&0.604&0.602&0.605&0.598&0.603&0.604&0.604 \\
2.5 $\times 10^{-2}$&  35.0&0.570&0.570&0.570&0.570&0.570&0.574&0.566&0.571&0.567&0.569&0.570&0.582&0.567&0.571&0.568&0.575&0.569&0.571&0.573&0.567&0.569&0.575&0.574&0.571&0.567&0.569&0.571&0.589&0.570&0.570 \\
8.1 $\times 10^{-2}$&  35.0&0.461&0.461&0.461&0.461&0.461&0.464&0.458&0.459&0.457&0.459&0.471&0.461&0.449&0.470&0.448&0.468&0.465&0.454&0.464&0.457&0.511&0.429&0.431&0.460&0.459&0.465&0.465&0.449&0.461&0.461 \\
1.0 $\times 10^{-3}$&  45.0&1.441&1.451&1.436&1.441&1.441&1.452&1.436&1.462&1.441&1.441&1.441&1.462&1.458&1.456&1.478&1.413&1.466&1.427&1.438&1.458&1.433&1.445&1.456&1.451&1.454&1.443&1.434&1.468&1.428&1.453 \\
1.6 $\times 10^{-3}$&  45.0&1.178&1.180&1.178&1.178&1.178&1.179&1.169&1.178&1.157&1.176&1.178&1.200&1.163&1.160&1.187&1.172&1.168&1.186&1.165&1.189&1.172&1.195&1.187&1.179&1.175&1.175&1.165&1.168&1.163&1.194 \\
2.5 $\times 10^{-3}$&  45.0&1.071&1.072&1.071&1.071&1.071&1.071&1.071&1.074&1.075&1.070&1.071&1.077&1.074&1.072&1.072&1.069&1.068&1.081&1.059&1.079&1.066&1.066&1.073&1.076&1.074&1.070&1.052&1.073&1.071&1.071 \\
4.0 $\times 10^{-3}$&  45.0&0.894&0.895&0.894&0.894&0.894&0.902&0.892&0.897&0.891&0.891&0.894&0.905&0.894&0.893&0.889&0.895&0.887&0.896&0.899&0.891&0.880&0.896&0.909&0.897&0.894&0.895&0.917&0.883&0.894&0.894 \\
6.3 $\times 10^{-3}$&  45.0&0.804&0.804&0.804&0.804&0.804&0.804&0.797&0.809&0.787&0.800&0.804&0.827&0.799&0.792&0.802&0.804&0.794&0.807&0.809&0.803&0.800&0.808&0.829&0.810&0.804&0.801&0.782&0.847&0.804&0.804 \\
1.0 $\times 10^{-2}$&  45.0&0.745&0.745&0.745&0.745&0.745&0.755&0.746&0.749&0.736&0.747&0.745&0.763&0.746&0.761&0.738&0.751&0.753&0.746&0.751&0.739&0.724&0.754&0.732&0.748&0.749&0.743&0.754&0.713&0.745&0.745 \\
1.0 $\times 10^{-3}$&  60.0&1.553&1.567&1.540&1.553&1.553&1.563&1.557&1.599&1.526&1.545&1.553&1.555&1.510&1.573&1.555&1.529&1.547&1.554&1.544&1.566&1.542&1.576&1.552&1.540&1.543&1.553&1.585&1.535&1.545&1.562 \\
1.6 $\times 10^{-3}$&  60.0&1.339&1.342&1.338&1.339&1.339&1.345&1.349&1.345&1.331&1.339&1.339&1.357&1.334&1.332&1.331&1.344&1.332&1.342&1.331&1.344&1.337&1.328&1.365&1.349&1.353&1.340&1.334&1.327&1.334&1.343 \\
2.5 $\times 10^{-3}$&  60.0&1.100&1.101&1.100&1.100&1.100&1.107&1.083&1.117&1.085&1.099&1.100&1.114&1.090&1.106&1.105&1.102&1.095&1.113&1.095&1.113&1.099&1.110&1.090&1.101&1.104&1.101&1.106&1.089&1.096&1.104 \\
4.0 $\times 10^{-3}$&  60.0&1.012&1.013&1.012&1.012&1.012&1.005&1.006&1.011&1.001&1.009&1.012&1.029&1.014&0.989&1.002&1.003&0.999&1.016&1.002&1.003&0.997&1.016&1.023&1.017&1.011&1.011&1.003&1.018&1.012&1.012 \\
6.3 $\times 10^{-3}$&  60.0&0.890&0.891&0.890&0.890&0.890&0.894&0.886&0.900&0.883&0.893&0.890&0.900&0.883&0.892&0.889&0.900&0.890&0.890&0.906&0.884&0.878&0.893&0.917&0.900&0.890&0.891&0.914&0.881&0.890&0.890 \\
1.0 $\times 10^{-2}$&  60.0&0.672&0.672&0.672&0.672&0.672&0.675&0.676&0.684&0.675&0.672&0.672&0.678&0.673&0.681&0.673&0.672&0.670&0.679&0.671&0.668&0.673&0.683&0.660&0.673&0.679&0.671&0.676&0.683&0.672&0.672 \\
1.6 $\times 10^{-2}$&  60.0&0.614&0.614&0.614&0.614&0.614&0.618&0.609&0.633&0.621&0.610&0.615&0.621&0.614&0.611&0.615&0.612&0.611&0.625&0.621&0.612&0.594&0.625&0.608&0.616&0.617&0.611&0.620&0.660&0.614&0.614 \\
2.5 $\times 10^{-2}$&  60.0&0.627&0.627&0.627&0.627&0.627&0.626&0.626&0.628&0.624&0.625&0.628&0.637&0.626&0.628&0.624&0.631&0.626&0.624&0.628&0.625&0.620&0.634&0.633&0.628&0.626&0.633&0.628&0.639&0.627&0.627 \\
8.1 $\times 10^{-2}$&  60.0&0.521&0.521&0.521&0.521&0.521&0.528&0.518&0.528&0.518&0.518&0.523&0.524&0.510&0.527&0.511&0.528&0.524&0.518&0.526&0.514&0.564&0.494&0.498&0.520&0.518&0.516&0.525&0.511&0.521&0.521 \\
1.6 $\times 10^{-3}$&  70.0&1.318&1.327&1.313&1.318&1.318&1.313&1.308&1.321&1.318&1.322&1.318&1.344&1.320&1.305&1.346&1.298&1.335&1.311&1.315&1.326&1.308&1.330&1.318&1.330&1.337&1.314&1.357&1.297&1.298&1.338 \\
2.5 $\times 10^{-3}$&  70.0&1.228&1.229&1.228&1.228&1.228&1.226&1.219&1.233&1.192&1.229&1.228&1.233&1.195&1.221&1.207&1.224&1.210&1.236&1.216&1.231&1.219&1.217&1.235&1.238&1.237&1.225&1.183&1.239&1.228&1.228 \\
4.0 $\times 10^{-3}$&  70.0&0.917&0.917&0.917&0.917&0.917&0.931&0.913&0.940&0.911&0.916&0.917&0.923&0.911&0.935&0.918&0.925&0.915&0.926&0.918&0.922&0.931&0.918&0.923&0.921&0.907&0.915&0.932&0.919&0.917&0.917 \\
6.3 $\times 10^{-3}$&  70.0&0.887&0.888&0.887&0.887&0.887&0.891&0.883&0.894&0.872&0.880&0.887&0.896&0.881&0.880&0.890&0.886&0.883&0.892&0.891&0.891&0.888&0.889&0.917&0.885&0.884&0.888&0.864&0.891&0.887&0.887 \\
1.0 $\times 10^{-2}$&  70.0&0.774&0.774&0.774&0.774&0.774&0.789&0.766&0.792&0.756&0.778&0.774&0.792&0.762&0.768&0.771&0.778&0.766&0.777&0.774&0.765&0.759&0.789&0.753&0.779&0.778&0.775&0.784&0.778&0.774&0.774 \\
1.6 $\times 10^{-2}$&  70.0&0.697&0.697&0.697&0.697&0.697&0.692&0.698&0.693&0.693&0.698&0.697&0.715&0.709&0.707&0.697&0.699&0.708&0.699&0.713&0.692&0.680&0.701&0.711&0.696&0.701&0.697&0.700&0.681&0.697&0.697 \\
2.5 $\times 10^{-2}$&  70.0&0.607&0.607&0.607&0.607&0.607&0.613&0.602&0.614&0.593&0.608&0.598&0.609&0.608&0.603&0.607&0.604&0.602&0.611&0.606&0.607&0.590&0.609&0.606&0.608&0.604&0.609&0.609&0.621&0.607&0.607 \\
8.1 $\times 10^{-2}$&  70.0&0.484&0.484&0.484&0.484&0.484&0.481&0.485&0.489&0.480&0.480&0.478&0.494&0.477&0.490&0.477&0.491&0.486&0.480&0.488&0.481&0.508&0.471&0.466&0.483&0.485&0.483&0.476&0.486&0.484&0.484 \\
1.6 $\times 10^{-3}$&  90.0&1.363&1.386&1.349&1.363&1.363&1.390&1.372&1.404&1.381&1.357&1.363&1.385&1.395&1.381&1.426&1.319&1.382&1.358&1.346&1.377&1.372&1.374&1.377&1.359&1.374&1.360&1.381&1.411&1.363&1.363 \\
2.5 $\times 10^{-3}$&  90.0&1.275&1.277&1.274&1.275&1.275&1.278&1.275&1.278&1.259&1.271&1.275&1.274&1.251&1.270&1.300&1.257&1.290&1.277&1.254&1.288&1.258&1.289&1.275&1.282&1.265&1.277&1.281&1.274&1.268&1.282 \\
4.0 $\times 10^{-3}$&  90.0&1.068&1.069&1.068&1.068&1.068&1.068&1.073&1.075&1.068&1.063&1.068&1.083&1.064&1.059&1.070&1.092&1.050&1.084&1.068&1.081&1.071&1.076&1.079&1.073&1.077&1.070&1.092&1.048&1.062&1.075 \\
6.3 $\times 10^{-3}$&  90.0&0.974&0.974&0.974&0.974&0.974&0.974&0.970&0.981&0.960&0.978&0.974&0.989&0.963&0.980&0.972&0.985&0.971&0.978&0.987&0.977&0.968&0.984&0.976&0.978&0.975&0.974&0.997&0.971&0.974&0.974 \\
1.0 $\times 10^{-2}$&  90.0&0.845&0.845&0.845&0.845&0.845&0.836&0.843&0.838&0.830&0.851&0.845&0.857&0.826&0.840&0.834&0.848&0.847&0.847&0.844&0.828&0.839&0.852&0.869&0.849&0.861&0.844&0.837&0.836&0.845&0.845 \\
1.6 $\times 10^{-2}$&  90.0&0.619&0.619&0.619&0.619&0.619&0.634&0.610&0.647&0.592&0.618&0.619&0.623&0.622&0.629&0.616&0.615&0.609&0.619&0.620&0.612&0.596&0.612&0.607&0.622&0.610&0.619&0.616&0.580&0.619&0.619 \\
2.5 $\times 10^{-2}$&  90.0&0.556&0.557&0.556&0.556&0.556&0.555&0.560&0.568&0.557&0.554&0.550&0.548&0.565&0.557&0.554&0.551&0.551&0.563&0.555&0.551&0.544&0.563&0.560&0.559&0.562&0.561&0.559&0.590&0.556&0.556 \\
4.0 $\times 10^{-2}$&  90.0&0.526&0.526&0.526&0.526&0.526&0.528&0.523&0.526&0.518&0.527&0.516&0.526&0.521&0.513&0.528&0.516&0.520&0.529&0.527&0.534&0.526&0.510&0.513&0.523&0.525&0.521&0.523&0.560&0.526&0.526 \\
8.1 $\times 10^{-2}$&  90.0&0.453&0.453&0.453&0.453&0.453&0.456&0.451&0.458&0.443&0.452&0.463&0.476&0.454&0.453&0.451&0.460&0.455&0.454&0.456&0.449&0.466&0.445&0.443&0.452&0.450&0.449&0.443&0.473&0.453&0.453 \\
2.0 $\times 10^{-1}$&  90.0&0.255&0.255&0.255&0.255&0.255&0.251&0.257&0.259&0.253&0.253&0.268&0.305&0.245&0.264&0.233&0.263&0.263&0.235&0.257&0.251&0.296&0.242&0.255&0.254&0.259&0.278&0.243&0.292&0.255&0.255 \\
2.0 $\times 10^{-1}$& 120.0&0.207&0.207&0.207&0.207&0.207&0.213&0.207&0.210&0.203&0.207&0.232&0.212&0.198&0.217&0.196&0.219&0.218&0.198&0.207&0.205&0.245&0.193&0.187&0.206&0.208&0.226&0.197&0.255&0.207&0.207 \\
2.0 $\times 10^{-1}$& 150.0&0.288&0.288&0.288&0.288&0.288&0.286&0.284&0.287&0.280&0.289&0.301&0.296&0.269&0.297&0.273&0.302&0.296&0.278&0.288&0.287&0.310&0.272&0.259&0.288&0.290&0.294&0.294&0.281&0.288&0.288 \\
\hline
\end{tabular}
\end{center}
\vspace{-0.5cm}
\caption{ \it  The extracted $F_2$ values for each systematic
check at values of $Q^2< 100$ {\rm GeV}$^2$ $or$ $y=\frac{Q^2}{xs} < 0.01$.
}
\label{t:allb}
\end{minipage}\end{sideways}
\end{table}
\normalsize

\renewcommand{\thetable}{\Alph{aptab}\arabic{table}}

\begin{table}
\begin{center}
\begin{tabular}{|r|c|r|r|r|} \hline
     &Checks contri- &           &           &            \\
Class&buting to class&$a_{class}$&$b_{class}$&$c_{class}$ \\
\hline
A&11;5;7&2.34&0.96&0 \\
--A&6;8;9&--1.94&--0.83&0 \\
B&1;29&0.126&0&0 \\
--B&2;28&--0.085&0&0 \\
C&12;14&0.153&4.4$\cdot 10^{-3}$&0 \\
--C&3;4;13;15;27&--0.145&--3.4$\cdot 10^{-3}$& 0 \\
D&10;16;26;25&0.132&4.3$\cdot 10^{-3}$& 0.69 \\
--D&17;22;24;23&--0.146&--7.4$\cdot 10^{-3}$&--1.29 \\
E&21&1.62&0.39&0 \\
--E&20&--3.05&--0.77&0 \\
F&19&1.05&0.83& 0 \\
--F&18&--1.00&--0.79& 0 \\ \hline
\end{tabular}
\end{center}
\caption{ \it The parameters of the functions describing the
different classes of systematic uncertainties of $F_2$ (see text).
}
\label{t:app}
\end{table}

\begin{figure}[p]
\begin{center}
\mbox{\epsfxsize=12cm\epsffile{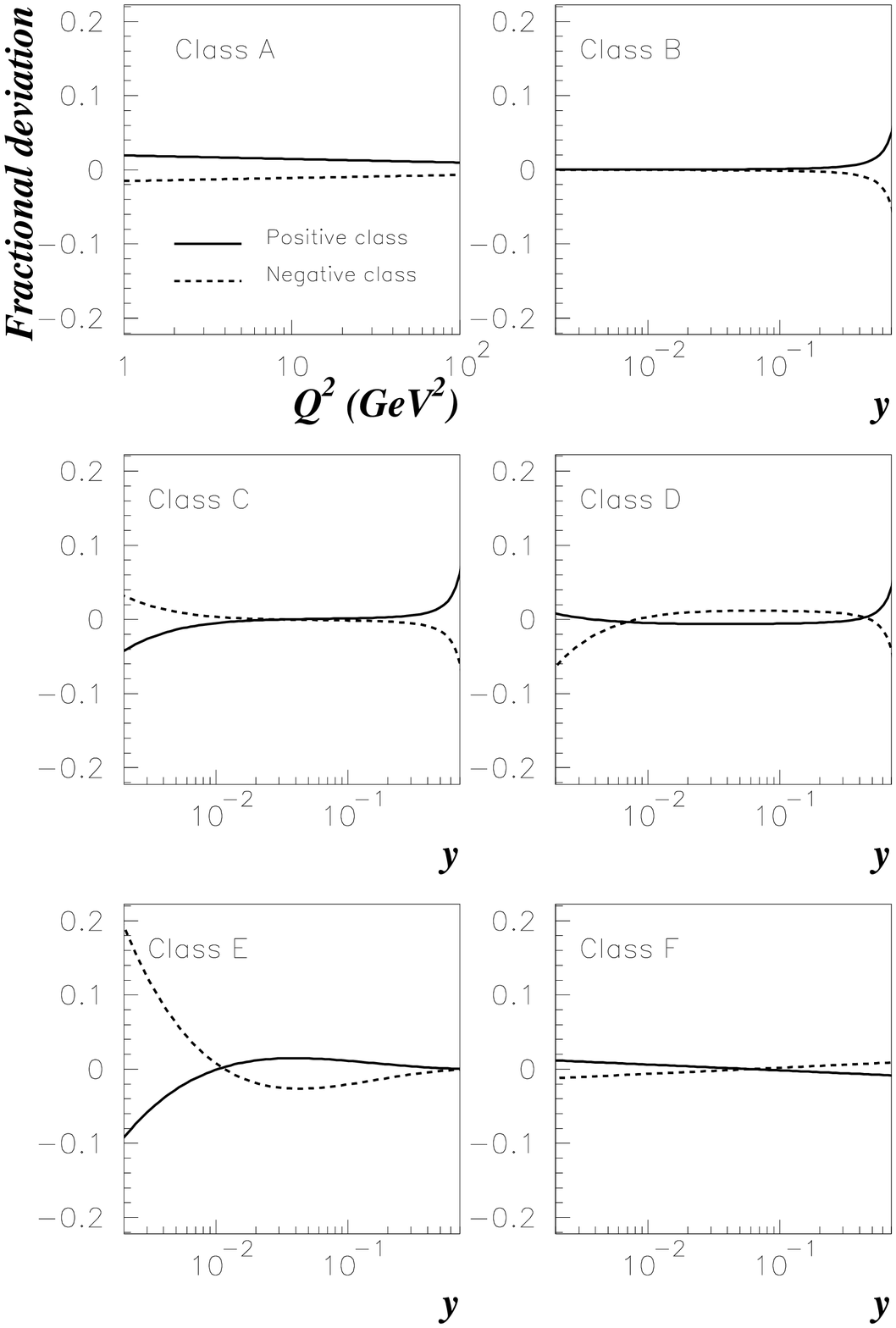}}
\vspace{0.3cm}
\end{center}
\caption[]{\it The $Q^2$ or $y$ dependence of $\frac{\Delta F_2}{F_2}$ for the six classes of
systematic uncertainties (see text).}
\label{f:app}
\end{figure}

\end{document}